\documentstyle[12pt,epsf,epsfig]{article}
\newfont{\thiplo}{msbm10 scaled\magstep 2}
\newfont{\gothic}{eufb10 scaled\magstep 2}
\newfont{\unc}{eurb10}  
\newskip\humongous \humongous=0pt plus 1000pt minus 1000pt
\def\caja{\mathsurround=0pt}
\def\eqalign#1{\,\vcenter{\openup1\jot \caja
        \ialign{\strut \hfil$\displaystyle{##}$&$
        \displaystyle{{}##}$\hfil\crcr#1\crcr}}\,}
\newif\ifdtup

\def\eqright #1\cr{\noalign{\hfill$\displaystyle{{}#1}$}}
\def\eqleft #1\cr{\noalign{\noindent$\displaystyle{{}#1}$\hfill}}
\def\oldreffmt#1{\rlap{[#1]} \hbox to 2\parindent{}}

\def\figfmt#1{\rlap{Figure {#1}} \hbox to 1in{}}

\def\sectioneq{\def\theequation{\thesection.\arabic{equation}}{\let
\holdsection=\section\def\section{\setcounter{equation}{0}\holdsection}}}%

\newcounter{holdequation}

\def\auto{\eqno(\refstepcounter{equation}\theequation)}
\def\begineq #1\endeq{$$ \refstepcounter{equation}\eqalign{#1}\eqno
	(\theequation) $$}
\def\contlimit{\,{\hbox{$\longrightarrow$}\kern-1.8em\lower1ex
\hbox{${\scriptstyle (a\rightarrow0)}$}}\,}
\def\centeron#1#2{{\setbox0=\hbox{#1}\setbox1=\hbox{#2}\ifdim
\wd1>\wd0\kern.5\wd1\kern-.5\wd0\fi
\copy0\kern-.5\wd0\kern-.5\wd1\copy1\ifdim\wd0>\wd1
\kern.5\wd0\kern-.5\wd1\fi}}
\def\centerover#1#2{\centeron{#1}{\setbox0=\hbox{#1}\setbox
1=\hbox{#2}\raise\ht0\hbox{\raise\dp1\hbox{\copy1}}}}
\def\centerunder#1#2{\centeron{#1}{\setbox0=\hbox{#1}\setbox
1=\hbox{#2}\lower\dp0\hbox{\lower\ht1\hbox{\copy1}}}}
\def\lsim{\;\centeron{\raise.35ex\hbox{$<$}}{\lower.65ex\hbox
{$\sim$}}\;}
\def\gsim{\;\centeron{\raise.35ex\hbox{$>$}}{\lower.65ex\hbox
{$\sim$}}\;}
\def\st#1{\centeron{$#1$}{$/$}}
\def\super#1{\ifmmode \hbox{\textsuper{#1}}\else\textsuper{#1}\fi}
\def\textsuper#1{\newcount\holdspacefactor\holdspacefactor=\spacefactor
$^{#1}$\spacefactor=\holdspacefactor}
\def\getcite#1,{\advance\citenumber by1
\def\getcitearg{#1}\def\lastarg{@}
\ifnum\citenumber=1
\ref{#1}\let\next=\getcite\else\ifx\getcitearg\lastarg\let\next=\relax
\else ,\ref{#1}\let\next=\getcite\fi\fi\next}
\def\pom{{\rm P\kern -0.53em\llap I\,}}
\def\spom{{\rm P\kern -0.36em\llap \small I\,}}
\def\sspom{{\rm P\kern -0.33em\llap \footnotesize I\,}}

\relax
\def\contlimit{\,{\hbox{$\longrightarrow$}\kern-1.8em\lower1ex
\hbox{${\scriptstyle (a\rightarrow0)}$}}\,}
\def\upon #1/#2 {{\textstyle{#1\over #2}}}
\relax
\renewcommand{\thefootnote}{\fnsymbol{footnote}} 

\def\mainhead#1{\setcounter{equation}{0}\addtocounter{section}{1}
  \vbox{\begin{center}\large\bf #1\end{center}}\nobreak\par}
\sectioneq
\def\subhead#1{\bigskip\vbox{\noindent\bf #1}\nobreak\par}

\def\til#1{\centeron{\hbox{$#1$}}{\lower 2ex\hbox{$\char'176$}}}
\def\tild#1{\centeron{\hbox{$\,#1$}}{\lower 2.5ex\hbox{$\char'176$}}}
\def\sumtil{\centeron{\hbox{$\displaystyle\sum$}}{\lower
-1.5ex\hbox{$\widetilde{\phantom{xx}}$}}}

\newcommand{\bit}{\begin{itemize}}
\newcommand{\eit}{\end{itemize}}

\newcommand{\beq}{\begin{equation}}
\newcommand{\eeq}{\end{equation}}
\newcommand{\beqa}{\begin{eqnarray}}
\newcommand{\eeqa}{\end{eqnarray}}

\begin{document} 

\begin{titlepage} 

\rightline{\vbox{\halign{&#\hfil\cr
&ANL-HEP-PR-02-033 \cr
&\today\cr}}} 
\vspace{0.25in} 

\begin{center} 
 
{\large\bf CHIRALITY VIOLATION IN  }

{\large \bf QCD REGGEON INTERACTIONS}\footnote{Work 
supported by the U.S.
Department of Energy, Division of High Energy Physics, \newline Contracts
W-31-109-ENG-38 and DEFG05-86-ER-40272} 
\medskip

Alan. R. White\footnote{arw@hep.anl.gov }

\vskip 0.6cm

\centerline{High Energy Physics Division}
\centerline{Argonne National Laboratory}
\centerline{9700 South Cass, Il 60439, USA.}
\vspace{0.5cm}

\end{center}

\begin{abstract} 
The appearance of the triangle graph infra-red axial anomaly
in reduced quark loops contributing to  
QCD triple-regge interactions is studied. 
In a dispersion relation formalism, the anomaly can only
be present in the contributions of 
unphysical triple discontinuities. In 
this paper an asymptotic discontinuity analysis is applied to
high-order feynman diagrams to show that the anomaly does indeed 
occur in sufficiently high-order reggeized gluon interactions. 
The reggeon states involved must contain 
reggeized gluon combinations with the quantum 
numbers of the anomaly (winding-number) current. 
A direct connection with the well-known U(1) problem
is thus established. Closely related diagrams that contribute 
to the pion/pomeron and 
triple pomeron couplings in color superconducting QCD
are also discussed.

\end{abstract}

\renewcommand{\thefootnote}{\arabic{footnote}} \end{titlepage}

\mainhead{1. INTRODUCTION} 

It is commonly believed that non-perturbative 
quark chirality transitions play an important role    
within the QCD bound-state S-Matrix.
Assuming that the theory can be quantized via a suitably defined 
euclidean path-integral\footnote{We note, though, that the elimination 
of unphysical degrees of freedom remains an unsolved problem\cite{vg}.}, 
the chirality transitions
are understood as originating from gauge-dependent
non-perturbative classical solutions with non-trivial topology.
Field configurations of this kind produce zero modes
of the Dirac operator which\cite{aj} 
prevent the gauge-invariant separation of massless fermion fields into  
right- and left-handed components that separately create 
particles and antiparticles. The resulting violation of  
axial charge conservation is described by the 
anomalous divergence equation for the U(1) axial current. 
While many consequences of chirality violation are understood, for example
the generation\cite{gth} of a mass for the $\eta'$, it's full significance  
in determining the non-perturbative 
massless S-Matrix is far from understood. In particular,
the role of chirality violation due to topological gauge fields in 
chiral symmetry breaking is the subject of much debate\cite{pet,smi}.

In this paper, and a companion paper\cite{arw02},
we provide a completely different 
understanding of chirality transitions in the massless, high-energy,
QCD bound-state S-Matrix. No mention is made 
of euclidean path-integral quantization or topological fields.
Rather, as we explain further below, 
our arguments are based directly on the singularity structure
of high-order feynman diagrams
that contribute to the high-energy scattering of bound-states.

It is well established\cite{fkl} 
that when the gauge symmetry of QCD is spontaneously broken 
general high-energy limits (multi-regge limits) of quark and gluon
amplitudes are described perturbatively by reggeon diagrams 
in which the reggeons are simply massive, reggeized, gluons and quarks.
Both $t$- and $s$- channel unitarity are satisfied.
Reggeon interactions are described, in general, 
by ``reduced'' feynman diagrams,  
obtained from underlying diagrams by placing some propagators on-shell.
It is important, however, to distinguish two kinds of interaction vertices. 
The simplest kind are those that describe the repeated 
interaction of reggeons ``propagating'' in a single reggeon channel (for 
which there is only one overall transverse momentum). The well-known BFKL
kernel is, essentially, an example of this kind of vertex. The second kind 
are the vertices that couple different reggeon channels, the simplest being
the triple-regge vertices\cite{gw} that couple three reggeon channels - each
carrying a separate transverse momentum. In the massless 
theory such vertices should contain the couplings of bound-state
reggeons (e.g. pions and nucleons) together with their couplings
to the physical pomeron. Effectively, therefore, vertices of this kind 
determine the bound-states of the theory and their high-energy scattering
amplitudes.

There are, of course, no axial-vector 
currents in the QCD interaction but in the reduced diagrams
providing the crucial triple-regge vertices,
components of an axial-vector interaction can appear. Therefore,
we have suggested\cite{arw01} that, in sufficiently high orders,
chirality violation due to the infra-red triangle anomaly 
should appear in reggeized gluon interactions of this kind.
The purpose of this paper is to finally establish that this is the case. 
It is necessary, however, to study very high-order diagrams. 

We have long believed\cite{arw84} that the massless, bound-state, 
multi-regge, S-Matrix 
should be obtainable from the massive reggeon diagrams 
once the infra-red role of the chiral anomaly is determined.
In previous papers we have outlined\cite{arw98,arw001} how 
(appropriately regularized) anomaly interactions can be 
the essential element that, in combination with 
the infra-red divergences of the massless limit, produce the
``non-perturbative'' 
properties of confinement and chiral symmetry breaking.
We argued that, while the anomaly interactions
cancel when the scattering states are perturbative quarks and gluons,
for compound multiregge states with an appropriate infra-red component
such interactions dominate and infra-red divergences
self-consistently produce the bound-state S-Matrix.
However, to demonstrate this via the 
construction of a full set of multi-regge 
amplitudes is a complicated project which, of necessity,
will involve much abstract multi-regge theory. While this construction 
is still our eventual goal, as an intermediate step, 
we have first developed, in the companion paper to this, 
a calculational method that demonstrates the dynamical role of the anomaly 
while avoiding the (little known)
multi-regge formalism as much as possible. 
Light-cone properties of the anomaly are heavily exploited and
we are able to show how  
both the U(1) and chiral flavor anomaly play essential roles. 

By studying the interaction of (infinite momentum)
axial currents we show\cite{arw02} that, when the 
SU(3) gauge symmetry is partially broken to SU(2), 
U(1) anomaly interactions combine with couplings due to the 
flavor anomaly to produce an infra-red divergent amplitude for
the scattering of Goldstone boson ``pions'' and ``nucleons''. 
The flavor anomaly produces the pion particle poles, 
while the U(1) anomaly produces the high-energy 
coupling of the pions to the exchanged pomeron. After the divergence is
factorized off, as a wee gluon condensate within the scattering states, 
the remaining amplitudes have both confinement and chiral
symmetry breaking. (The wee gluon condensate can be identified directly
with the infra-red component of multi-regge states that 
appears in the multi-regge program.)
It is apparent that the nature of the pomeron
is crucially dependent on chiral symmetry breaking. We anticipate
that restoration of
full SU(3) gauge symmetry will result from randomization
of the SU(2) condensate within SU(3) and that the Critical 
Pomeron\cite{cri} will appear.

The main focus of this paper will be on multigluon reggeon inteactions 
that are most directly
relevant to the general multi-regge program and the pomeron interactions
that emerge. However, as we briefly describe at the end of
this paper, the multiquark/gluon interaction that provides the pion/pomeron 
coupling in \cite{arw02} is very closely related 
to the reggeized gluon interactions that we study.
We will establish, remarkably perhaps, that for the anomaly 
to appear the reggeon states involved must contain 
gluon combinations with the quantum 
numbers of the anomaly (winding-number) current.
The conventional U(1) problem is, therefore, clearly encountered.
We will concentrate on isolating the anomaly via infra-red properties.
Nevertheless, although we will discuss this only briefly at a few key 
points, we expect the infra-red phenomena we discuss 
to be connected to ``ultra-violet'' reggeon interaction 
problems (involving momenta flowing 
around an internal quark loop that are comparable in magnitude to  
large external momenta) where 
short-distance interactions of the winding number current appear
directly. That the anomaly 
is a high-order, many gluon, phenomenon
is not surprising if the anomaly current, containing a 
product of three gluon fields, has to be involved.

Properties of the triangle diagram are discussed 
in detail in the companion paper\cite{arw02},
where a complete set of the the relevant references is given.
For our present purposes we note that 
the massless axial-vector graph has an infra-red divergence  
that involves a  zero four-momentum fermion 
propagator. Both the ``particle'' and ``antiparticle'' poles  
of this propagator contribute to the 
divergence. The coupling at one end of the propagator can be viewed 
as the vertex for production of the particle while simultaneously (and
symmetrically) that
at the other end describes the production of the antiparticle.
If the zero momentum propagator describes a physical
transition it implies that there is, necessarily, an accompanying
``spectral flow'' of the fermion energy spectrum so that
the production of the antiparticle (or the particle)
corresponds to the production of a Dirac hole state, i.e.
the absorption of a particle (antiparticle). In this way, 
the transition is understood as a ``chirality transition''

In Minkowski space the Dirac zero modes due to topological gauge fields 
do indeed produce\cite{aj} spectral flow (with time) of the eigenvalues of
the corresponding (gauge-dependent) ``Hamiltonian''. However, since there 
is no complete non-perturbative Hamiltonian formalism for massless QCD, 
there is no understanding of the full  
consequences of spectral flow\footnote{The conventional wisdom is probably 
that strong coupling confinement effects overwhelm such phenomena
altogether. As we have
emphasized elsewhere, we expect our discussion to apply to a weak coupling
version of massless QCD in which there is, effectively at least, 
an infra-red fixed point.}. The phenomenon we see
is, arguably, the minimum spectral flow that could be present (if there is 
any). Zero momentum fermion states identified initially 
as a particle (within a boundstate) can evolve with time into 
a filled vacuum state of the corresponding Dirac sea 
and, similarly, filled vacuum states can evolve into particles. (The existence
of stable bound states and physical scattering processes in such an
environment is clearly far from trivial!).
In our analysis spectral flow of this kind is directly
introduced by the appearance
of the triangle graph infra-red divergence 
in reggeized gluon interactions. It is interesting that a 
related phenomenon has already been encountered
in next-to-leading order calculations\cite{fl0} of the high-energy
scattering of massless gluons. A massless gluon triangle diagram
occurs in the effective vertex for reggeized gluon exchange and produces 
a ``particle/antiparticle transition'' that for gluons is simply an
unanticipated helicity transition.

A reggeon interaction vertex can be obtained by calculating
the contribution of Feynman diagrams to the simplest multi-regge limit 
in which the vertex appears. In \cite{arw01} we distinguished
two methods for calculating multi-regge amplitudes - the direct calculation
of diagrams in light-cone co-ordinates and the calculation of multiple
asymptotic discontinuities with the subsequent use of an asymptotic dispersion
relation. Although, the two methods should ultimately produce the same 
results, direct light-cone calculations 
are impractical for the problem we are discussing. This is because of the 
large number of diagrams that could contribute
and because the complexity of the diagrams
makes a full discussion of 
whether or not integration contours are truly trapped, in the asymptotic  
limits involved, very difficult. Consequently the asymptotic dispersion 
relation method has to be used. In this paper, therefore, we develop
methods aimed at directly calculating multiple asymptotic discontinuities.

The form of the asymptotic dispersion relation for a given multi-regge 
process is determined by the possible
asymptotic multiple discontinuities that satisfy the Steinmann relation
property (that the discontinuities occur in non-overlapping 
invariant channels).
Such discontinuities are explicitly reflected in the analytic structure of 
asymptotic amplitudes provided by multi-regge theory and, conversely, 
using the dispersion relation, multi-regge
amplitudes can be calculated directly 
from the discontinuities. In \cite{arw02} we 
described how the appearance of the anomaly pole in the elementary three 
current amplitude could be understood as due to an unphysical triangle 
landau singularity appearing (from an unphysical sheet) at the edge of the
physical region. Correspondingly, 
the crucial feature of the high-order amplitudes that produce reggeon 
interactions containing the anomaly
is the presence of unphysical multiple discontinuities that 
satisfy the Steinmann relation property and
approach physical scattering regions 
only asymptotically. 
(This implies that they correspond to contour trappings that would be very 
difficult to demonstrate using direct light-cone calculations.)
Discontinuities of this kind are present in 
complex (imaginary momentum) parts of the asymptotic region for 
sufficiently complicated many-particle multi-regge 
processes, the simplest of which is the full triple-regge region\cite{gw}
that we study in this paper.
Because they
are in non-overlapping channels these discontinuities can 
(and must) consistently 
appear in the asymptotic amplitudes that describe also the 
real physical region behavior. 

The familiar amplitudes that appear in multi-regge
production processes (such as 
those that contribute to the BFKL equation\cite{fkl}) 
do not contain unphysical multiple discontinuities.
Rather they contain only multiple discontinuities
that are naturally interpreted as due to a succession of physical
region on-shell scattering processes. (The necessity for a
physical time-ordering of such processes then determines the absence of
overlapping channel discontinuities.) Because physical region multiple
discontinuities involve only physical amplitudes and physical intermediate
states, when they are calculated using the perturbative amplitudes
of the massless theory, they can not contain chirality transitions
associated with particle/antiparticle ambiguities. Therefore,
when only production processes are 
involved (i.e. at what we might call the BFKL level of multi-regge theory) 
there is no possibility for ``chirality violation''. 

A priori, there is no reason why unphysical multiple discontinuities 
should not contain potential chirality
transitions when calculated perturbatively. Nevertheless, 
the occurence of the infra-red anomaly within such discontinuities
is very subtle.
The divergence is produced by a quark loop that reduces to a triangle
by the placing of many propagators 
on-shell. Of the three propagators associated with the 
triangle diagram, one must  
carry the zero momentum that allows a chirality
transition while the other two carry the same light-like momentum. 
The additional on-shell propagators have to be associated with a 
triple discontinuity in such a way that (when all transverse momenta are 
zero) they also can
all carry the same light-cone momentum (relative to the direction of
the loop momentum). It is obvious that this requirement
can not be satisfied by a physical triple discontinuity and, in fact, it is 
very difficult to satisfy. (As we briefly discuss towards the end of 
this paper, this difficulty is likely to be closely related to
the complexity involved in having local interactions of the
anomaly current appear in the ultraviolet region of reggeon interaction
vertices.) Indeed, we will see that
by itself this requirement is sufficient to
ensure that (in the obtained reggeon interaction)
at least three reggeons are present in each reggeon channel.
Requiring that the spin structure that generates 
the anomaly also be 
present, then further restricts the contributing triple discontinuities 
to those originating from a small class of feynman diagrams.
The discontinuities involved are truly unphysical in that they
correspond to three ``asymptotic pseudothresholds''
(or, in
more technical S-Matrix language, ``mixed $\alpha$ singularities'')
each of which contains particles (effectively)
going in opposite time directions. Not surprisingly, though, this provides
just the right circumstances for the anomaly to appear.

As we noted above, the obtained reggeon interactions 
are of such high order that
the minimum circumstances in which they can occur (between color zero
reggeon states) is when each of the states involved carries
the quantum numbers of the U(1) anomaly current. 
The lower-order diagrams considered in \cite{arw01}
remain valuable to discuss for illustrative processes but the analysis
within this paper shows that they are essentially irrelevant. We
do not give any detailed discussion of further cancelations 
amongst the diagrams we consider. We note, however, that
the signature rule of \cite{arw01} implies that the full vertex 
for three reggeon states, each of which carries
the quantum numbers of the U(1) anomaly current, must vanish.
In the pion/pomeron vertex obtained in \cite{arw02} there
are, in addition to the three gluon reggeons, a 
quark/antiquark pair in the pion, and an additional reggeon in the pomeron.
In the triple pomeron vertex, which we also briefly discuss,  
there is an additional reggeon in each channel.
In following papers we hope to lay out the details 
of the construction of the full multi-regge S-Matrix alluded to above.
For the moment we note only that triple-regge interactions of
the kind we consider here will contribute generally to the vertices 
and interactions of the reggeon bound states that emerge 
and refer to the brief discussion 
in \cite{arw01}, and also to the outline in
\cite{arw001}, for more details. 

\newpage

\mainhead{2. MULTIPLE DISCONTINUITIES AND THE STEINMANN RELATIONS.} 

\subhead{2.1 Physical Region Discontinuities}

The Steinmann relations originated in axiomatic field theory\cite{ste}. They 
(essentially) describe the restrictions that the time-ordering of interactions
places on the combinations of intermediate states that can occur in a
scattering process. For on-shell S-Matrix amplitudes their significance is
most immediately appreciated in the approximation that we ignore higher-order
Landau singularities and consider only the normal threshold branch points (and
stable particle poles) that occur in individual channel invariants. The
Steinmann relations then say that simultaneous thresholds (and/or poles) can
not occur in overlapping channels. (Channels overlap if they contain a common
subset of external particles.) As a result an $N$-point amplitude has at most
$N-3$ simultaneous cuts (or poles) 
in distinct invariants. The possible combinations of
cuts can be described by tree diagrams with three-point vertices in which each
internal line corresponds to a channel invariant in which there is a cut due
to intermediate state thresholds - as illustrated in Fig.~2.1 for the 7-point
amplitude.
\begin{center}
\epsfxsize=4in
\epsffile{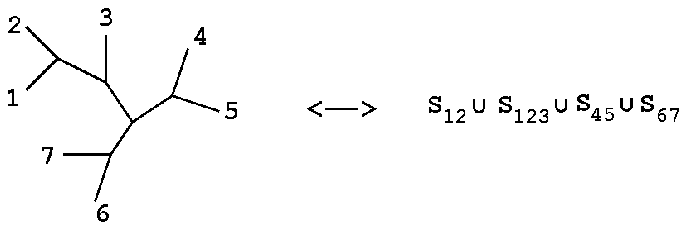}

Fig.~2.1 A Tree Diagram Representing Simultaneous Invariant Cuts.
\end{center}
(As usual, $s_{12}=(P_1+P_2)^2~, s_{123}=(P_1+P_2+P_3)^2~,$
etc.) The set of all combinations of thresholds (and poles) allowed by the 
Steinmann relations is the basic singularity structure of all 
scattering amplitudes. The higher-order Landau singularities are 
believed\cite{arw00} to emerge from the normal
thresholds in a manner that, for most purposes, makes them a secondary effect.

Conversely, the combination of cuts represented 
by a particular tree diagram can
be directly associated with a set of physical scattering processes. As
illustrated in Fig.~2.2, this is the set of all processes (involving all the
external particles of the diagram as either ingoing or outgoing particles) in
which it is kinematically possible for all of the internal lines to be
replaced by physical 
multiparticle states\footnote{We do not distinguish processes in
which ingoing and outgoing particles are interchanged via CPT conjugation}.
\begin{center}
\epsfxsize=6in
\epsffile{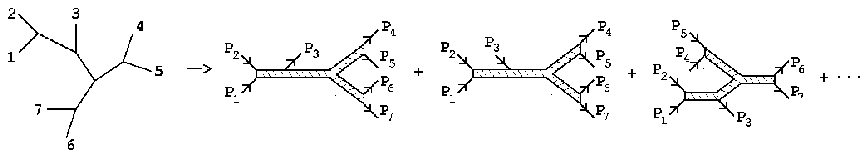}

Fig.~2.2 Physical Scattering Processes Corresponding to Fig.~2.1.
\end{center}
The hatched segments represent physical intermediate states that, if they are 
all placed on shell, give (essentially) the associated multiple discontinuity.

The Steinmann relations play a fundamental role in multi-regge theory. It is 
possible to show\cite{arw00} that in a physical multi-regge asymptotic region 
the analytic structure of scattering amplitudes can be treated as if only 
normal thresholds satisfying the Steinmann relations were present. In effect,
higher-order Landau singularities are suppressed. This has the very important
consequence that only the normal threshold cuts in individual channel 
invariants need be represented by multi-regge
asymptotic formulae. Furthermore, if we consider only the multi-regge limits
accessible in $2 \to M$ production processes, it can be shown that the maximal
number (M-1) of simultaneous thresholds is encountered asymptotically only in
physical regions. This is a generalization of the cut-plane analyticity
property familiar from elastic scattering.

\subhead{2.2 Unphysical Multiple Discontinuities}

If we consider the multi-regge regions of $M \to M'$ scattering amplitudes
($M,M' \geq 3$) there is a significant change. To understand the point 
involved consider the simplest case of the tree diagram of
Fig.~2.3. At first sight this diagram corresponds only to the 
$2 \to 4$ production processes shown. 
\begin{center}
\epsfxsize=5in
\epsffile{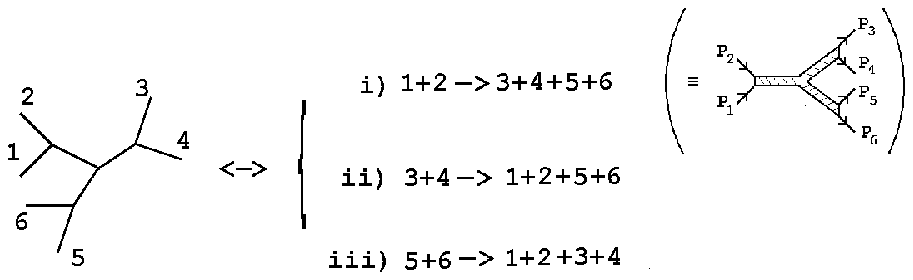}

Fig.~2.3 A Tree Diagram and Corresponding Physical Scattering Processes.
\end{center}
The three distinct scattering processes are distinguished 
by different constraints on the invariants, i.e.
$$
\eqalign{\hbox{i)}& ~\sqrt{s_{12}} ~> ~\sqrt{s_{34}}+\sqrt{s_{56}}~,~~ \cr
\hbox{ii)}&~ \sqrt{s_{34}} ~> ~\sqrt{s_{12}}+\sqrt{s_{56}}~,~~\cr
\hbox{iii)}&~ \sqrt{s_{56}} ~>~ \sqrt{s_{12}}+\sqrt{s_{34}} }
\auto\label{mag}
$$
We can also regard the three processes involved as distinguished by the 
selection of one
pair of particles as incoming, which then must have energy larger than the
sum of the subenergies of the other two pairs, which are necessarily in the 
outgoing state.

We may wonder about the symmetric asymptotic region in which
$$
\sqrt{s_{12}} ~\sim ~\sqrt{s_{34}}~\sim~ \sqrt{s_{56}} ~~\to~~ \infty
\auto\label{mag1}
$$
There are no physical scattering processes in this region. However,
the three processes of (\ref{mag}) are described by the same (analytically
continued) amplitude and so analytic continuation from each of the 
physical regions implies that such cuts must be present. 
It is, perhaps, natural 
that a triple discontinuity should exist that is 
symmetric with respect to the three processes of Fig.~2.3. Apparently, though,
the symmetry requirement could only be satisfied if all the external
particles are in the 
final, or initial, state. In fact, as we discuss further 
in the next Sections, 
if we allow particles to carry complex momenta, a positive 
value for a two-particle energy invariant can  
be achieved by a combination of an ``incoming'' and an 
``outgoing'' particle in that they carry opposite sign, but imaginary, 
energies. Therefore, in the symmetric region it is possible for the 
three cuts of Fig.~2.3 
to be present if each is associated with such a combination.    
We will show in the following that
there are unphysical processes (with imaginary momenta) in this region that
do produce a triple discontinuity of this kind and we will refer to it as an 
``unphysical triple discontinuity''.

Since the external particles for each cut are both  
ingoing and outgoing it is, perhaps, not surprising that intermediate states 
appear that also involve
such combinations. Indeed, we will see that 
this is how a triple discontinuity can contain the 
``particle - antiparticle'' transitions that ultimately provide  
the massless chirality transitions that we are looking for. 
Since the complex momentum
part of (\ref{mag1}) is contained in the triple-regge
asymptotic region, a triple 
discontinuity of this kind is just what we are looking for.
The importance of the triple-regge region is that it is the simplest
multi-regge limit in which the vertices appear that provide the couplings
of bound-state regge
poles such as the pomeron or the pion. For higher-point $M \to M'$
amplitudes there is a wide range of unphysical multiple discontinuities 
satisfying the Steinmann relations. Bound-state scattering amplitudes can thus
appear in which the anomaly is a crucial element.

\newpage

\mainhead{3. THE PHYSICAL REGION ANOMALY AND THE 
TRIPLE-REGGE DISPERSION RELATION}

\subhead{3.1 The Triple Regge Limit and Maximally Non-Planar Diagrams} 

In our previous paper\cite{arw01} we studied the full triple-regge
limit\cite{gw} of 
three-to-three quark scattering. 
If we denote the initial momenta as $P_i~, ~i=1,2,3$, and the final momenta 
as $- P_{i'} = P_i + Q_i, ~i=1,2,3$,
the triple-regge limit can be realized, within the physical region, by taking 
each of $P_1,~P_2$ and $P_3$ large 
along distinct light-cones, with the momentum transfers $Q_1, Q_2$ and $Q_3$
kept finite, i.e.
\newline \parbox{3.1in}{
$$
\eqalign{ P_1~\to&~ P_{1^+}~= ~(p_1,p_1,0,0)~,~~p_1 \to \infty \cr
P_2~\to&~ {P_2^+}~= ~(p_2,0,p_2,0)~,~~p_2 \to \infty \cr
P_3~\to&~ {P_3^+}~= ~(p_3,0,0,p_3)~,~~p_3 \to \infty  }
$$}
\parbox{2.9in}{
$$ \eqalign{
~~~q_1=Q_1/2~\to&~~ (\hat{q}_1,\hat{q}_1,q_{12},q_{13})\cr
~~~q_2=Q_2/2~\to&~ ~(\hat{q}_2,q_{21},\hat{q}_2,q_{23})\cr
~~~q_3=Q_3/2~\to&~~(\hat{q}_3,q_{31},q_{32},\hat{q}_3)}
\auto\label{np3}
$$}
Momentum conservation gives 
a total of five independent $q$ variables which, along  
with $p_1, p_2$ and $p_3$, give the necessary eight variables. The definition
of the triple-regge limit in terms of angular variables is 
given in \cite{arw01}. For our present purposes the above definition in 
terms of momenta will be sufficient. This will alow us to avoid the
extra complication of defining helicity angles, helicity-pole limits
etc. The asymptotic behavior involved must hold 
for all complex values of the large momenta,
including the additional physical regions reached by reversing the 
signs of the $p_i$.

In \cite{arw01} we also studied feynman diagrams that contain a 
closed quark loop and generate
triple-regge reggeized gluon interactions containing the loop.
To set the context for the present paper we briefly review the results.
We considered the lowest-order amplitudes in which 
the anomaly could 
potentially appear and, in particular, studied ``maximally non-planar'' 
diagrams of the kind shown in Fig.~3.1(a). 
\begin{center}
\leavevmode
\epsfxsize=4.5in
\epsffile{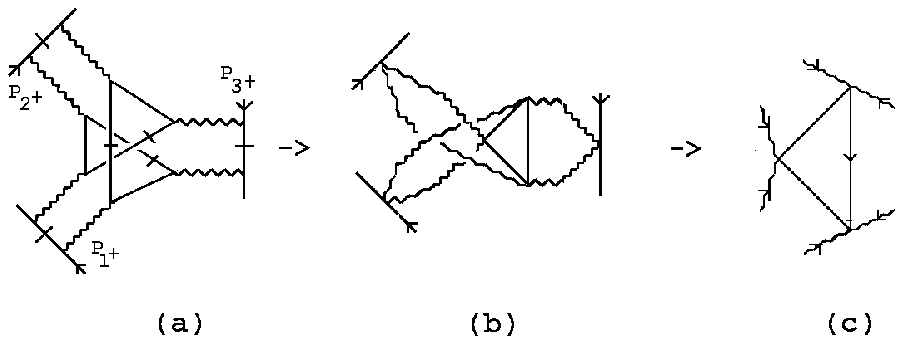}

Fig.~3.1 A maximally non-planar diagram and 
the triangle diagram reggeon interaction produced.   
\end{center}
(Throughout this paper we adopt the usual convention that solid and wavy lines 
respectively represent a quark and a gluon.
We have reversed the direction of $P_3$ relative 
to the notation of \cite{arw01} in order to have a completely symmetric
notation.) The leading asymptotic contributions come from 
regions of gluon loop integrations where some of the 
propagators in the quark loop and the scattering quark systems
are on-shell. We discuss the determination of which propagators can be
on-shell below. For the moment we consider the possibility, discussed at 
length in \cite{arw01}, that the 
on-shell lines are those that are hatched in Fig.~3.1(a). We will eventually 
conclude that this combination of on-shell propagators can not produce a 
reggeon interaction with a physical region anomaly divergence, 
even though it does produce a triangle
diagram interaction. As we will see, the crucial issue is not just which 
propagators are placed on-shell but also which pole (``particle'' or 
``antiparticle'') is involved. (As the discussion in the previous 
Section suggested,
for the unphysical discontinuities, with which we will ultimately 
be concerned, the answer to this question is not unambiguous.)
In the following we initially 
ignore this subtlety. As it emerges in 
our discussion it will become clear that it is a vital part 
of the search for further diagrams which do produce an interaction 
containing the anomaly.

If the hatched on-shell propagators are used to carry out
light-like longitudinal momentum integrations the integrals over gluon loop
momenta reduce to two-dimensional ``transverse'' integrals over
spacelike momenta, as illustrated by Fig.~3.1(b). The transverse 
plane (and orthogonal light-like momenta)
can, in general, be chosen differently in each $t$-channel. 
If $Q_{i\perp}$ is the projection of $Q_i$ on the corresponding 
transverse plane, the leading asymptotic contribution then has the form
$$
\eqalign{ ~~~~~P_{1^+}~ P_{2^+}~ P_{3^+}~
\prod_{i=1}^3 \int & { d^2 k_{i1}d^2 k_{i2}\over  k_{ i1}^2  k_{i2}^2}  
~~ \delta^2 (Q_{i\perp} -  k_{i1} -  k_{i2})~G^2_i(k_{i1},k_{i2},\cdots)
\cr &~~~~~~~~\times ~ R^6(Q_1,Q_2,Q_3,
k_{11}, k_{12}, \cdots )} \auto \label{211}
$$
where $ R^6(Q_1,Q_2,Q_3,k_{11}, k_{12}, \cdots )$  
can be identified with the ``reduced'',
or ``contracted'', feynman diagram of Fig.~3.1(c). If we write
$$
k_{i1} ~= ~q_i + k_i~, ~~~~ k_{i2} ~= ~q_i - k_i~,
\auto\label{dki}
$$
then we showed in \cite{arw01}
that (with a particular choice of transverse planes)
$$
\eqalign{ &R^6(q_1,q_2,q_3,k_1,k_2,k_3) ~=\cr
& \int d^4 k  {  Tr \{ 
\gamma_5 \gamma^{-,-,+} (\st{k}+ \st{k}_1 + \st{q}_2 +\st{k}_3) 
\gamma_5 \gamma^{-,-,-} ~\st{k}~ 
\gamma_5 \gamma^{-,-,-}(\st{k}- \st{k}_2 + \st{q}_1 + \st{k}_3 ) \}
\over  (k + k_1 + q_2 + k_3 )^2  
~k^2 ~
 (k - k_2 + q_1 + k_3)^2 }  ~+ ~ \cdots }
\auto\label{580}
$$
where
$$
\gamma^{\pm,\pm,\pm} ~=~ \gamma^{\mu}\cdot n^{\pm,\pm,\pm}_{ \mu} ~,~~~~
n^{\pm,\pm,\pm}_{\mu} ~= ~ (1,\pm1,\pm1,\pm1)
\auto\label{g64}
$$
The contributions to $R^6$ not shown explicitly in (\ref{580}) do not have
a $\gamma_5$ at all three vertices of the triangle diagram. 
As we will discuss again in Section 5, the particular
$\gamma$-matrix projections appearing depend on the choice of transverse
co-ordinates. If the anomaly is present in $R^6$, however, we expect it to be
independent of this choice. We should
emphasize that while we have written (\ref{580}) as a function of
four-dimensional momenta, the $k_i$ are restricted to be 
two-dimensional spacelike momenta (plus longitudinal components
determined by the mass-shell conditions for the on-shell quarks) and 
the $q_i$ have the restricted form given by (\ref{np3}). These restrictions 
play a crucial role in determining whether the anomaly can occur in a physical
region reggeon interaction.

\subhead{3.2 A Reggeon Diagram Amplitude}
 
For completeness, we give a brief description (full details can be found in
\cite{arw01}) of how a reggeon vertex is extracted from (\ref{211}).
A reggeon diagram amplitude 
that represents right-hand cuts in the unphysical triplet 
$\{s_{13'}, s_{32'}, s_{21'}\}$ and has two reggeons in each $t$-channel, 
each with trajectory $\alpha(t) = 1 +~O(g^2)$, has the form\cite{arw01}
$$
\eqalign{ &~~\prod_i\int { d^2k_i \over 
sin \pi \alpha (k_i^2)  sin \pi \alpha ((Q_i -k_i)^2)  } 
~~~~\beta(k_1,k_2,k_3,Q_1,Q_2,Q_3)\cr 
&\biggl[ ~(s_{13'})^{[\alpha (k_1^2)+\alpha ((Q_1 -k_1)^2) + 
\alpha (k_3^2)+\alpha ((Q_3 -k_3)^2) -
\alpha (k_2^2)-\alpha ((Q_2 -k_2)^2) -1]/2} \cr
& ~~~~~~~~ (s_{32'})^{[\alpha (k_3^2)+\alpha ((Q_3 -k_3)^2) + 
\alpha (k_2^2)+\alpha ((Q_2 -k_2)^2) -
\alpha (k_1^2)-\alpha ((Q_1 -k_1)^2) -1]/2} \cr
& ~~~~~~~~~~~~(s_{21'})^{[\alpha (k_1^2)+\alpha ((Q_1 -k_1)^2) + 
\alpha (k_2^2)+\alpha ((Q_2 -k_2)^2) -
\alpha (k_3^2)-\alpha ((Q_3 -k_3)^2) -1]/2} \biggr]~  \cr
&~~~~\biggl[[sin {\pi \over 2} [\hbox{${\scriptstyle\alpha (k_1^2)
+\alpha ((Q_1 -k_1)^2) + 
\alpha (k_3^2)+\alpha ((Q_3 -k_3)^2) -
\alpha (k_2^2)-\alpha ((Q_2 -k_2)^2)}$}] \cr
&~~~~~~~~~~ sin {\pi \over 2} [\hbox{${\scriptstyle \alpha (k_3^2)
+\alpha ((Q_3 -k_3)^2) + 
\alpha (k_2^2)+\alpha ((Q_2 -k_2)^2) -
\alpha (k_1^2)-\alpha ((Q_1 -k_1)^2)}$} ] \cr
& ~~ ~~~~~~~~~~~~sin {\pi \over 2} [\hbox{${\scriptstyle \alpha (k_1^2)
+\alpha ((Q_1 -k_1)^2) + 
\alpha (k_2^2)+\alpha ((Q_2 -k_2)^2) -
\alpha (k_3^2)-\alpha ((Q_3 -k_3)^2)}$} ] \biggr]^{-1}}
 \auto \label {2ra1}
$$
$$
\centerunder{$\sim$}{\raisebox{-6mm}{$ g^2 \to 0$}}~
(s_{13'})^{1/2}(s_{32'})^{1/2}(s_{21'})^{1/2}
~\prod_i\int{ d^2k_i \over 
k_i^2  (Q_i -k_i)^2  }
~~\beta_(k_1,k_2,k_3,Q_1,Q_2,Q_3) ~~
\auto\label{2ra}
$$
(The generalization of this formula to include more reggeons in any of
the channels should be obvious.)
Taking the triple discontinuity in $s_{13'}$, $s_{32'}$
and $s_{23'}$ of (\ref{2ra1})
removes the poles due to the sine factors in the second
square bracket, but leaves the $g^2 \to 0$ limit unchanged. Since   
the triple discontinuity is unphysical and
of the kind discussed in the previous Section,
according to the discussion in
\cite{arw01}, the ``six-reggeon interaction vertex''
$\beta_(k_1,k_2,k_3,Q_1,Q_2,Q_3)$ could contain the anomaly.

Writing
$$
P_{1^+} P_{2^+} P_{3^+}~\equiv ~(s_{13'})^{1/2}(s_{32'})^{1/2}(s_{21'})^{1/2}
\auto\label{p+inv}
$$
and comparing with (\ref{2ra}) we see that it would be straightforward
to identify (\ref{211}) as a lowest-order contribution to such a
reggeon diagram amplitude if the reduced feynman
diagram amplitude of Fig.~3.1(c) is identified as a reggeon vertex, i.e.
$$
R^6(Q_1,Q_2,Q_3, k_{1},Q_1- k_{1},\cdots)~\equiv ~
\beta(k_1,k_2,k_3,Q_1,Q_2,Q_3)
\auto\label{6rv}
$$
Note, however, that while the right-side of (\ref{p+inv})
clearly has a triple discontinuity in $\{s_{13'}, s_{32'}, s_{21'}\}$,
the left-side does not. The equivalence of the two sides is only determined
if higher-order terms in (\ref{2ra1}) appear and add to (\ref{211}) in 
the appropriate manner. Such terms are contributed by what we refer to 
in the following as reggeization diagrams. Note, also, that for parts of $R^6$ 
(not including $\gamma_5$ couplings) higher-order terms would be expected to  
appear implying that, one or more of,  the transverse integrals in (\ref{211})
should be interpreted as arising from the trajectory function terms in
(\ref{2ra1}). Such parts of $R^6$ would then be interpreted as providing
interaction vertices for fewer reggeons. 

The amplitude (\ref{580}) representing Fig.~3.1(c) 
is the full four-dimensional
triangle diagram amplitude except that special $\gamma$-matrices
appear at the vertices and
only combinations of (essentially) two-dimensional transverse  
momenta flow through the diagram.  
It is shown in \cite{arw01} that the $\gamma$-matrix
couplings are appropriate to produce the anomaly but, as we discuss next, 
whether the necessary 
momentum configuration can occur within a physical region 
and provide a physical region infra-red divergence is a non-trivial
and subtle question that depends crucially on the choice of propagator
poles used to put lines on-shell.

\subhead{3.3 The Basic Anomaly Process}

As we discuss further in Section 5, the divergence of 
the (massless) triangle diagram occurs\cite{arw02} 
when a single light-like momentum flows through the diagram and
all other momenta are spacelike and scaled to zero.
Such a momentum configuration for the reggeon interaction $R$ is 
realized by that of the 
full feynman diagram shown in Fig.~3.2(a).
If we label the momenta entering the reggeon interaction as in Fig.~3.2(b)
an explicit configuration for Fig.~3.2(a), discussed in \cite{arw01}, is
$$
q_1-k_1~=~(2l,2l,0,0)~, ~~~
q_2-k_2~=~(-2l,0,-2l,0)
\auto\label{chm1}
$$ 
together with
$$
\hat{q}_1=- \hat{q}_2 =l ~~~~q_{13}=-q_{23} ~~~~ q_{12}=q_{21}=0
\auto\label{chm2}
$$
This determines $k_1$ and $k_2$ and also gives
$$
q_3~=~- (q_1+q_2)~= ~(0,-l,l,0)
\auto\label{chm3}
$$
If we then take
$$
k_3~=~l(0, 1 -2~ \cos{\theta}~, 1 - 2~\sin{\theta}~, 0) 
\auto\label{chm4}
$$
the light-cone momentum 
$$
-~2l(1, \cos{\theta},\sin{\theta}, 0)
\auto\label{chm30}
$$
flows along the two vertical non-hatched lines in Fig.~3.2(b). 
It is straightforward
to check that all three of the hatched lines are on mass-shell.
\newline \parbox{4.3in}{
\begin{center}
\epsfxsize=4.1in
\epsffile{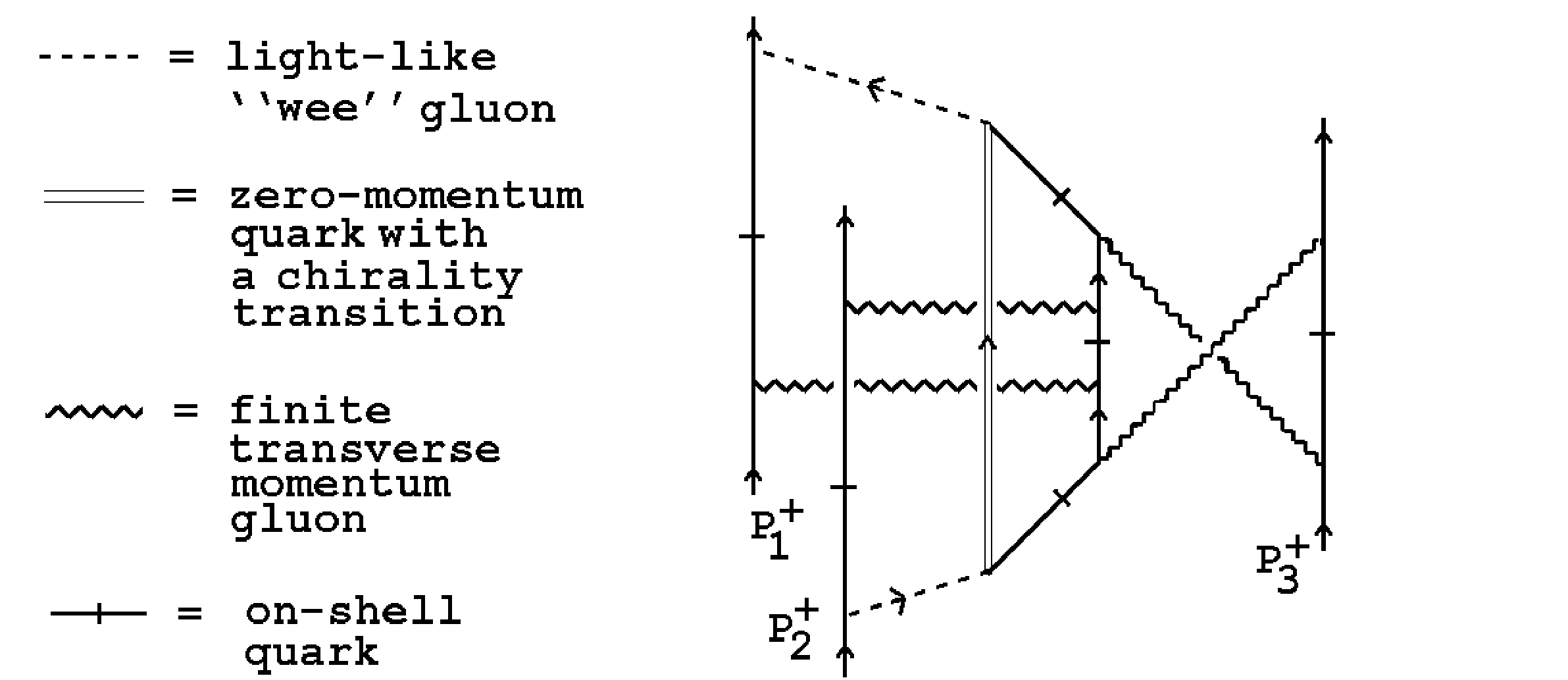}
\newline (a)
\end{center}}
\parbox{1.7in}{
\begin{center}
\leavevmode
\epsfxsize=1.5in
\epsffile{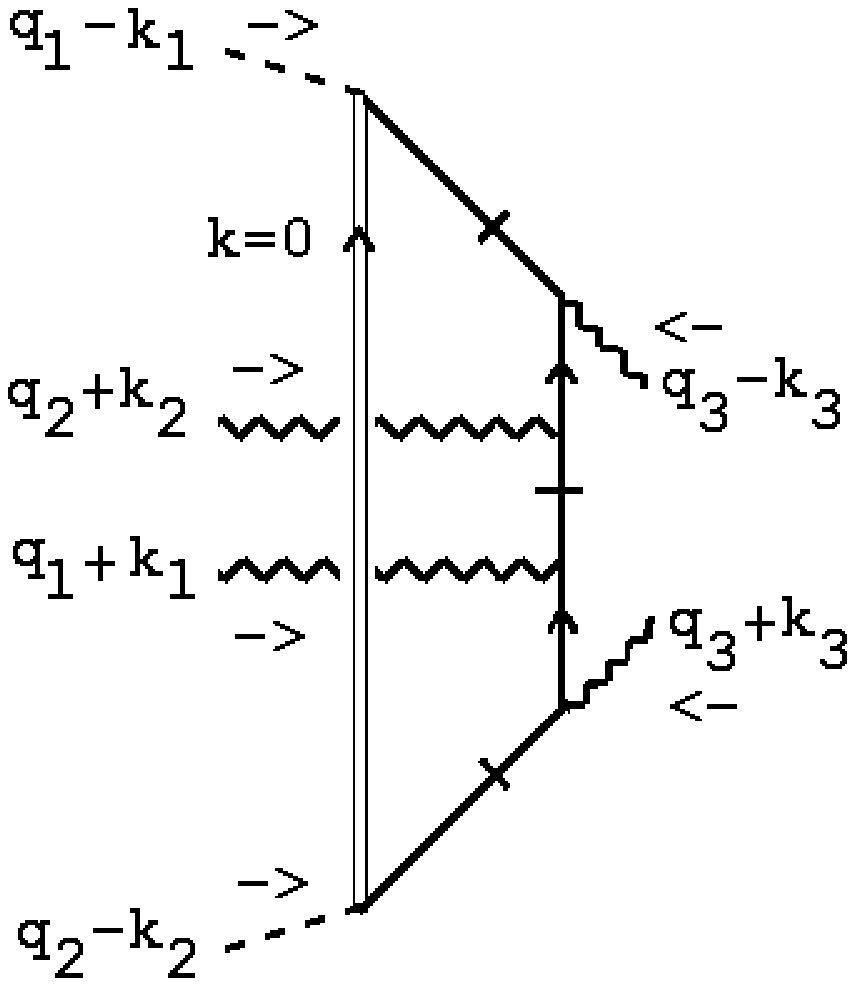}
\newline (b)
\end{center}}
\begin{center}
Fig.~3.2 The basic anomaly process.
\end{center}
If spacelike momenta of $O(q)$ are added to the momentum 
configuration (\ref{chm1})-(\ref{chm30}) and the limit $q \to 0$ is taken,
the anomaly divergence occurs. (We will discuss this
in more detail in Section 5.)

Apart from the reversal of direction for $P_3$,  the process represented by 
Fig.~3.2(a) 
is what we called ``the basic anomaly process'' in \cite{arw01}.
The zero momentum quark is produced by one ``wee
gluon'' and absorbed by the other, allowing the chirality
transition produced by the anomaly
to compensate for a spin flip of the antiquark.  
Note, however, that when the wee gluons are massless,
the scattering processs represented by Fig.~3.2 is physical only when the 
quark and antiquark involved are also massless. In addition, as we 
noted in the Introduction (and discussed in more detail in \cite{arw02}, 
the anomaly infra-red divergence involves both poles of 
the zero momentum quark propagator. Moreover,
the vertices coupling to the propagator 
should, a priori, be symmetrically interpreted
as describing either the simultaneous production of the two states 
in the propagator or their simultaneous 
absorption. When, as is the case in \cite{arw02}, 
the infra-red divergence analysis  
used to define physical states and amplitudes 
requires that the massless scattering enter the physical region
with the time ordering implied by Fig.~3.2,
the presence of a non-perturbative background gauge field is effectively
implied. The background field would be needed to
produce the necessary spectral flow at one vertex 
that is required to interpret the process as a chirality transition. 
 
While the required mass-shell
conditions are indeed satisfied by (\ref{chm1})-(\ref{chm30}), 
there is a problem.
With the momenta given by  (\ref{chm1})-(\ref{chm30}),
the energy component of each of the three hatched lines in Fig.~3.2(b)
has the same sign.
Since the exchanged gluons carry only spacelike momenta, it is clear that 
this must be the case. We will see that this is a very difficult
configuration to obtain within a reggeon interaction. We can emphasize
the problem by 
letting $l \to 0$ while simultaneously making a boost $a_z(\zeta)$
such that $ l \cosh\zeta =n$ is kept finite. 
(This is what is done in \cite{arw02}.)
If we then take all transverse momenta to zero, we obtain 
$$
q_1-k_1~\to ~(2n,0,0,2n)~, ~~~
q_2-k_2~\to~(-2n,0,0,-2n)
\auto\label{chm101}
$$ 
and all the on-shell propagators carry the same light-like momentum
(with respect to the direction of the loop momentum). 
Effectively, then, the on-shell states in the loop must be in a 
symmetric  light-like situation. 
(This implies that if the zero momentum state is an antiquark (quark), 
all hatched lines must be quarks (antiquarks).)

As we already remarked on in the Introduction, and as is discussed at length 
in \cite{arw01}, the only practicable calculational method
to determine whether
a given combination of on-shell lines contributes to the triple-regge behavior
(after all diagrams are added) is the dispersion relation method that 
we outline very briefly below. In this 
approach all on-shell lines 
in a reggeon interaction result directly from the taking of a 
triple asymptotic 
discontinuity. ``Real part'' interactions with the 
same on-shell lines may be
generated when the full dispersion relation is written
or, equivalently, multi-regge theory is used\cite{arw01} 
to convert the triple discontinuity to a full amplitude. 

To have all on-shell lines carry the same
light-like momentum (around a loop) in a 
multiple discontinuity is a very restrictive requirement. 
The essential point becomes clear if we consider a physical
region double discontinuity which gives the cut lines of Fig.~3.1(a),
as in Fig.~3.3(a).
\begin{center}
\epsfxsize=4.5in
\epsffile{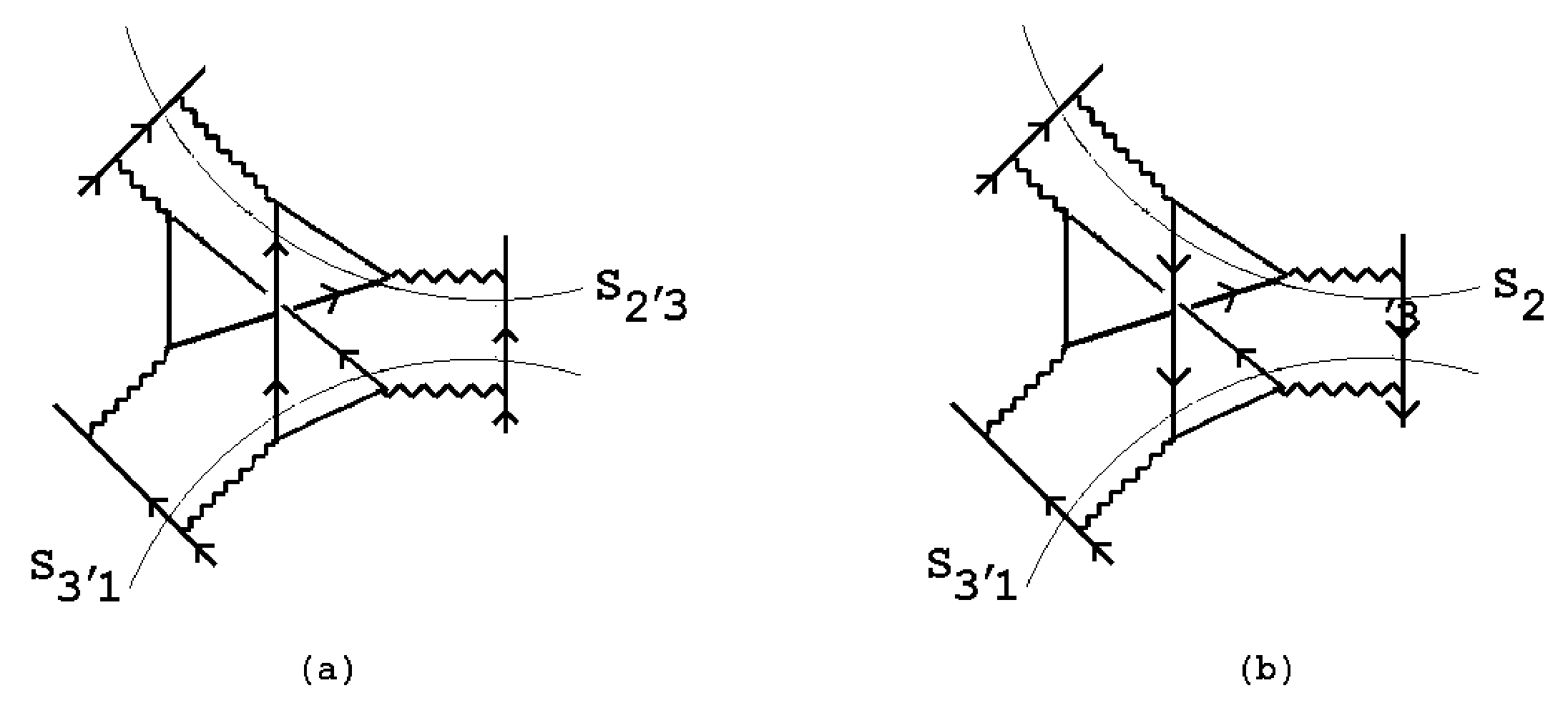}

Fig.~3.3 (a) A physical threshold double discontinuity (b) 
A pseudothreshold double discontinuity.
\end{center}
If all cut lines carry light-like
momenta, the positive direction for the energy component
must be as indicated by the arrows in Fig.~3.3(a). Obviously  
the direction can not be the same, relative to the internal
loop momentum, for all cut lines. Nevertheless, this is an essential
requirement if a reggeon interaction is 
to contain a physical region divergence
produced by the anomaly (i.e. some variant of the ``basic anomaly process''
must be involved). We obtain what we require if we reverse one internal 
line and one external line 
(to make the cutting completely symmetric) as in Fig.~3.3(b). 
Since both cuts now involve both forward and backward going (in time)
particles it is clear that we must have a combination of pseudothresholds,
just as suggested in the previous Section, that can occur only in an 
unphysical region of momentum space.

We already recognized in \cite{arw01} that 
the necessary triple discontinuity is not present in the diagram of Fig.~3.1
but we suggested that nevertheless it may be present in related 
higher-order reggeization
diagrams that produce the reggeization of the gluons. In which
case, the basic anomaly 
process of Fig.~3.2 would be required as a real part interaction.
In fact, we will show in the remaining part of
this paper that this is not the case. Instead, the requirement that all 
cut lines are treated symmetrically will require more
wee gluons and ultimately will require that reggeon
interactions with the quantum numbers of the winding number current must
be involved. Also, as we already anticipated in the previous Section
the discontinuities involved must be unphysical.
 
As we discussed in \cite{arw01},
we do not expect the anomaly divergence to be present in the scattering of
elementary quarks and/or gluons after all diagrams are summed. Rather,
we expect it to be present when  
the basic process is generalized to describe the scattering of 
the particular multi-regge states that ultimately form
bound states, and then only in color superconducting QCD. 
In \cite{arw02} it is clear that the relevant bound-states are just the
Goldstone bosons produced by chiral symmetry breaking. 
The corresponding $G_i$ will then appear in 
a generalization of (\ref{211}) and the 
wee gluons involved will be a crucial characteristic of  
scattering states. Also the chirality transitions produced (and the implicit 
spectral flow) will be an essential part of scattering processes.

\subhead{3.4 The Triple-Regge Dispersion Relation}

In general, an asymptotic dispersion relation\cite{arw00} gives the leading 
multi-regge behavior of an amplitude as
a sum over multiple discontinuity contributions 
allowed by the Steinmann relations. For the particular case (described in
detail in \cite{arw01}) of 
the triple-regge 
behavior of a six-point amplitude we can write 
$$
M_6(P_1,P_2,P_3,Q_1,Q_2,Q_3)~ =~ 
\sum_{\cal C} M_6^{\cal C}(P_1,P_2,P_3,Q_1,Q_2,Q_3)
~+~M_6^0~,\auto\label{dis}
$$
where $M_6^0$ contains all non-leading 
triple-regge behavior, double-regge behavior, etc. and the sum is
over all triplets ${\cal C}$ of 
asymptotic 
cuts in non-overlapping (large) invariants. For each triplet ${\cal C}$, 
say ${\cal C}= (s_1,s_2,s_3)$, we can write 
$$  
\eqalign{M_6^{\cal C}(P_1,P_2,P_3,Q_1,Q_2,Q_3)~=~{1\over (2\pi i)^{3}}  &~~\int
ds'_1 ds'_2 ds'_{3} ~~{\Delta^{\cal C} \over
(s'_1-s_1)(s'_2-s_2)(s'_{3}-s_{3})} }
\auto\label{dis2}
$$
where $\Delta^{\cal C}$ is the triple discontinuity.

The triple discontinuities are of three
kinds, described by the tree diagrams of Fig.~3.4. 
There are 24 corresponding to Fig.~3.4(a),
12 to Fig.~3.4(b), 
and 12 of the  Fig.~3.4(c) kind - including those described by Fig.~2.3. 
Those of Fig.~3.4(a) and (b), occur in the physical
regions, while those corresponding to Fig.~3.4(c) are all unphysical 
triple discontinuities of the kind discussed in the last Section.
\begin{center}
\leavevmode
\epsfxsize=2.2in
\epsffile{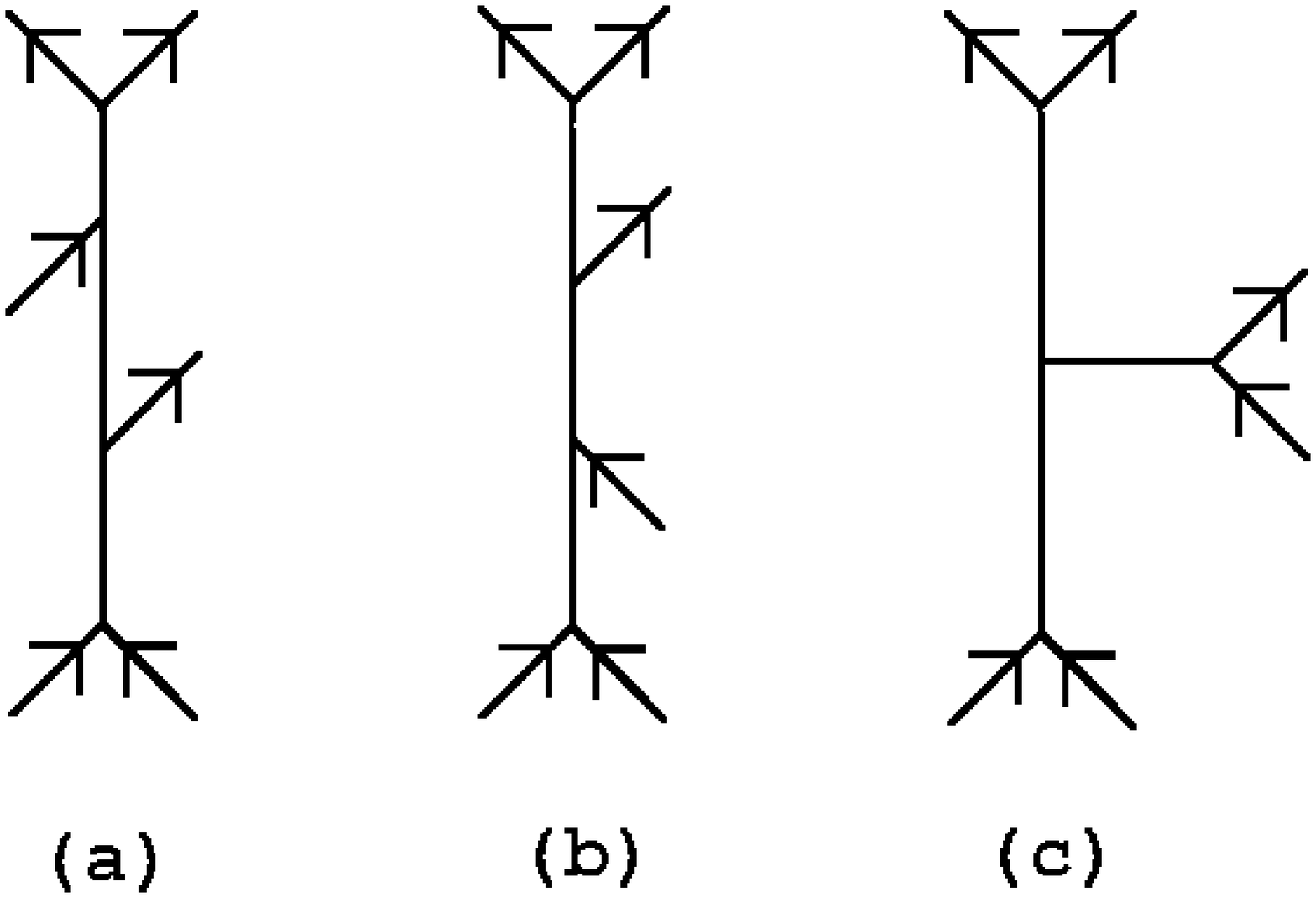}

Fig.~3.4 Tree Diagrams for triple discontinuities.
\end{center}

\subhead{3.5 Unphysical Triple Discontinuities and Reggeization}
 
As we discussed in \cite{arw01}, the diagram of Fig.~3.1(a) has physical region
triple discontinuities of both the Fig.~3.4(a) and (b) kinds, although 
neither gives leading triple-regge behavior. Unphysical discontinuities
are more complicated to discuss. If the usual cutting rules hold,
the diagram of Fig.~3.1(a) has no asymptotic triple 
discontinuities corresponding to Fig.~3.4(c), but rather has
only double discontinuities. To see this, 
consider cutting the diagram as in Fig.~3.5, superficially giving an
$\{s_{13'}, s_{32'}, s_{21'} \}$  triple discontinuity.
\begin{center}
\epsfxsize=2in
\epsffile{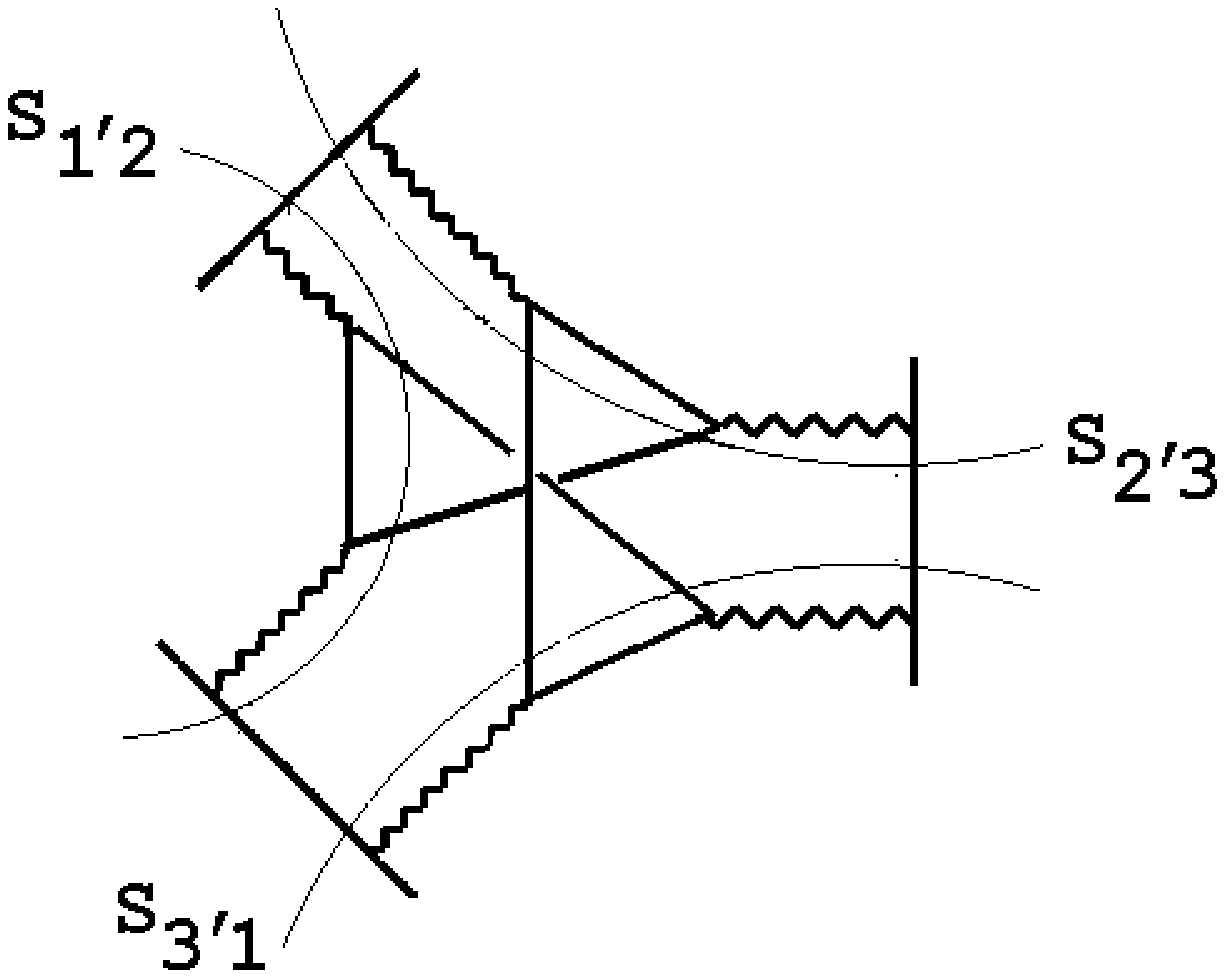}

Fig.~3.5 An unphysical triple discontinuity?
\end{center}
In fact, just taking a double discontinuity, as in Fig.~3.3, 
cuts all the available lines, implying 
that there is no independent third discontinuity that can be taken.

It is not clear, a-priori, 
that the cutting rules do apply to unphysical discontinuities. However, we will
show directly in the next Section that, indeed, there is no symmetric
triple discontinuity present giving the desired common energy 
component sign in the diagram of Fig.~3.1. 
Therefore, as we described above, whether there is an anomaly contribution
from diagrams of this kind depends on whether the necessary triple
discontinuities are present when  
reggeization effects appear.
In \cite{arw01} we noted only that such discontinuities 
appeared to be present in reggeization diagrams 
but did not discuss the structure of such diagrams in any detail. 

As an example of a diagram that should produce reggeization, 
consider that shown in Fig.~3.6  
\begin{center}
\epsfxsize=4.5in
\epsffile{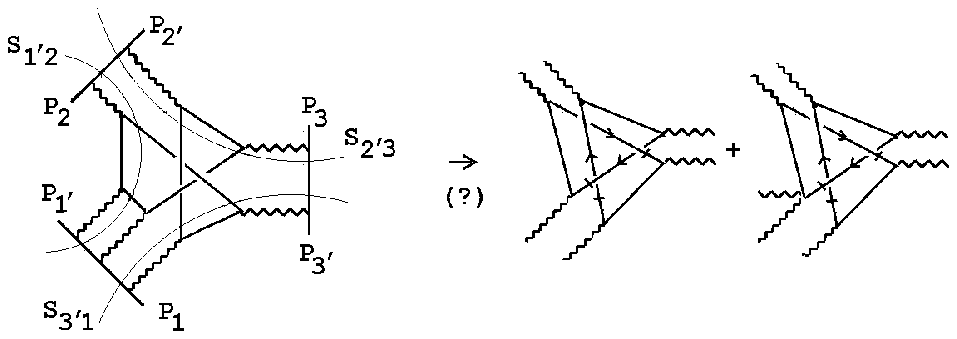}

Fig.~3.6 A diagram with an unphysical triple discontinuity.
\end{center} 
in which one of the gluons in the diagram of Fig.~3.5 is replaced by 
two-gluon exchange - potentially  
giving the one-loop contribution to the trajectory
function of the original gluon. The thin lines again indicate how an
unphysical $\{s_{13'}, s_{32'}, s_{21'} \}$ discontinuity would be taken.
The corresponding six reggeon interaction, together with a remnant 
seven reggeon interaction, would be generated by putting the cut
lines on-shell. The discontinuity is clearly not symmetric and
in the next Section we will confirm by direct calculation that there 
is no triple discontinuity giving the anomaly. This will
be sufficient to determine that the anomaly process of Fig.~3.2 is not 
generated as a ``real part interaction'' when higher-order 
reggeization effects are included. 

\subhead{3.6 A Symmetric Triple Discontinuity}

To obtain a symmetric triple discontinuity 
in which the normal cutting rules could potentially give the anomaly amplitude 
associated with  Fig.~3.2 , we consider the
high-order diagram shown in Fig.~3.7(a)
in which there are three gluons in each
$t$-channel. A triple discontinuity in $\{s_{1'2}, s_{2'3}, s_{3'1} \}$ 
is obtained by cutting the diagram as indicated in Fig.~3.7(b). 
The closed loops involving two-gluon exchange could give
both one loop contributions to the one reggeon trajectory
function and the leading contribution of a two reggeon state. A-priori,
therefore, we expect the diagram to contribute to the six-, seven-,
eight- and nine-reggeon interaction as illustrated.
\newline \parbox{1.7in}{
\begin{center}
$~$
%\newline $~$
\newline $~$
\newline $~$
\newline \epsfxsize=1.3in
\epsffile{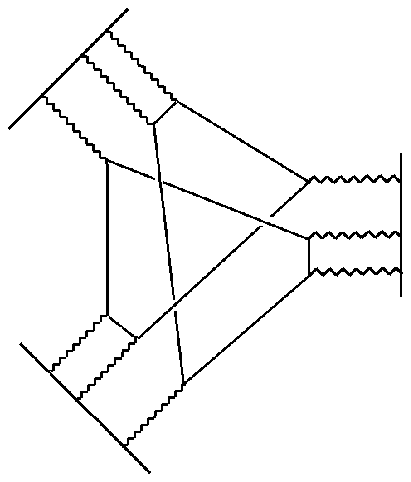}
\newline $~$
\newline $~$
\newline(a)
\end{center}}
\parbox{4.3in}{
\begin{center}
\epsfxsize=3.8in
\epsffile{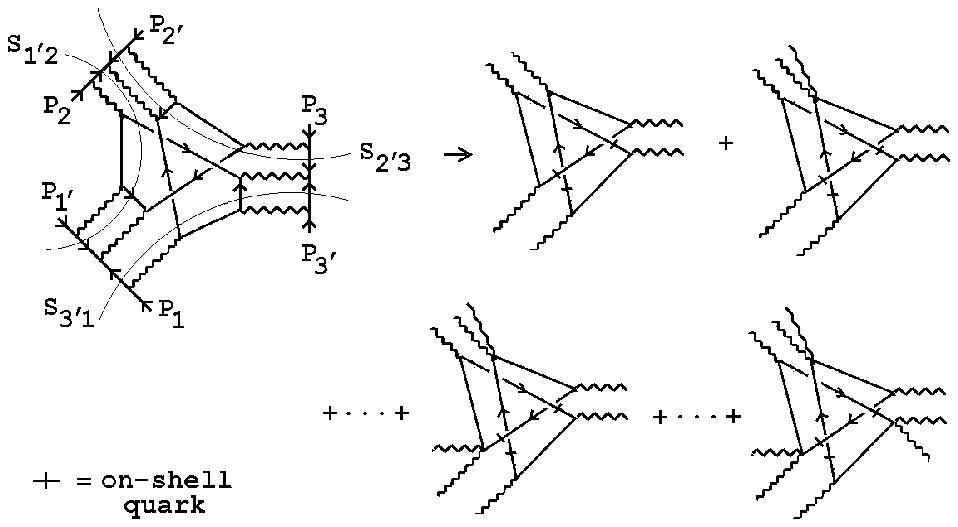}
\newline $~$
\newline(b)
\end{center}}
\begin{center}
Fig.~3.7 (a) A diagram with a symmetric 
unphysical triple discontinuity
\newline (b) expected reggeon interactions.
\end{center}

Since the triple discontinuity of Fig.~3.7(b)
is manifestly symmetric we again might 
expect the symmetric configuration giving the anomaly 
to appear in the six-reggeon interaction. 
However, for consistency with our previous discussion,
the anomaly should not (and does not) appear quite so simply.
After we carry out the explicit 
evaluation of asymptotic discontinuities in the next Section,  
it will be clear that the triple discontinuity of Fig.~3.7(b) does not 
contain the required symmetric momentum configuration.
In fact, the anomaly does occur within 
a reggeon interaction generated by the diagram of Fig.~3.7(a), but  
only when the unphysical discontinuities are actually taken 
as shown in Fig.~3.8. 
\begin{center}
\epsfxsize=3.2in
\epsffile{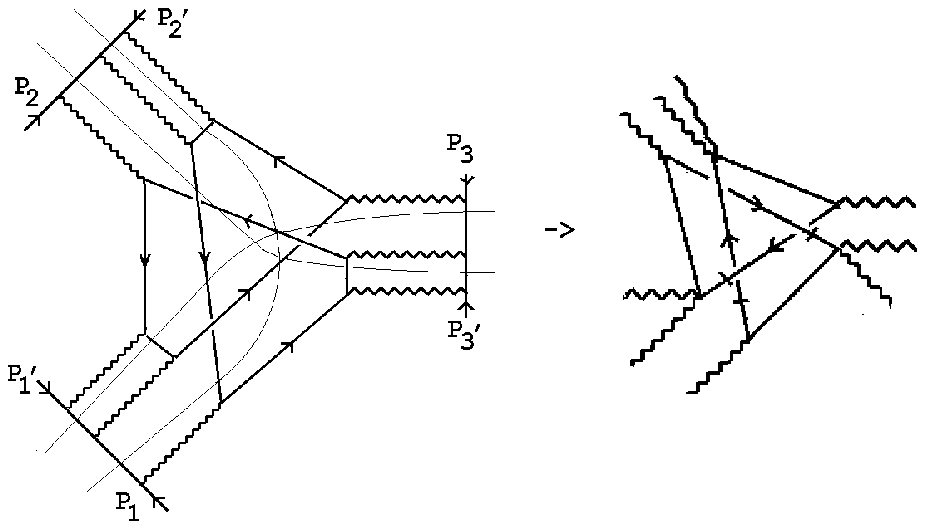}

Fig.~3.8 Another cutting of Fig.~3.7(a).
\end{center} 
However, 
we will postpone until  Section 5 a discussion of which
reggeon interaction is involved.
Note that the discontinuity lines in Fig.~3.8 cross each other. We will
see that this is 
possible because, as anticipated in the previous Section, 
the particles contributing to each discontinuity will not
all have the same time direction. 
To evaluate a multiple discontinuity of this kind
we must develop direct methods to compute asymptotic discontinuities.

\newpage

\mainhead{4. UNPHYSICAL TRIPLE DISCONTINUITIES AND HIGHER-ORDER GRAPHS}

In this Section we generalize the single asymptotic
discontinuity analysis described in the Appendix to asymptotic 
triple discontinuities. The essential idea is that there is a well-defined 
leading-log result for each triple discontinuity (just as there is for
the single discontinuity calculated in the Appendix) that can be found
from the leading-log calculation of amplitudes by keeping the $i \epsilon$
dependence of all logarithms. 

\subhead{4.1 A Physical Region Discontinuity}

We begin by considering again the maximally non-planar graph
shown in Fig.~3.1. To understand 
how asymptotic discontinuities arise, we first consider
a physical region discontinuity. For this we interchange 
$P_1$ and $P_{1'}$ in (\ref{np3}) so that $P_{1'}$ and $P_{2}$ are the
momenta of incoming particles. For simplicity, we also set 
$Q_i=0,~ i=1,2,3$. This could cause confusion as to which invariants
discontinuities actually occur in. However, for the discontinuities
that interest us, we will be able to
avoid this issue. (As is the case for our
discussion in the Appendix, adding both transverse 
momenta and masses to our discussion would not change the essential features 
of the analysis, but would eliminate gluon 
infra-red divergences. We will discuss,
at some points, the general effect of adding transverse momenta.)
Therefore we write, asymptotically,  
$$
\eqalign{ P_{1'}~\to &~- P_{1}~= ~(p_{1'},p_{1'},0,0)~,~~p_{1'} \to \infty \cr
P_2~\to &~- P_{2'}~= ~(p_2,0,p_2,0)~,~~p_2 \to \infty \cr
P_3~\to &~ -P_{3'}~= ~(p_3,0,0,p_3)~,~~p_3 \to \infty  }
\auto\label{pas}
$$

For the reasons given in the last Section,
we will ultimately be looking for a symmetric triple discontinuity.
Therefore, we consider only routes for the 
internal loop momenta of Fig.~3.1 that are completely symmetric with 
respect to the three external loops. There is essentially only one possibility.
The two apparently distinct possibilities   
illustrated in Fig.~4.1 are related by interchanging the primed and 
unprimed external momenta.
We will also want to make a symmetric choice for 
the quark lines we place on shell. Although we will not discuss the anomaly 
in detail until the next Section, in anticipation of this we will demand 
that a product of three orthogonal $\gamma$-matrices
be associated with the process of putting on-shell 
each internal quark line.
To achieve this it is necessary to put on-shell, symmetrically,  
the internal lines in Fig.~4.1(a) 
along which a single loop momentum flows. Therefore, we 
consider only such lines in the following.
\newline \parbox{3in}{ 
\begin{center}
\epsfxsize=2.2in
\epsffile{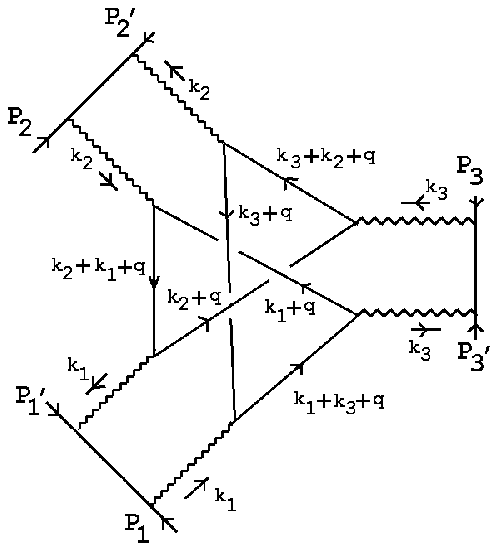}
\newline (a)
\end{center}}
\parbox{3in}{ 
\begin{center}
\epsfxsize=2.2in
\epsffile{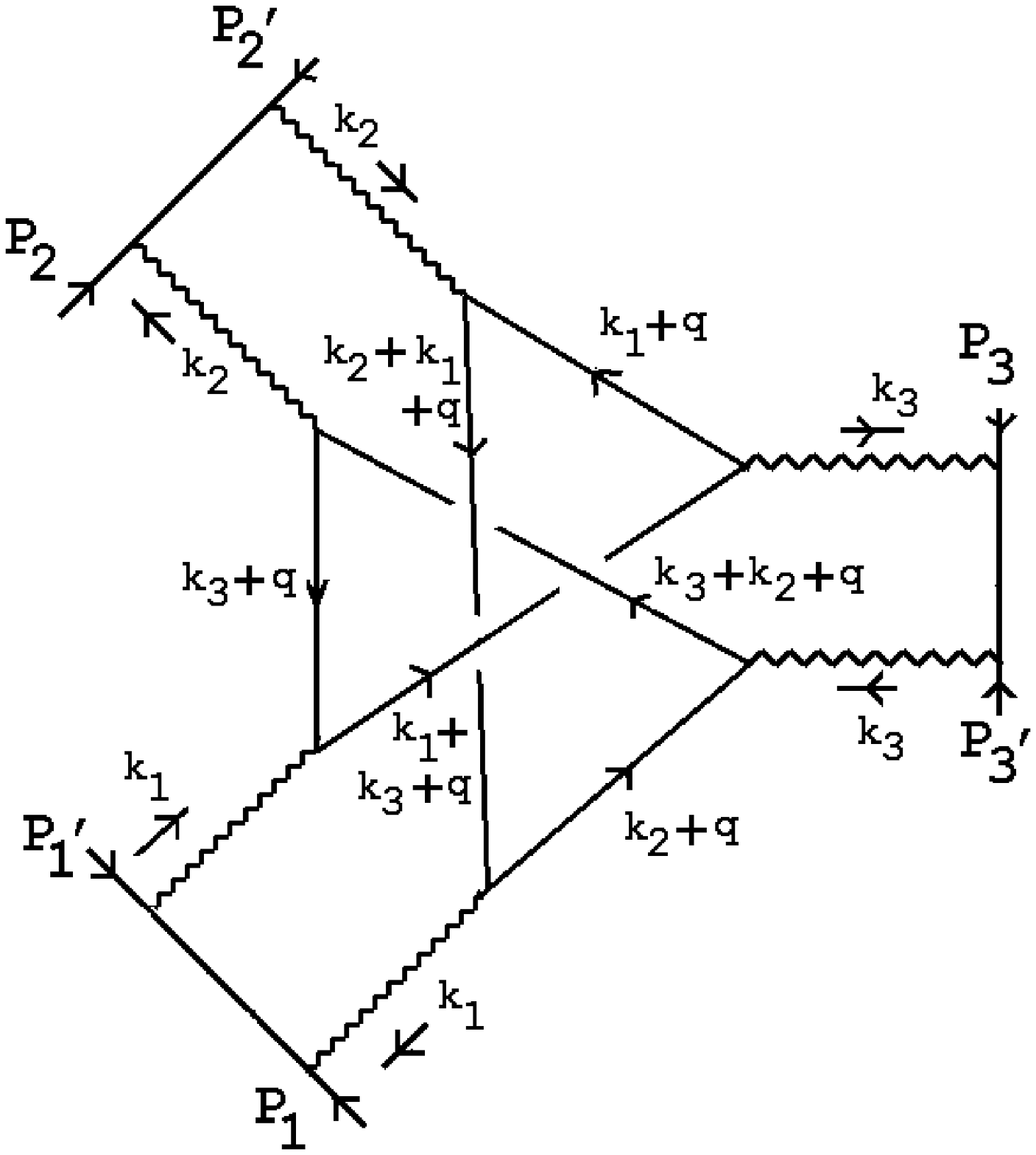}
\newline (b)
\end{center}}
\newline \centerline{Fig.~4.1 Routing Loop Momenta for Fig.~3.2.}

Using the momentum routing of Fig.~4.1(a)
and the analysis of the Appendix
we consider logarithms generated by the $k_1$ and $k_2$ 
integrations.
The $k_1$ and $k_2$ loops are shown in Fig.~4.2.
\begin{center}
\epsfxsize=4in
\epsffile{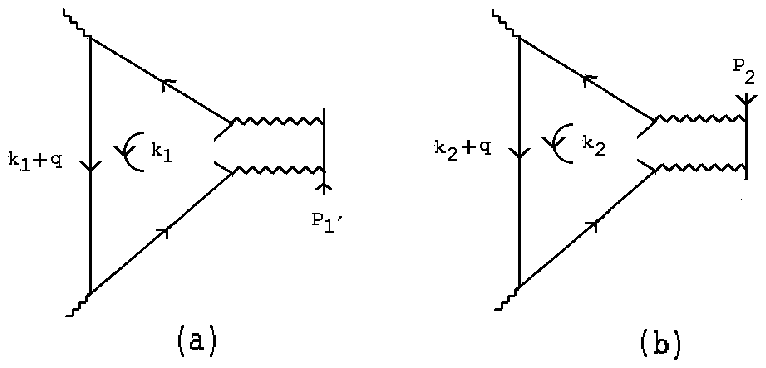}

Fig.~4.2 (a) The $k_1$ Loop (b) The $k_2$ Loop.
\end{center}
For the moment, we omit the propagators in the sloping lines
and all propagator numerators. (The omitted propagators will, nevertheless,
play an important role below. They are also
relevant if we wish to consider the other kinds of discontinuities
that appear in Fig.~3.4.)
In this case, the two loops differ only in the light-cone direction of
$P_1'$ and $P_2$.

We consider Fig.~4.2(a) first. We 
can directly apply the discussion following (\ref{lcan4})
if we identify $P_{1'}$ with $p$, 
$q$ with $p'$, 
$k_1$ with $k$, and consider the propagator pole at 
$(k_1+q)^2 = 0$. We then obtain 
$$
\eqalign{ I(p_{1'}q_{1^-})~\sim~&i\int 
d^2\underline{k}_{1\perp} \left[ -k^2_{1\perp} + 
i\epsilon \right]^{-2} ~\int_0^{\lambda q_{1^-}} dk_{1^-}  
\left[k_{1^-} -q_{1^-} \right]^{-1}\left[ p_{1'} k_{1^-}
- \underline{k}^2_{1\perp}
+i\epsilon \right]^{-1} \cr 
 \sim~&
{1 \over p_{1'}q_{1^-}}~\log{[p_{1'}\lambda q_{1^-} + i\epsilon]} } 
\auto\label{l50}
$$
We have used the notation (used extensively in the following) that for any
four-momentum $k$
$$
k_{i^{\pm}}~=~k_0 \pm k_i ~~~~~
\underline{k}_{i\perp}~=~(k_j,k_k)~~j\neq k \neq i
~~~~~ ~~i,j,k~ = 1,2,3 
\auto\label{not-}
$$
The $q_{1^-}$ 
dependence indicates that the logarithm is a reflection of a threshold
in the invariant $P_{1'}\cdot q$ . This dependence plays an important 
role in the following discussion. We also retain the $\lambda$-dependence, 
for technical reasons that will become apparent later. The final
result will be independent of $\lambda$, as it must be. 
From Fig.~4.2(b) we analagously obtain
$$
I(p_2q_{2^-})~\sim~~
{1 \over p_{2}q_{2^-}}~\log{[- p_{2}\lambda q_{2^-} + i\epsilon]}
\auto\label{l51}
$$
The minus sign (which is very important in the following)
appears relative to (\ref{l50}) because of the 
opposite direction of $P_2$.

Next we consider how the logarithmic branch cuts
generated by the  $k_1$ and $k_2$ integrations can trap the internal loop
integration over $q$ to produce an overall
discontinuity in $s_{1'2} \sim p_{1'}p_{2}$. 
For simplicity, we consider the region where
$$
\underline{k}_{i \perp}^2 ~\sim~ q^2~~\sim ~ 0 ~~~~~i~=~1,2,3
\auto\label{5an0}
$$
Appealing to (\ref{5an}) we can then, for our present purposes, effectively 
ignore the remaining 
$k_i$ dependence of the quark loop (including the propagators that
we ignored in the above discussion). If we parameterize $q$ as 
$$
q~=~\biggl(q_0,~q_{1^-},~q_{2^-},~q_{3^-} \biggr)
\auto\label{l52}
$$
we can treat the $q_{i^-}$ as independent variables, 
with $q_0$ essentially determined by the constraint $q^2 \sim 0$.
The logarithmic cuts of (\ref{l50}) and (\ref{l51}) appear, respectively,
in the $q_{1^-}$ and $q_{2^-}$ planes and if we make a further change of 
variables to
$$
q_{1^-}~=~x_2x_3~~, ~~~~q_{2^-}~=~x_3x_1 ~ ~, ~~~~~q_{3^-}~=~x_1x_2
\auto\label{l53}
$$
the two branch points appear in the $x_3$-plane, for fixed, positive, 
$x_1,x_2$, 
as illustrated in Fig.~4.3(a).
\begin{center}
\epsfxsize=5.8in
\epsffile{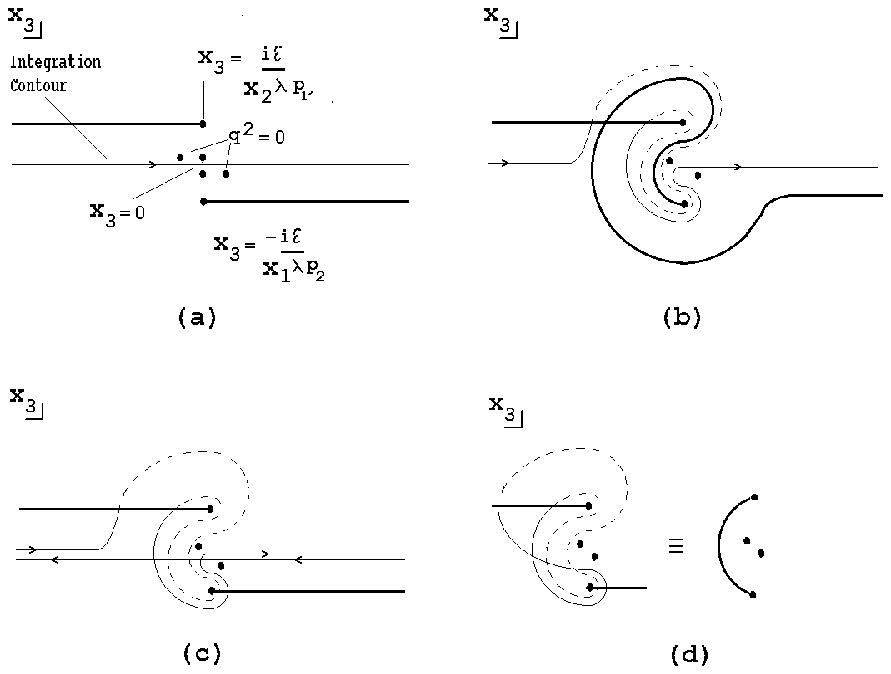}
\newline Fig.~4.3 Contours in the $x_3$-plane (a) the initial contour 
(b) $p_{2} \to e^{2\pi i}p_{2}$ 
\newline (c)  the discontinuity   
(d) the discontinuity as a line integral. 
\end{center} 
(The branch points also appear, separately, 
in the $x_2$ and $x_1$ planes. To focus on the
$s_{1'2}$ discontinuity and avoid any complication
from discontinuities involving a
logarithm of $p_3$ in these planes 
we can take the $\lambda$ for this logarithm to be much smaller.) 
The propagator poles that are not on-shell, that we ignored in the 
above discussion, combine to give a multiple pole at
$q^2=0$ (on both sides of the contour, as determined by
the presence of $i\epsilon$ in all propagators). 
If we continue to ignore propagator
numerators then the factors of $1/q_{1^-}$ and  $1/q_{2^-}$, 
in (\ref{l50}) and (\ref{l51}) respectively, will also contribute poles at 
$x_3=0$ (that will partly be compensated by the jacobian due to 
the change of variables).
However, in the anomaly contribution we will ultimately consider, these poles 
will be directly canceled by numerator factors.

The threshold we are interested in occurs when the two branch points collide
(at $x_3 = 0$ for $\epsilon = 0$). To extract the discontinuity we
consider a full-plane rotation of $p_{2}$, with $p_{1'}$ fixed, so that  
the logarithmic branch-cut
(\ref{l51}) deforms the contour as shown in Fig.~4.3(b) 
- the dashed line indicates that the contour is on the 
second sheet of the branch-point (\ref{l50}). (In this figure 
we have omitted the poles at $x_3=0$.) Note that 
the continuation path we have chosen isolates the discontinuity
around the $s_{1'2}$ branch cut, since it avoids the pinching of the 
integration contour with the singularity at 
$q^2=0$ that would give other discontinuities. 
The desired discontinuity is obtained by adding 
the original contour in the opposite direction, as shown in Fig.~4.3(c).
Combining both contours 
we obtain Fig.~4.3(d) which, as illustrated, can be written as a line integral 
between the two branch 
points of the double-discontinuity due to both cuts. As $\epsilon \to 0$, 
or in the asymptotic
limit $p_{1'}, p_2 \to \infty$, the branch points approach each other
and the result is a closed contour integral around the 
singularity at $q^2=0$ which is independent of the position of the end points 
and remains finite in the asymptotic limit. This is the asymptotic 
discontinuity and the singularity at $q^2 = 0$ 
is clearly crucial in producing a non-zero result. 

In Fig.~4.4(a) we have illustrated the last stage of the contour contraction
as $\epsilon \to 0$ and have also included 
the effect of adding 
(external and internal) transverse momenta in the 
the foregoing analysis. 
The integral between the branch points, of the double discontinuity,
is still obtained, while the 
singularity at $q^2 =0$ separates into a set of poles at both 
positive and negative $x_3$.
\begin{center}
\epsfxsize=5in
\epsffile{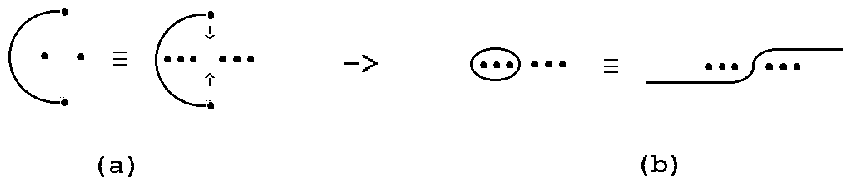}

Fig.~4.4 (a) The $x_3$
contour with finite transverse momenta (b) equivalence of the
asymptotic contour to the original contour.  
\end{center} 
In Fig.~4.4(b) we have  shown the asymptotic discontinuity.
Since the branch points are
logarithmic, the double discontinuity involved is simply $4 \pi^2$ and so 
no longer contains either branch cut.
Consequently, the asymptotically finite integral around 
the poles to the left can be opened up to give the original contour, 
as illustrated. (If there is a singularity at $x_3=0$, the contour is 
constrained to pass through this point although, as we noted above, for the
anomaly contribution to graphs, this will not be the case). 
The final result shown in Fig.~4.4(b) is just what
would be given by the normal cutting rules for a discontinuity in $s_{1'2}$ ,
i.e. the original integral with the four propagators involved in generating 
the discontinuity placed on-shell. Note that the same result 
is obtained if the discontinuity is evaluated by 
varying $p_{1'}$. An integral around the positive $x_3$ poles appears at the
intermediate stage, which can then 
be opened up to give the same final contour as in Fig.~4.4(b).  

An obvious, but essential, requirement in the
origin of the asymptotic discontinuity, which we want to emphasize, 
is that the branch-cuts due to the logarithms
in $p_{1'}$ and $p_2$ must lie on opposite sides of the $x_3$ contour.
(This is the sign difference between (\ref{l50}) and (\ref{l51}) 
that we emphasized above.) In a physical region this requirement 
is normally straightforward for a loop integration producing a threshold
due to two massive states since the loop momentum will flow oppositely
through the two states and the $i\epsilon$ prescription will place
the states on opposite sides of the energy integration contour. In the 
variables we are using the generation of the threshold is a little more 
subtle.
Note, for example, that when $x_1 < 0$ the branch-point (\ref{l50}) appears 
in the upper half-plane (moving through infinity as $x_1$ moves through zero)
and there is no discontinuity. Therefore, the signs of the $x_{i}$ play an 
essential role in the occurrence of the discontinuity. 
A further requirement, which clearly holds in the case just discussed, 
is that the trapping (pinching)
of the contour that we have discussed 
must combine with the pinching associated with the logarithms
to give a complete cut through the diagram. That is to say, the complete set of
pinchings must correspond to an overall invariant cut.

\subhead{4.2 Maximally Non-Planar Unphysical Discontinuities}

We consider next the unphysical discontinuities that are our 
principal interest.
According to the discussion in Section 3, we are looking for a triple 
discontinuity of the form of Fig.~3.5 that treats the three cut lines of the
quark loop symmetrically so that, in a  
physical region, the sign of the energy component can be 
the same for all three on-shell states. We will, therefore, confine 
our discussion to a search for a symmetric triple discontinuity. 
As we noted, if the normal cutting rules apply 
there is no triple discontinuity (symmetric or not) of the Fig.~3.5  kind.
We consider whether the direct evaluation of discontinuities 
gives the same result. 

The discontinuity we 
discussed above occurred in a physical region that is unsymmetric 
in that $P_2$ is the momentum of an incoming particle while 
$P_1$ is the momentum of an outgoing particle.
To look for a symmetric discontinuity we will 
use an analysis that treats the complete graph symmetrically throughout.
To this end, we start in the symmetric asymptotic region (\ref{np3})
where all momenta are real and 
$$
s_{i'j}~\sim ~-p_i p_j ~~<~0
\auto\label{ninv}
$$
In this region, the diagram is defined by the usual $i\epsilon$ prescription.
Since all three invariants must be positive, the triple discontinuity of
Fig.~3.5 can only be present in the triple-regge limit if we allow the large
momenta involved to be unphysical. A symmetric way to do this is to start 
from the real physical region and take
$$
p_i ~\to ~ e^{-i\pi /2} p_i~= i p_i~,~~i=1,2,3 ~~~~~
=> s_{i'j}~\sim ~ (- ip_i)(i p_j) ~~>~0  
\auto\label{pinv}
$$

Given the symmetry of the present discussion, 
it is immediately apparent that there will not be a (symmetric)
triple discontinuity, as we now argue.
Using the above analysis, logarithms will be generated by each of the $k_i$ 
integrations. If we consider again the region where the 
transverse momenta are close to zero then, from (\ref{5an}), the requirement 
that the energy component of each on-shell line in the loop have the same sign
is equivalent to requiring that the $q_{i^-}$ all have the same sign. 
This, in turn, requires 
that the $x_i$ should all have the same sign.
However, in the symmetric real physical region, 
if $x_1$ and $x_2$ have the same sign, 
the logarithmic branch cuts in $P_1$ and $P_2$ lie on the same side 
of the $x_3$ contour as illustrated in Fig.~4.5.
\begin{center}
\epsfxsize=2in
\epsffile{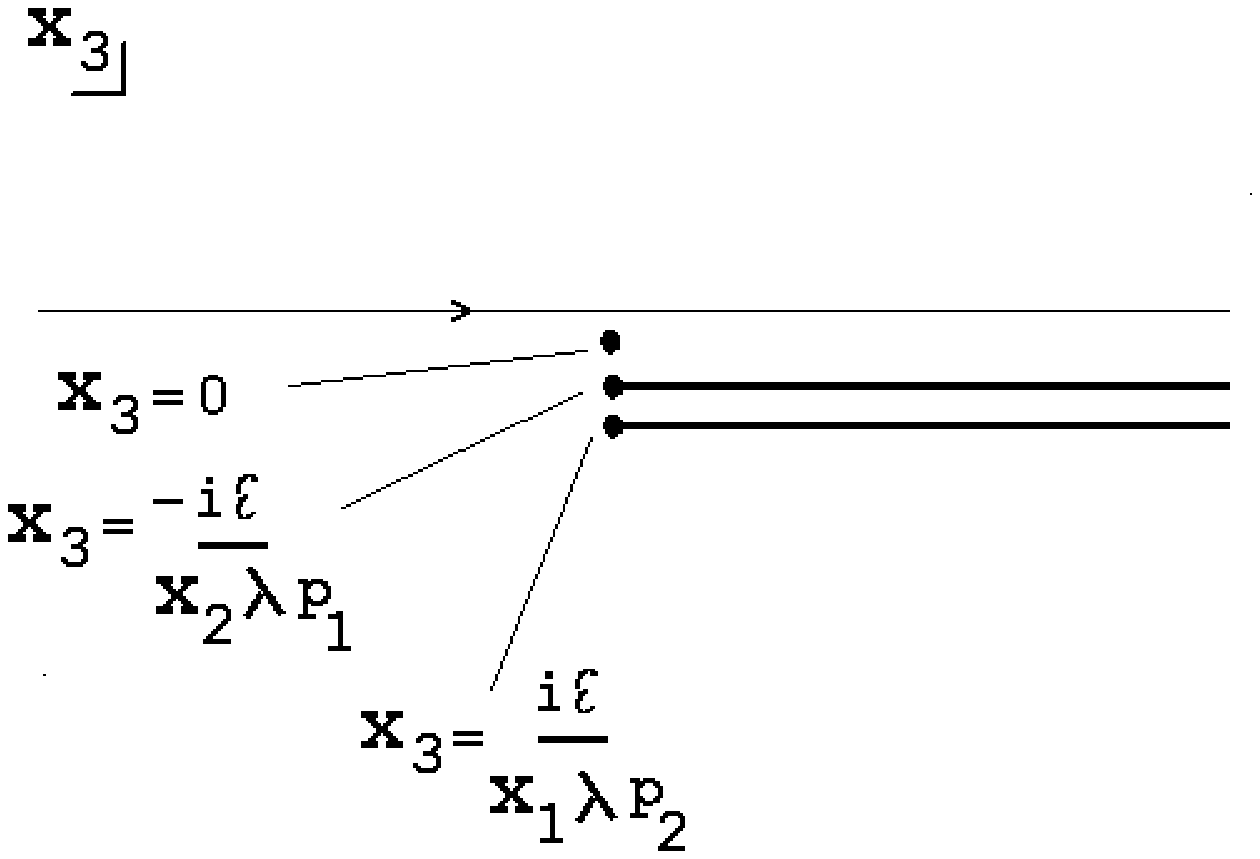}

Fig.~4.5 The Symmetric Location of Branch-Cuts in the $x_3$-plane. 
\end{center}
Since the continuation
(\ref{pinv}) is symmetric they will remain on the same side after the
continuation. 
As a consequence, in the symmetric $x_i$ region, 
the contour will not be trapped and distorted as 
one branch point moves aound the other, as it was in Fig.~4.3, and no
discontinuity will result. We conclude therefore that, 
for the graph we are discussing, 
discontinuities can only be generated in asymmetric 
regions of the $x_i$ that can 
not provide the symmetric triple discontinuity that we are looking for.
The foregoing analysis also precludes the occurence of a triple discontinuity,
that is appropriately symmetric, in the diagram of Fig.~3.6. 

\subhead{4.3 A Symmetric Unphysical Triple Discontinuity}

To obtain a symmetric triple discontinuity we look for a graph that has the
appropriate overall symmetry and also, for each $i \neq j \neq k$,
has logarithmic branch
cuts on both sides of the $x_i$ contour
in a symmetric region of $x_j$ and $x_k$. With these requirements in mind, 
an obvious graph to consider is that of Fig.~3.7. To discuss this graph we 
continue, for simplicity, to take $Q_1=Q_2=Q_3=0$. Two symmetric (distinct)
routes for the internal momenta are shown in Fig.~4.6. 
To be consistent with our previous notation we have used the notation that 
we direct the $k_i$ momenta in the opposite direction to the $P_i$,
$k_i'$ momenta in the opposite direction to the $P_i'$ (i.e. in the same
direction as the $P_i$) and direct the internal loop momentum in the
same direction as the $k_i'$ momenta.

For a threshold corresponding to the cutting of particular
lines of the internal quark loop to be generated the  
external loop momentum generating the relevant logarithms
must pass through at least one of the lines. With this constraint,
only the routing shown in 
Fig.~4.6(a) will give both discontinuities of the kind we are looking for
and the $\gamma$-matrix structure for on-shell contributions
that we show, in the next Section, gives the anomaly.
The routing of Fig.~4.6(b) 
would be appropriate for discussing 
the triple discontinuity of Fig.~3.7(b). However, as we noted in the previous
Section, and will explain further below,
this triple discontinuity does not contain the symmetric momentum configuration
needed for the anomaly.

Using the momentum routing of Fig.~4.6(a) 
we consider the logarithms generated by both the $k_i$ and $k_i'$ loop 
integrations. Extracting all logarithms places on-shell all the hatched lines
of Fig.~4.6(a), and 
gives leading behavior of the form of (\ref{211}) multiplied by double 
logarithms of each of the $P_{i^+}$.
\newline \parbox{3in}{ 
\begin{center}
\epsfxsize=2.4in
\epsffile{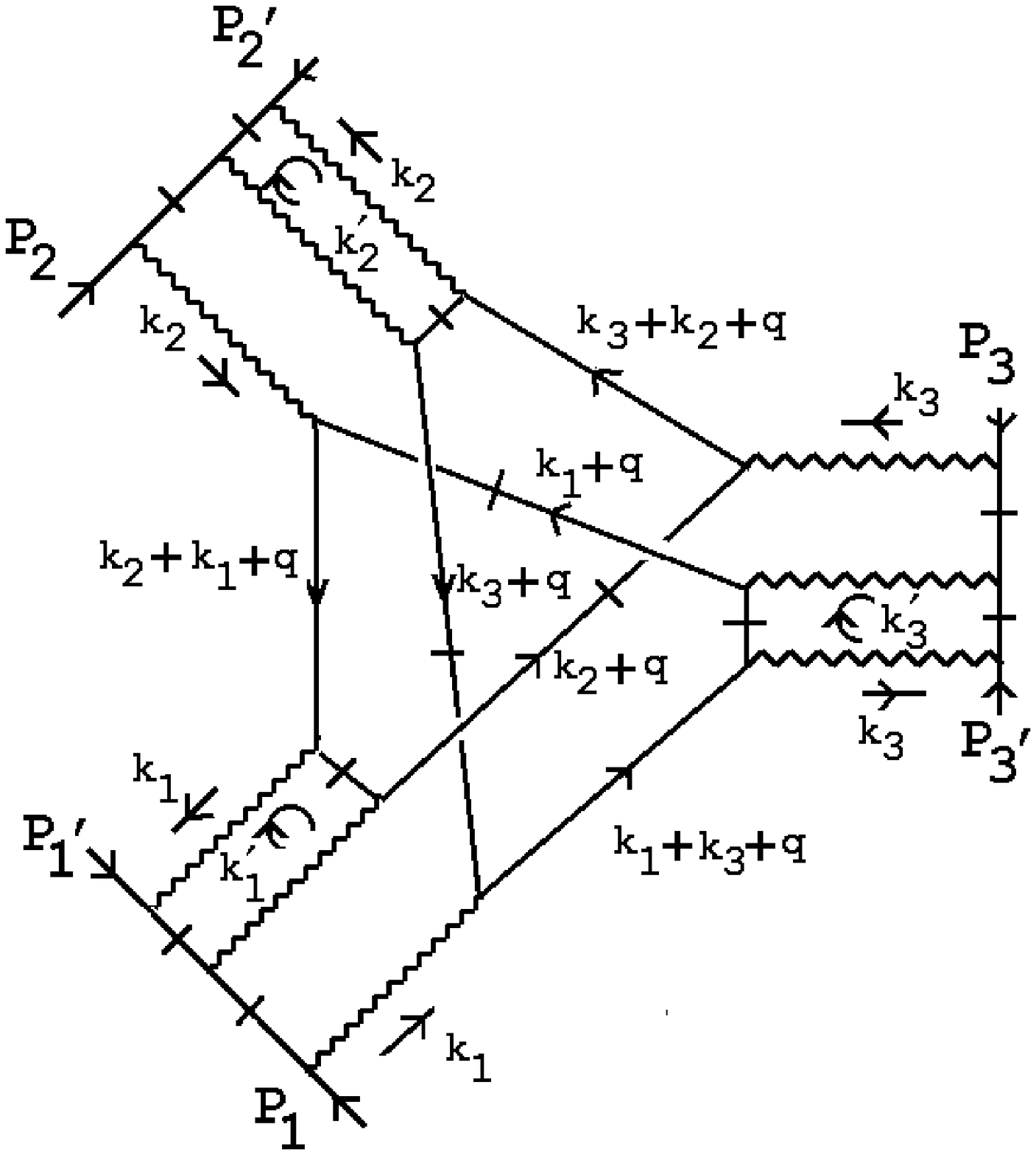}
\newline (a)
\end{center}}
\parbox{3in}{ 
\begin{center}
\epsfxsize=2.4in
\epsffile{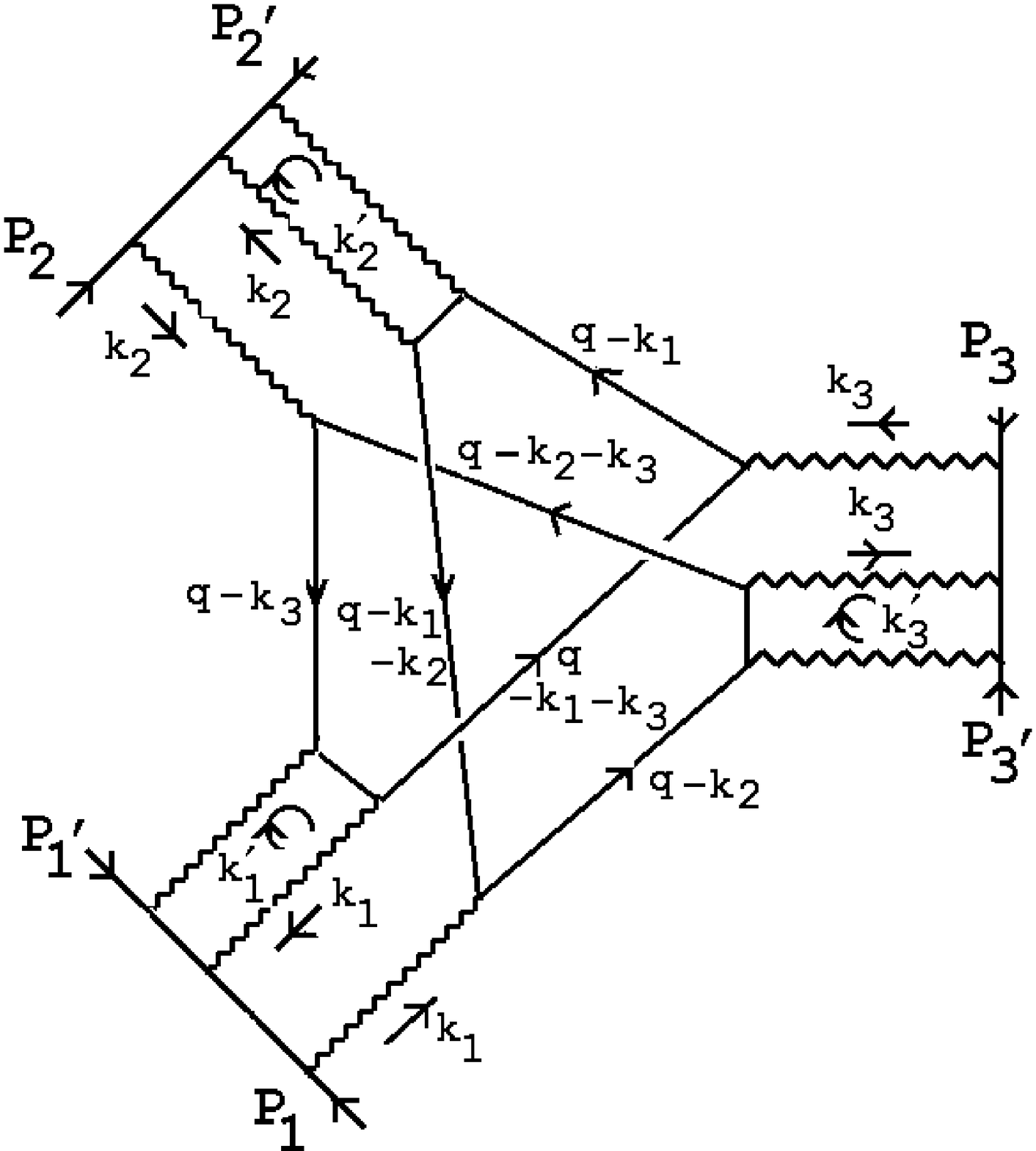}
\newline (b)
\end{center}}
\begin{center}
Fig.~4.6 Labeling Momenta for Fig.~3.7(a).
\end{center}
How the logarithms cancel or combine with other diagrams is, of course, 
a very complicated problem.
As we have emphasized, to discuss this 
systematically we must 
consider all the multiple asymptotic discontinuities that occur 
rather than the behavior of full diagrams. Our present concern is, however,
the much narrower purpose of determining only whether there is 
a symmetric triple discontinuity in which the anomaly can occur.

As above, to study discontinuities
we keep the $q$-dependence of all logarithms together with all
$i\epsilon$ dependence.
We consider specifically the logarithms generated by
the $k_1$ and $k_1'$ loops, but the symmetry of the diagram 
obviously determines that the others can be treated identically. The loops,
extracted from Fig.~4.6, are shown in Fig.~4.7.
The $k_1$ loop is identical to those of Fig.4.2 and 
can be evaluated analagously.
\begin{center}
\epsfxsize=3.5in
\epsffile{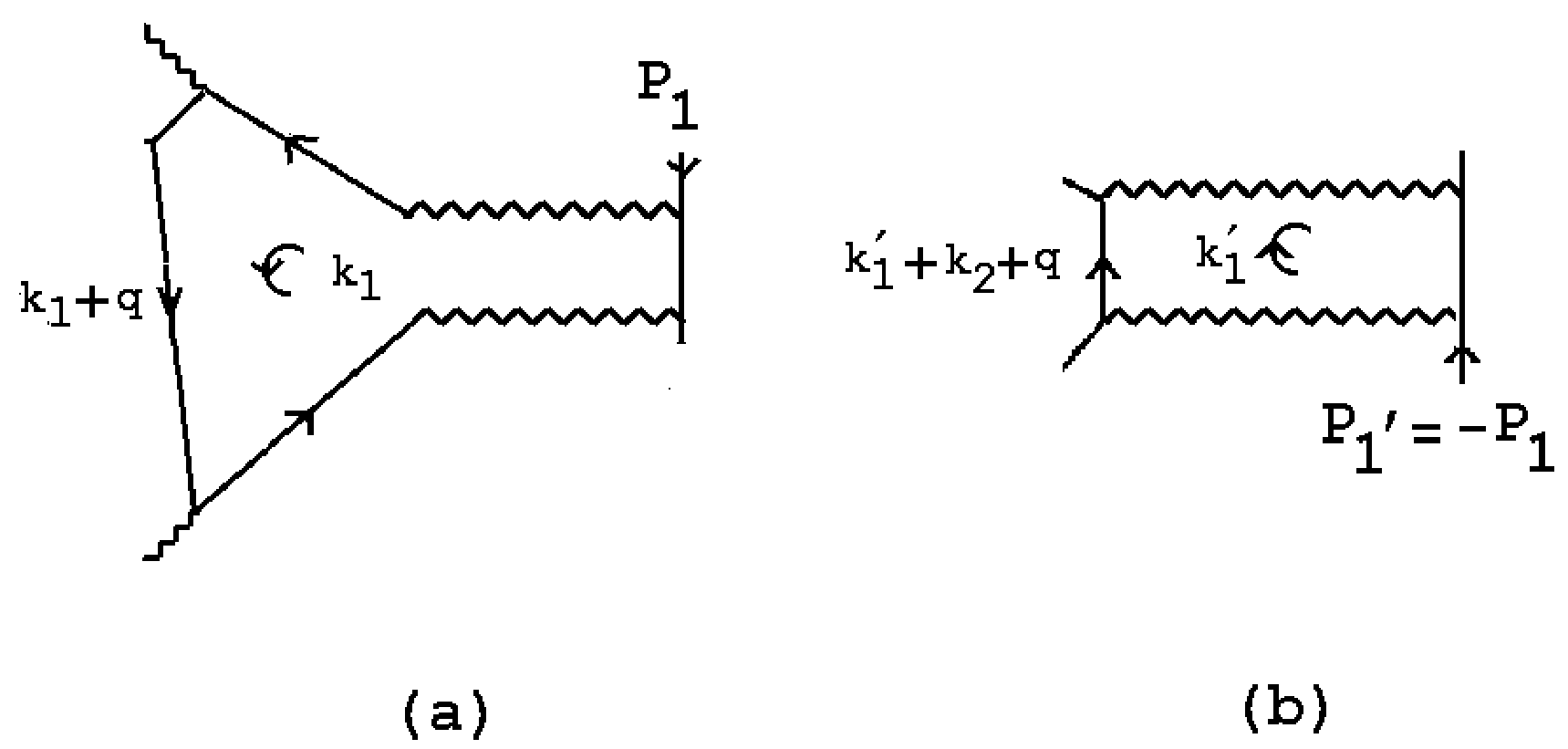}
\newline Fig.~4.7 (a) The $k_1$ Loop (b) The $k_1'$ Loop.
\end{center}
Using a similar analysis, the $k_1'$ loop gives an integral of the form
$$
\int_0^{(k_2+q)_{1^-}}~dk_{11^-}' ~~\cdots
\auto\label{l54}
$$

If we again go to the region where all transverse momenta are close 
to zero then,We have drawn the diagrams as basic anomaly processes in Fig.~6.1 
using (\ref{5an}), it follows that after the $k_2$ integration
$$
k_{21^-} ~ \sim ~k_{20} ~\sim ~q^2/q_{2^-}~ << q_{1^-}
\auto\label{l55}
$$
Therefore, we can take the upper end-point in (\ref{l54}) to be $q_{1^-}$. In 
this case both the $k_1$ and $k_1'$ integrations give logarithms with $q_{1^-}$
in the argument - but with opposite signs.
We then have branch-cuts located as in Fig.~4.8(a) in each of the 
$x_1,x_2$ and $x_3$ planes.
\begin{center}
\epsfxsize=5.5in
\epsffile{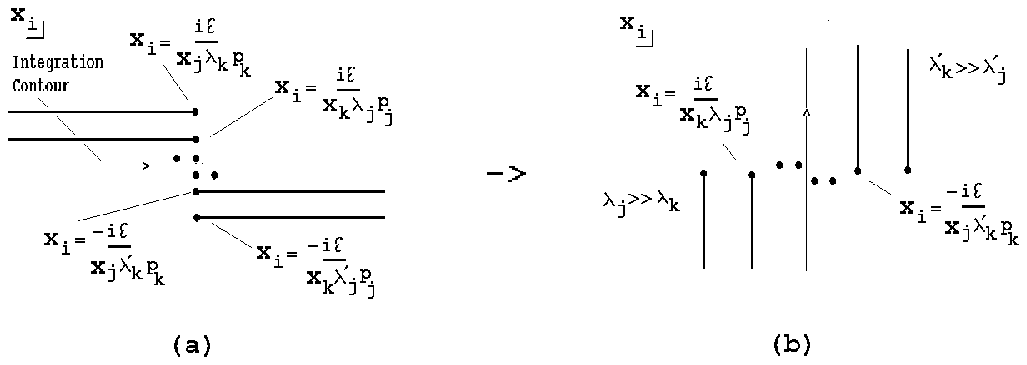}
\newline Fig.~4.8 (a) Branch Points in the $x_i$-plane 
(b) $p_i ~\to ~ e^{-i\pi /2} p_i~= i p_i~,~~i=1,2,3$ 
\end{center}
We have included the poles at $q^2=0$ and $x_i=0$
and have used different $\lambda_i$ and $\lambda_i'$
for each branch-cut to allow us to separate the branch points in our 
discussion. 

With values of the $\lambda_i$ and $\lambda_i'$ implied 
by Fig.~4.8(a),
we could clearly obtain a discontinuity in $s_{jk'}$ (due to the two
closest branch points) by repeating the 
discussion illustrated by Fig.~4.3. The discontinuity would similarly be an 
integral between the two branch points involved, as in Fig.~4.3(d), 
but because of the additional
branch points that are present, the contour could not be opened up as in 
Fig.~4.4. Therefore, having taken $x_j, x_k > 0$ so that the branch 
cuts lie  as in Fig.~4.8(a),
the discontinuity would involve only pure imaginary or negative 
real part values of $x_i$. Consequently,
any further discontinuity obtained by the collision of 
branch points in the $x_j$ or $x_k$ planes would have to involve mixed 
real part signs for
the $x_i$. We conclude (not surprisingly) that 
in the physical region a triple discontinuity can not be obtained that 
involves only positive values of all three $x_i$.

This brings us to the central point of the paper. If
we go to the unphysical
region (\ref{pinv}), where we expect to encounter
an unphysical triple discontinuity, the last analysis changes in a crucial
manner. The 
resulting location of branch cuts is now as shown in Fig.~4.8(b), allowing
the integration contour to be rotated as illustrated. 
In Fig.~4.8(b) we have also, for emphasis, 
chosen significantly different values of the $\lambda_i$ and $\lambda_i'$.
If we again determine the discontinuity associated with the collision of the 
two nearest branch points, as above, the result will be the contour 
integral of the double discontinuity shown in Fig.~4.9.
\begin{center}
\epsfxsize=4in
\epsffile{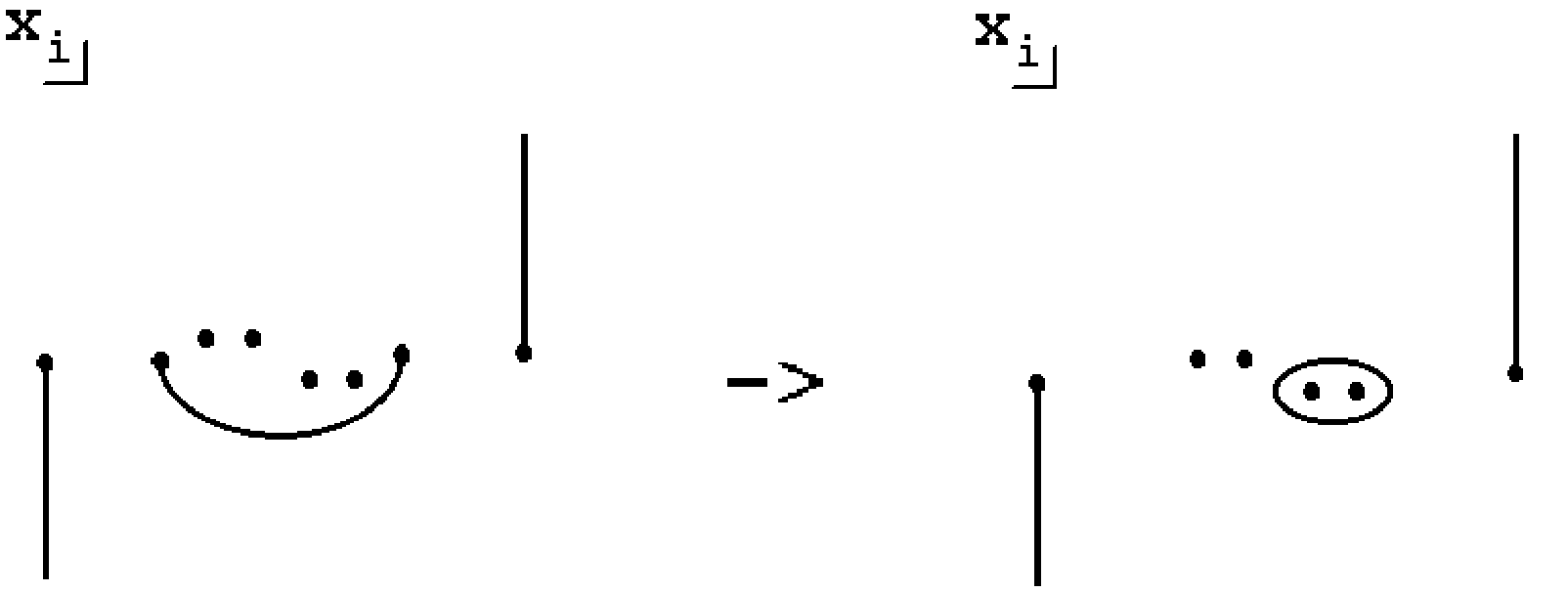}

Fig.~4.9 The unphysical region discontinuity.
\end{center}
Now the integral involves positive real values of $x_i$ and, as illustrated,
the asymptotic limit gives a loop integral over just positive values.
The contour integral can not be opened up, however, since the other branch cuts
remain.

Having derived a first discontinuity from two branch points in the 
$x_i$ plane, as in Fig.~4.9, it is straightforward
to keep the remaining branch points and move on to the $x_j$ and $x_k$ planes
where, in each case, only two branch cuts now appear. In both planes,
discontinuities of the form of Fig.~4.9 occur, provided the 
$x_i$ integration is restricted to positive real values. Therefore, we obtain 
a triple discontinuity in which each of the $x_i$, $x_j$ and $x_k$
integrations is consistently over positive values and the asymptotic 
contour is obtained as illustrated by the first two contours in Fig.~4.10.
\begin{center}
\epsfxsize=4in
\epsffile{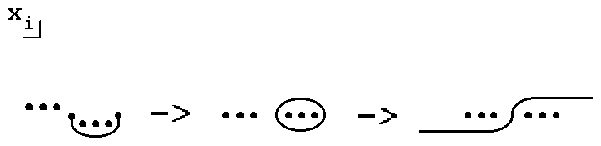}

Fig.~4.10 Contours for the $x_i$, $x_j$ and $x_k$ integrations.
\end{center}
Since all logarithmic branch cuts are now removed, all three contours can 
be opened up to obtain the last contour of Fig.~4.10 which is, once again the
original contour of integration for each of $x_i$, $x_j$ and $x_k$.
We thus obtain a triple discontinuity which, at first sight, corresponds to
the usual cutting rules since all cut lines are on-shell. 
However, the triple discontinuity is truly symmetric and as a result
each discontinuity is, necessarily, a pseudothreshold. There is also
a very important further subtlety.

If we consider the discontinuity arising from the pinching of logarithms 
of $p_1\lambda_1$ and $p_2\lambda_2'$, for example, then the lines  
put on-shell in the discontinuity are those that have thick hatches in
Fig.~4.11(a).
\begin{center}
\epsfxsize=4.5in
\epsffile{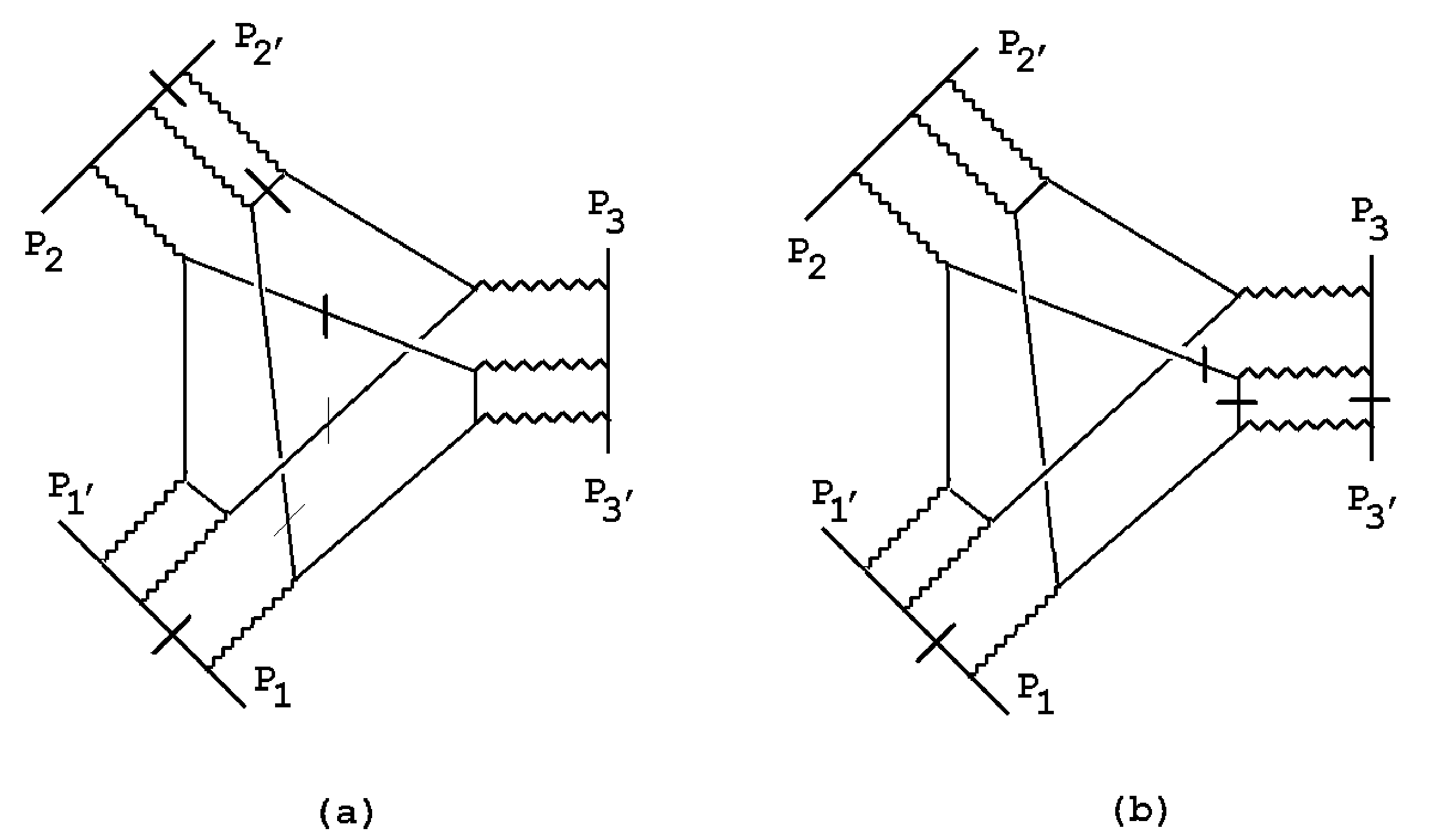}

Fig.~4.11 On-shell lines for (a) an $s_{12'}$ discontinuity (b) a potential
$s_{13'}$ discontinuity.
\end{center}
These lines are only a subset of those required to obtain a 
complete cut of the diagram. This implies that the corresponding
pinching does not, by itself, give a singularity of the complete integral and
a-priori the integration contour could be deformed away from the pinched
region.
To obtain a complete cut we must add the lines that have thin hatches in
Fig.~4.11(a). When 
these lines are on shell the pinching does give an overall singularity.
But, if we require a common sign for the $x_i$
the two thin-hatched lines again have the wrong momentum direction
to straightforwardly combine with the asymptotic
pinching to give what would be a physical sheet 
``asymptotic normal threshold''. However,
each of the two thin hatched lines is separately placed on shell 
by one of the additional discontinuities. Therefore, the full 
triple discontinuity we have found does correspond to a triplet 
$\{s_{12'},s_{23'},s_{32'}\}$ of invariant (pseudothreshold) cuts. 

If we consider instead 
the discontinuity arising from the pinching of logarithms 
of $p_1\lambda_1$ and $p_3\lambda_3'$ then the lines put on shell are 
those hatched in Fig.~4.11(b). In this case there is no simple way to 
include additional lines and obtain an invariant cut. Therefore, this pinching
can not be extended to a complete cut of the diagram.
We conclude that the triple discontinuity in $\{s_{12'},s_{23'},s_{32'}\}$
that is illustrated in Fig.~3.8 is the only combination that exists, 
as an extension of the above analysis. It
is symmetric, with each of the internal quark lines 
that are put on shell by $k_i$ integrations treated symmetrically. 
All three of these lines contribute to each invariant cut but, as we 
have just discussed, two of them always have the 
wrong $i\epsilon$ prescription, relative to the third,
to give a physical normal threshold. Singularities associated with 
combinations of forward and backward going particles are 
``mixed-$\alpha$'' solutions of the Landau equations\cite{arw00}.
In general, such ``pseudothresholds'' are not 
singular on the physical sheet because of the conflicting 
$i\epsilon$ prescriptions. However, they are generally singular on 
unphysical sheets and can appear in multiple 
discontinuities. They would be particularly expected to appear
in unphysical multiple discontinuities. 

Finally, we return to the triple discontinuity of Fig.~3.7(b),
using the momentum routing of Fig.~4.6(b). Consider, for example, the 
discontinuity in $s_{2'3}$. This will be due to the pinching of the
$x_1$ integration contour by the logarithms generated from the $k_2'$ 
and $k_3$ integrations. The relevant sub-part of Fig.~4.6(b) is shown in
Fig.~4.12.
\begin{center}
\epsfxsize=2.5in
\epsffile{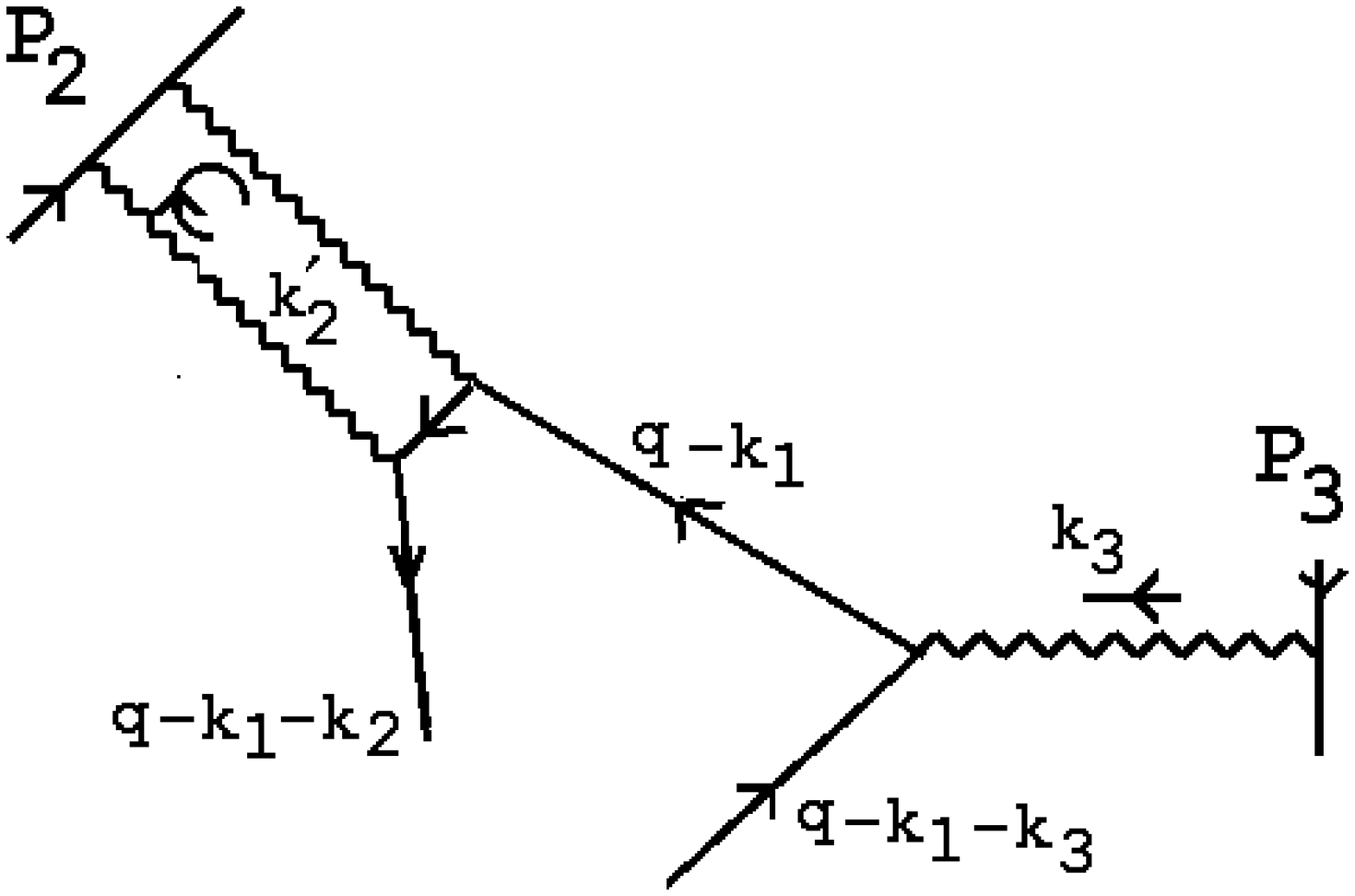}

Fig.~4.12 Part of the $q$ momentum flow within Fig.~4.6(b) 
\end{center}
That the direction of the $q$-momentum flow is opposite, 
relative to $k_2'$ and $k_3$, along the relevant internal lines 
implies that $x_2$ and $x_3$ must have opposite signs in order for the
branch cuts to be on opposite sides of the $x_1$ contour.
As a result no symmetric triple discontinuity exists.

There are clearly two criteria for the existence of a symmetric asymptotic
triple discontinuity - that 
we will appeal to further in the next Section. The first 
is that the $q$ momentum flow must be in the same relative direction along 
the relevant internal lines for each discontinuity. The second is that
all internal loop lines, besides those in the remaining triangle, must be
put on shell 
by the combination of the three pinches of the $x_i$ integrations.

\newpage

\mainhead{5. THE TRIANGLE ANOMALY AND OTHER DIAGRAMS}

In this Section we discuss how the anomaly occurs in a reggeon vertex 
obtained from the triple discontinuity of Fig.~3.8. 
We will also consider other diagrams that can contribute
and discuss how color quantum numbers
determine which reggeon interactions are involved.

\subhead{5.1 The Asymptotic Amplitude}

We can briefly describe the calculation of the asymptotic amplitude
obtained from Fig.~3.8 (in which all the cut lines 
are put on-shell as described in the last Section) as follows.
Additional background description of the method used
can be found in \cite{arw01}. We begin by 
adding in the numerator dependence that we essentially ignored in the 
previous Section. For the external lines, additional powers of the external
momenta are generated as in (\ref{lcan61}) and (\ref{coup}). As a result,
inverse external momentum factors, such as ${p_{1'}}^{-1}$ in (\ref{l50}) and 
 ${p_2}^{-1}$ in (\ref{l51}) are eliminated and the 
factor of $P_{1^+}P_{2^+}P_{3^+}$ that appears in (\ref{211}) is produced.
Also, if we use the natural transverse momenta
given by (\ref{not-}), the light-like 
$\gamma$-matrix couplings that appear at each of the vertices 
of the internal loop
(after the triple-regge limit is taken) are as illustrated in Fig.~5.1(a). 
\begin{center}
\epsfxsize=5.8in
\epsffile{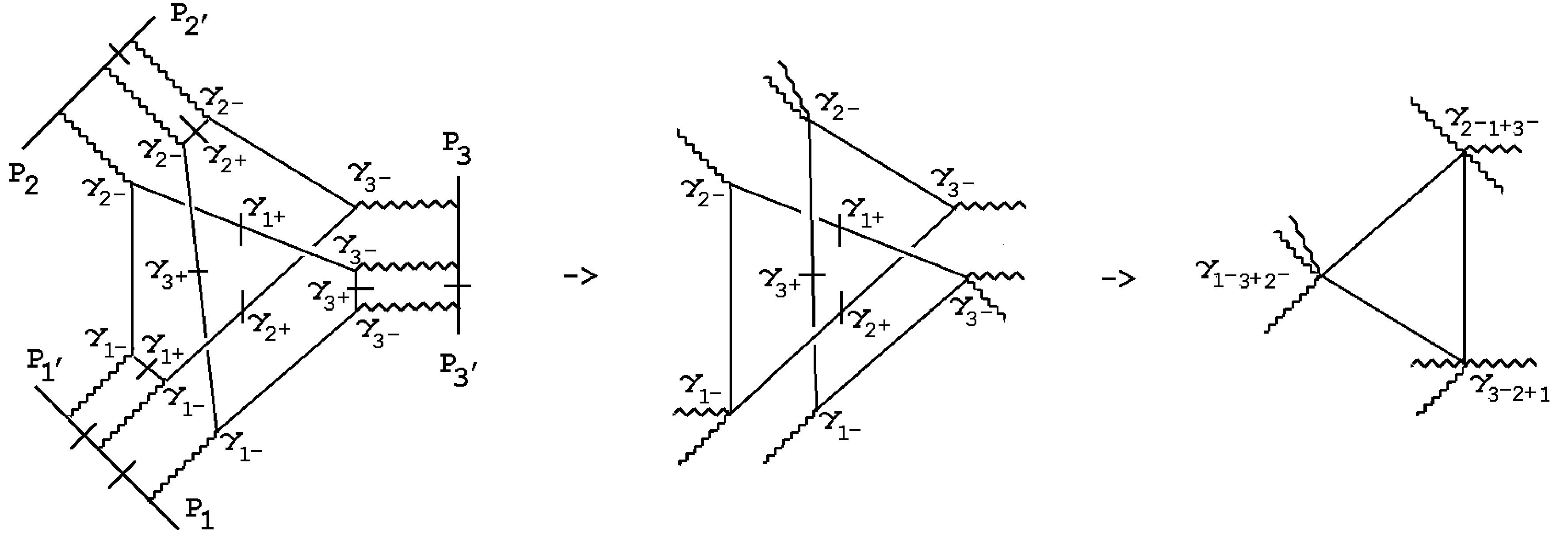}
\parbox{2.1in}{\centerline{(a)}}
\parbox{2in}{\centerline{(b)}}
\parbox{1.8in}{\centerline{(c)}}

Fig.~5.1 $\gamma$-matrix structure 
for the reggeon interaction extracted from Fig.~3.7.
\end{center}
For the hatched lines that appear in both
Fig.~5.1(a) and (b), we keep the $\gamma$ matrices shown. 
These are the ``local couplings'' (see \cite{arw01}) that appear when that 
part of the associated numerator is kept that cancels the 
internal momentum factors that arise from the longitudinal
loop momentum integrations
(such as ${q_{1^-}}^{-1}$ in (\ref{l50}) and 
${q_{2^-}}^{-1}$ in (\ref{l51}) ). To justify this procedure 
we appeal to the (``infra-red non-renormalization'') 
argument of Coleman and Grossman\cite{cg} 
that only a fermion triangle diagram, with particular helicities for  
the couplings, can produce the anomaly infra-red divergence.

We introduce external transverse momenta 
(that we essentially ignored in the previous Section)
using the notation illustrated in Fig.~5.2(a).
\begin{center}
\parbox{2.9in}{
\epsfxsize=2.5in
\epsffile{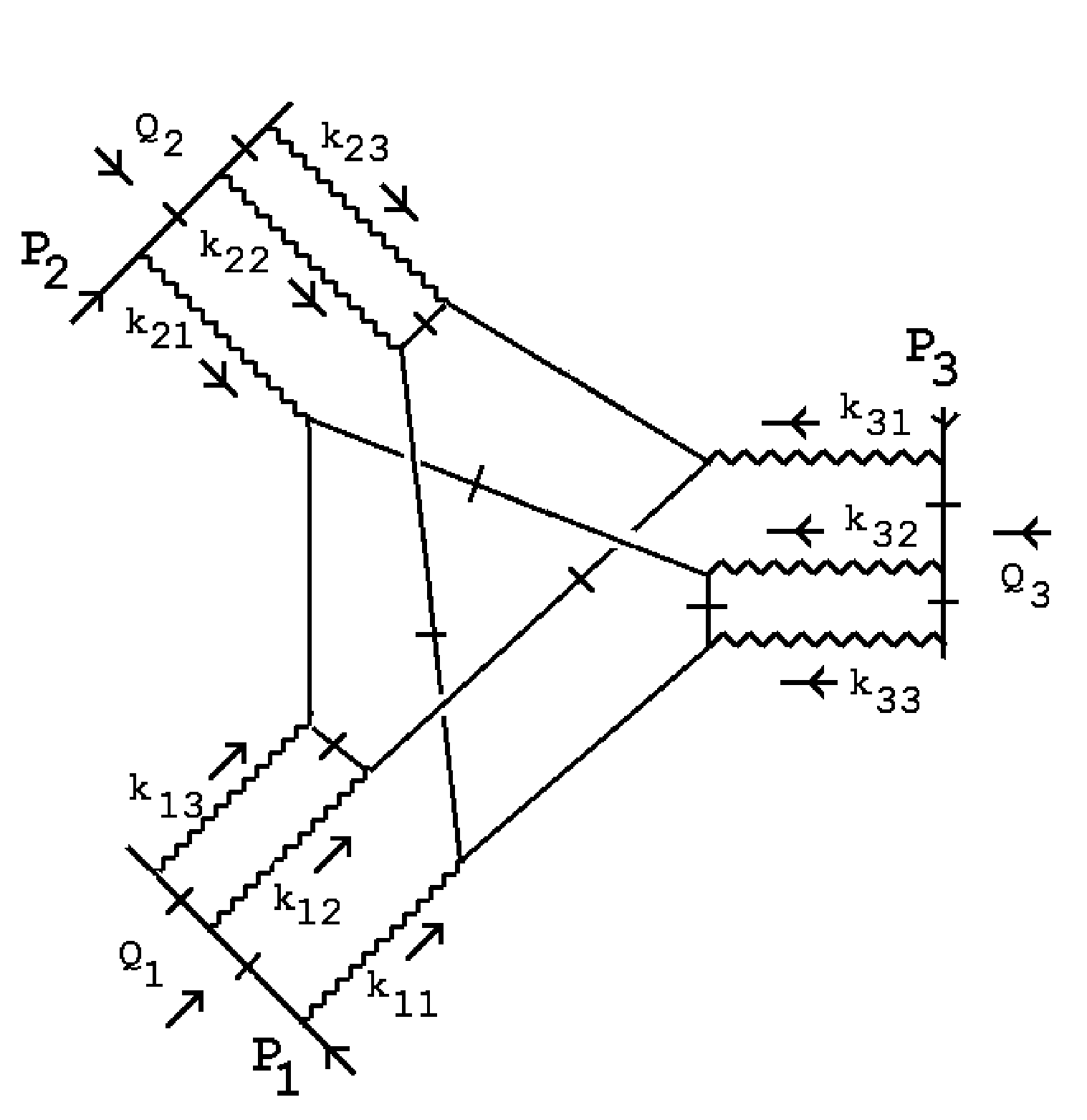}
}
\parbox{2.9in}{
\epsfxsize=2.5in
\epsffile{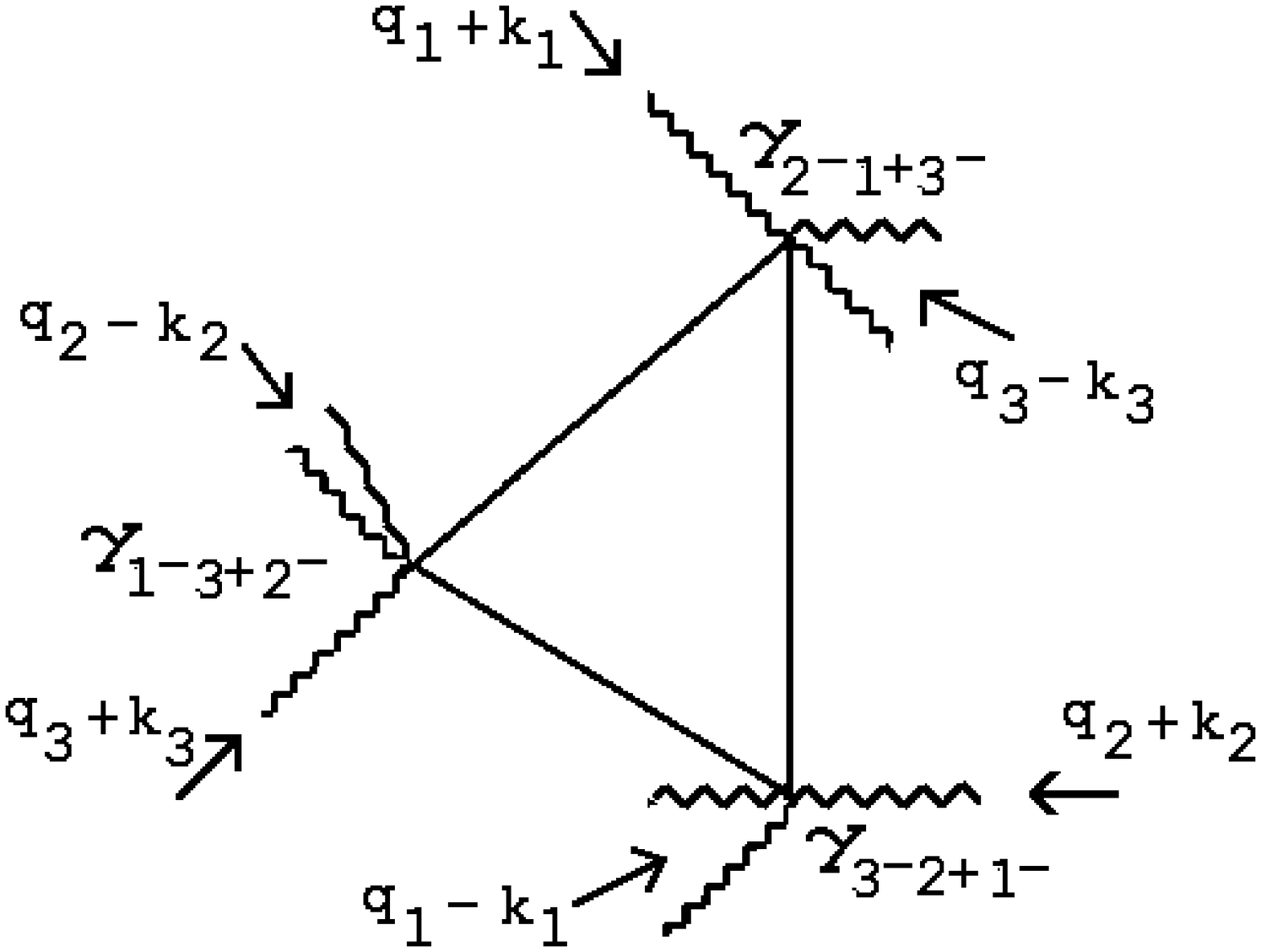}
}
\newline \centerline{(a)\hspace{3in}(b)}

Fig.~5.2 (a) Notation for (\ref{611})~~ (b) Notation for (\ref{612})
\end{center}
The resulting asymptotic behavior then has the form
$$
\eqalign{ ~~~~~P_{1^+}~ P_{2^+}~ P_{3^+}~
\prod_{i=1}^3 \int & { d^2 k_{i1}d^2 k_{i2} d^2 k_{i3}\over  
k_{ i1}^2  k_{i2}^2 k_{i3}^2 }  
~~ \delta^2 (Q_{i\perp} -  k_{i1} -  k_{i2}  - k_{i3})~G^3_i(k_{i1},k_{i2},k_{i3} 
\cdots) \cr 
&~~~~~~~~\times ~ R^9(Q_1,Q_2,Q_3,
k_{11}, k_{12},k_{13} \cdots )} \auto \label{611}
$$
where $R^9$ is the triangle diagram illustrated in Fig.~5.1(c). Note that
this diagram depends only on $k_{i2} + k_{i3}$ (i.e. it is
independent of $k_{i2} - k_{i3}$).

\subhead{5.2 The Reggeon interaction Anomaly}
 
By comparing with the generalization\cite{arw01} 
of (\ref{2ra1}) and (\ref{2ra}) to three reggeons in each $t$-channel,
we can directly interpret $R^9$ as a nine-reggeon interaction. 
If we write 
$$
k_{i1} ~= ~q_i + k_i~, ~~~~ k_{i2} ~= ~(q_i - k_i)/2 -k_i'~, ~~~~
k_{i3} ~=~(q_i - k_i)/2 +k_i'~,
\auto\label{dki6}
$$
then the momentum flow into the triangle diagram of Fig.~5.1(c) is as 
shown in Fig.~5.2(b). Using momentum conservation, i.e. 
$$
q_1~+~q_2~+~q_3~=~0
\auto\label{mcon}
$$
$R^9$, which does not depend on the $k_i'$, 
can be written (very similarly to (\ref{580})) as 
$$
\eqalign{R^9(q_1,q_2,q_3&,k_1,k_2,k_3) ~=
\int d^4 k Tr \{ 
\gamma_5 \gamma^{1^-3^+2^-}(\st{k}+ \st{k}_2 + \st{q}_3 - \st{k}_1 )\cr
&{ 
\gamma_5 \gamma^{2^-1^+3^-}(\st{k}-\st{q}_2 + \st{q}_3 -\st{k}_2 -\st{k}_3) 
\gamma_5 \gamma^{3^-2^+1^-}
(\st{k}+ \st{k}_1 - \st{q}_2 -\st{k}_3)\} 
\over  (k + k_2 + q_3 - k_1 )^2  
(k -q_2 +q_3 -k_2 -k_3)^2 
 (k + k_1 - q_2 - k_3)^2 } }
\auto\label{612}
$$
where 
$$ 
\eqalign{\gamma^{1^-3^+2^-}~&=~
\gamma_{1^-}\gamma_{3^+}\gamma_{2^-} ~=~\gamma^{-,-,-}~-~ i~
\gamma^{-,-,+}  ~\gamma_5 \cr
\gamma^{2^-1^+3^-}~&=~\gamma_{2^-}\gamma_{1^+}\gamma_{3^-}
 ~=~\gamma^{-,-,-}~-~ i~
\gamma^{+,-,-}  ~\gamma_5 \cr
\gamma^{3^-2^+1^-}~&=~\gamma_{3^-}\gamma_{2^+}\gamma_{1^-}
 ~=~\gamma^{-,-,-}~-~ i~
\gamma^{-,+,-}  ~\gamma_5 }
\auto\label{g63}
$$
and $\gamma^{\pm,\pm,\pm}$ is defined by (\ref{g64}). Again we obtain
a particular component of the tensor that describes the triangle diagram
contribution to a three current vertex function, i.e. we can write
$$
\eqalign{R^9(q_1,q_2,q_3,k_1,k_2,k_3)~=~&
(n^{-,-,-}_{\mu} -i~ n^{-,-,+}_{\mu})
(n^{-,-,-}_{\alpha} -i~ n^{+,-,-}_{\alpha})
(n^{-,-,-}_{\beta} -i~ n^{-,+,-}_{\beta})\cr
& ~~~~~~\times~~T^{\mu\alpha\beta}(k_1-k_3-q_2,k_2-k_1--q_3) }
\auto\label{tric}
$$
where $T^{\mu\alpha\beta}$ is the triangle diagram three current amplitude.

To discuss the occurence of the anomaly in (\ref{g63}) we first recall
the general invariant decomposition of 
$T_{\mu \alpha \beta}$ as discussed in \cite{arw02}. With the notation
illustrated in Fig.~5.3  
\begin{center}
\leavevmode
\epsfxsize=2.2in
\epsffile{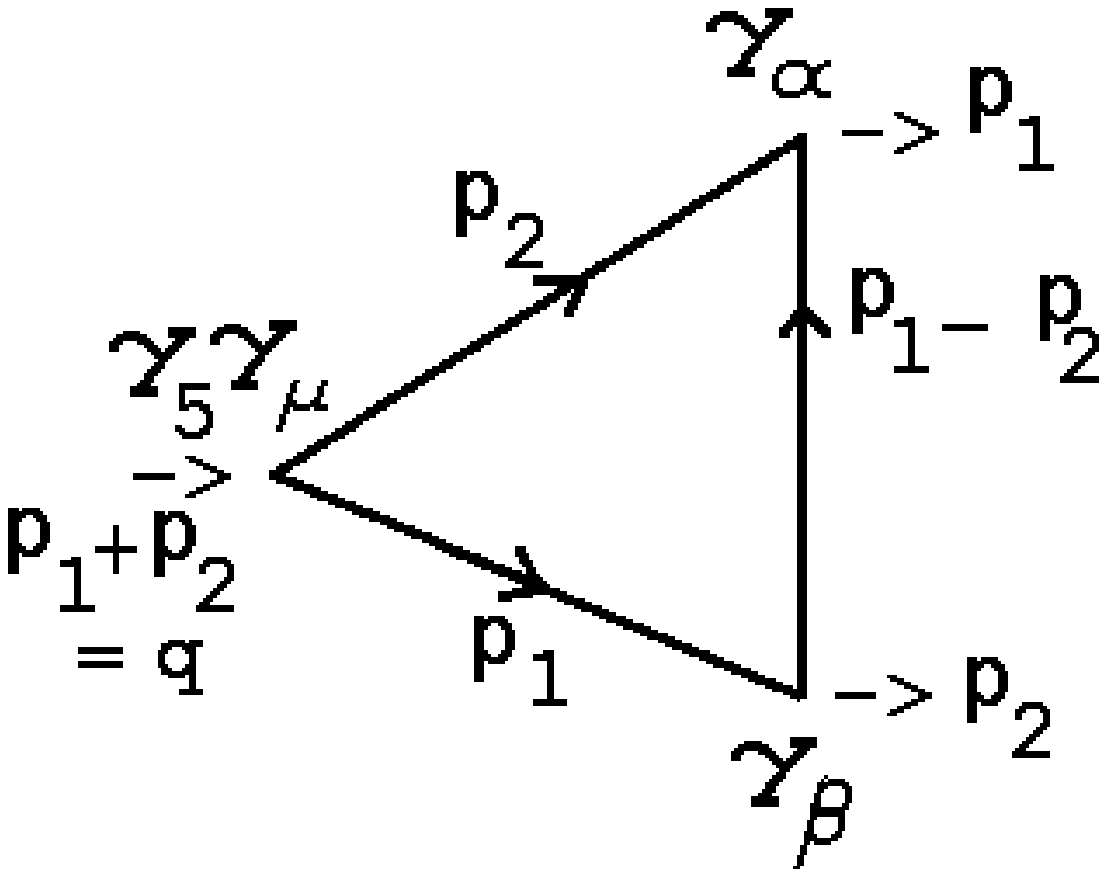}

Fig.~5.3 Triangle Diagram Notation
\end{center}
we can write 
$$
\eqalign{T_{\mu \alpha \beta}(p_1,p_2) ~&= ~ A_1~
{\hbox{\large $\epsilon$}}_{\sigma\alpha\beta\mu}~ p_1^{\sigma}  ~+~ A_2~ 
{\hbox{\large $\epsilon$}}_{\sigma\alpha\beta\mu} ~p_2^{\sigma} 
~+~A_3~
{\hbox{\large $\epsilon$}}_{\delta \sigma\alpha\mu}~ 
p_{1\beta}p_1^{\delta} p_2^{\sigma}  \cr
~~~& +~A_4~  {\hbox{\large $\epsilon$}}_{\delta \sigma\alpha\mu}
~ p_{2\beta}p_1^{\delta}
p_2^{\sigma}~+~A_5~  {\hbox{\large $\epsilon$}}_{\delta \sigma\beta\mu}
~p_{1\alpha}p_1^{\delta}
p_2^{\sigma}~+~A_6~ {\hbox{\large $\epsilon$}}_{\delta \sigma\beta\mu} 
~ p_{2\alpha}p_1^{\delta}
p_2^{\sigma} }
\auto\label{inde}
$$
The ultra-violet anomaly occurs in the first two terms of (\ref{inde}),
i.e. 
$$
T_{\mu \alpha \beta}(k_1,k_2) ~= ~ {1\over 4 \pi^2}~
{\hbox{\large $\epsilon$}}_{\sigma\alpha\beta\mu}~ p_1^{\sigma}  ~+~ 
{1\over 4 \pi^2}~ 
{\hbox{\large $\epsilon$}}_{\sigma\alpha\beta\mu} ~p_2^{\sigma} ~~+~~\cdots
\auto\label{uvco}
$$ 
leading to the well-known divergence equation 
$$
(p_1 + p_2)^{\mu}~T_{\mu \alpha \beta}~=~
{1\over 2 {\pi}^2 }~{\hbox{\Large $\epsilon$}}_{\delta\sigma\alpha\beta} 
~p_1^{\delta} p_2^{\sigma}
\auto\label{awi}
$$
The ultra-violet anomaly can therefore   
appear only in a tensor component with three orthogonal Lorentz indices.
If we keep just the $\gamma_5$ parts of the
three vertices in (\ref{g63}) we obtain a non-zero projection on 
such a tensor component. In fact this contribution to $R^9$ retains the full
symmetry of the original feynman diagram of Fig.~3.7(a) and, as a result, 
has the necessary symmetry to contain the ultra-violet anomaly. 

The infra-red ``anomaly pole'' occurs in  $A_3$ and $A_6$. When $p_1^2=0$
$$
A_3~=-A_6~=~{1 \over 2{\pi}^2 }{1 \over p_2^2 -q^2}
 \biggl({p_2^2 \over p_2^2 - q^2}~ln{p_2^2
\over q^2} ~-~1 \biggr) 
\auto\label{k1m0}
$$
and when $p_2^2 \to 0$ 
$$
A_3~\to ~ {1 \over 2 {\pi}^2}~  {1 \over q^2}
\auto\label{k20}
$$
That is, a pole appears in $A_3$ ($= - A_6$) and, as a consequence of all
divergence equations, the coefficient is also given by the anomaly.
If, instead, we integrate over spacelike values of $q^2$, we obtain 
$$
\int~ dq^2~A_3(q^2,p_2^2)~f(q^2,p_2^2)~
\centerunder{$\longrightarrow$}{\raisebox{-6mm}{$k_2^2 \to 0$}} ~ ~
{1\over \pi}~f(0,0)~=~ \int~ dq^2~{1\over \pi}\delta(q^2)~f(q^2,0)
\auto\label{dfn}
$$
(provided $f(q^2,p^2_2)$ is regular at $q^2,p_2^2 = 0$). 
As we discussed in \cite{arw02}, the pole (\ref{k20}) is responsible
for the appearance of a Goldstone boson pole in amplitudes containing
the chiral flavor anomaly. For the reggeized gluon interactions 
that we are discussing it is the
$\delta$-function property that is important. 

The tensor factors multiplying $A_3$ and $A_6$ in $T_{\mu \alpha \beta}$
potentially suppress the $q^2 \to 0$ divergence due to the anomaly pole.
To describe this, we consider a specific momentum configuration, e.g.
$$
\eqalign{~~~~&p_1 ~= ~(p/\sqrt{2},p /\sqrt{2},0,0)  \cr
& p_2 ~= ~(- p/\sqrt{2},- p\cos{\theta}/\sqrt{2}, 0,
- p \sin{\theta}/\sqrt{2})\cr
&\centerunder{$\sim$}{\raisebox{-4mm}{$\theta \to 0$}} ~-~p_1 ~-~ 
(0,0, p \theta/\sqrt{2}, 0)~= ~-~p_1 ~-~(0,0,q,0)   }
\auto\label{k+k-2}
$$
where
$$
q^2 ~=~ (p_1 +p_2)^2 ~
~\centerunder{$\sim$}{\raisebox{-4mm}{$\theta \to 0$}}
\auto\label{qth}
$$ 
In this configuration, 
we obtain the largest numerator if we consider
the anomaly contribution of $A_3$ to, say, $ T_{--3}$. This has the form
$$ 
 T_{--3}~=~ {\hbox{\Large $\epsilon$}}_{\sigma\delta - 3}~ 
{p_1^{\sigma} p_2^{\delta}~p_{1-} \over q^2} 
~~=~{~p^2 [p \theta /\sqrt{2} ] \over q^2}~ ~~
\centerunder{$\sim$}{\raisebox{-4mm}{$\theta \to 0$}}~~~
{\sqrt{2}p \over \theta}
\auto\label{A3an11}
$$
and so the divergence is suppressed, but only partially. A 
divergence of the form (\ref{A3an11}) is the strongest that can be obtained.

In general, to obtain the maximal infra-red divergence
we must have a component of $T_{\mu\alpha\beta}$  with
$\mu= \alpha $ and with $\mu$ having a light-like projection.
The corresponding
light-like momentum must also flow through the diagram. $\beta $ must 
have an orthogonal spacelike projection and the 
transverse momentum that vanishes, as $q^2 \to 0$,
must be in the remaining orthogonal
spacelike direction. If we choose the $\gamma_5$
component from all three vertices in (\ref{g63}) 
the first requirement is not met. However, if we choose the 
$\gamma_5$ component from one 
vertex and choose the vector coupling from the other
two vertices, it is met.
The finite light-like momentum involved must then have a projection on 
$n^{-,-,- \mu}$ and the orthogonal spacelike momentum must be distinct
in each case. There is then a divergence of the form of 
(\ref{A3an11}).

The three possibilities for the infra-red anomaly divergence 
to occur are associated 
with the three distinct hexagraphs described in \cite{arw01}, 
and hence with three distinct 
helicity amplitudes.
In the analysis of \cite{arw01} the co-ordinates used were asymmetric and
were chosen to isolate one anomaly configuration. These co-ordinates were
naturally associated with a particular hexagraph and the corresponding 
helicity amplitudes and limits. We could equally well use these 
co-ordinates in discussing Fig.~3.8. In which case, the $\gamma_5$
and non-$\gamma_5$ components in two of the three 
$\gamma$-matrices in (\ref{g63}) are interchanged.  
The anomaly pole contribution then comes from the three $\gamma_5$ components.
In either case, the result is the same. 
We anticipate, but will not attempt to demonstrate here,
that for each hexagraph amplitude 
the ultraviolet anomaly and anomaly pole components are
related by reggeon Ward identities, just as corresponding components
in (\ref{inde}) are related by normal vector Ward identities. (Note that the
``ultraviolet'' region for (\ref{612}) is actually the region $k 
~\centerunder{\raisebox{0.5mm}{$<$}}{\raisebox{0mm}{$\sim$}}~ P_{1^+}
\sim P_{2^+} \sim P_{3^+}$, rather than $k \sim \infty$.)
This implies that the occurrence of the infra-red and ultra-violet anomalies
in diagrams will be closely correlated. We will exploit this in the following.

As discussed at length in \cite{arw01}, while the triple discontinuity
giving Fig.~5.1 occurs in an unphysical region, there will be a
corresponding ``real'' reggeon interaction
in physical regions. In particular,
the anomaly infra-red divergence can occur in the physical-region
configuration shown in Fig.~5.4. (The dots indicate that a local
interaction is involved.)
The $\gamma_5$ interaction is at the intermediate vertex and the 
light-like momenta are as in(\ref{chm1})-(\ref{chm30}).  
Fig.~5.4 can then be identified with
the basic anomaly process of Fig.~3.2 except
that there is an additional wee gluon involved.
There are also additional gluons with finite transverse 
momentum. 
\begin{center}
\epsfxsize=2.2in
\epsffile{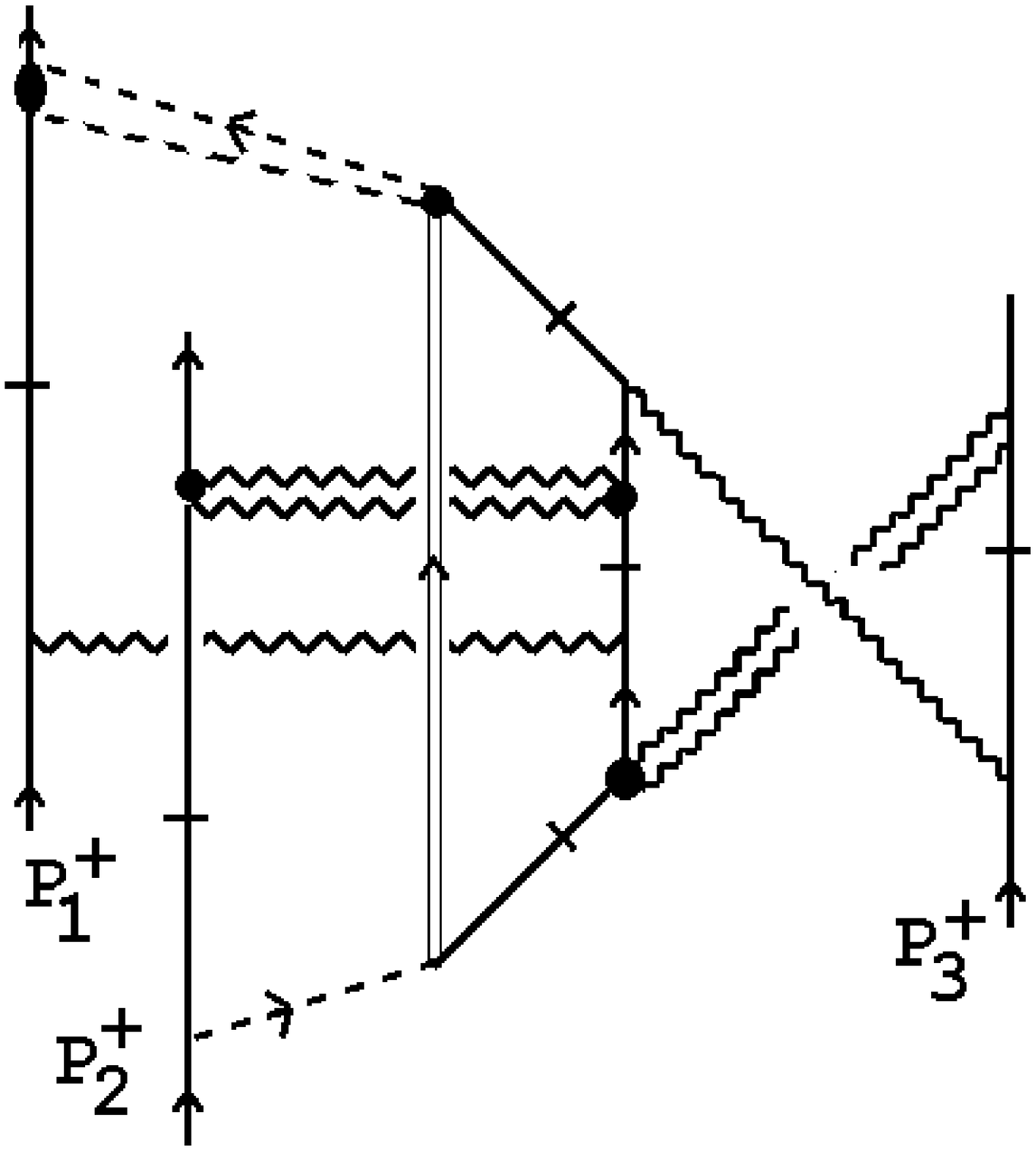}

Fig.~5.4 A basic anomaly process associated with Fig.~5.1.
\end{center}

\subhead{5.3 Other Diagrams}

We now consider whether, based on the discontinuity analysis
of the previous Section, there are other diagrams besides that of 
Fig.~3.7(a) that could contain the anomaly. We will not consider 
all possible diagrams - there are simply too many. We will make
the simplifying assumption that 
only diagrams that are completely symmetric (with respect to the three
$t$- channels) are relevant. There are two 
justifications ons for this assumption. First there is
the infra-red light-cone argument discussed in Section 3. Secondly, we 
anticipate, as we have just discussed, 
that infra-red and ultraviolet anomalies should occur together
so that reggeon Ward identities
are satisfied. It seems that at this ``simplest'' level, where it first
emerges, the ultraviolet anomaly is very likely to require a symmetric
diagram.

If we begin from the diagram of Fig.~3.7(a) and retain only the exterior
lines of the internal loop we obtain the ``bare'' diagram of Fig.~5.4(a).
The exterior lines give the triangle diagram in the reggeon vertex. 
Since they must remain uncut when a triple discontinuity is taken they
must remain on the exterior, as in the bare diagram. If we then 
add further lines such that a complete loop is formed
within a symmetric diagram, and there is no sub-loop, the only 
new possibilities (up to reflections) are shown in Figs.~5.4(b)-(d).  
\newline \parbox{1.45in}{
\begin{center}
\epsfxsize=1.1in \epsffile{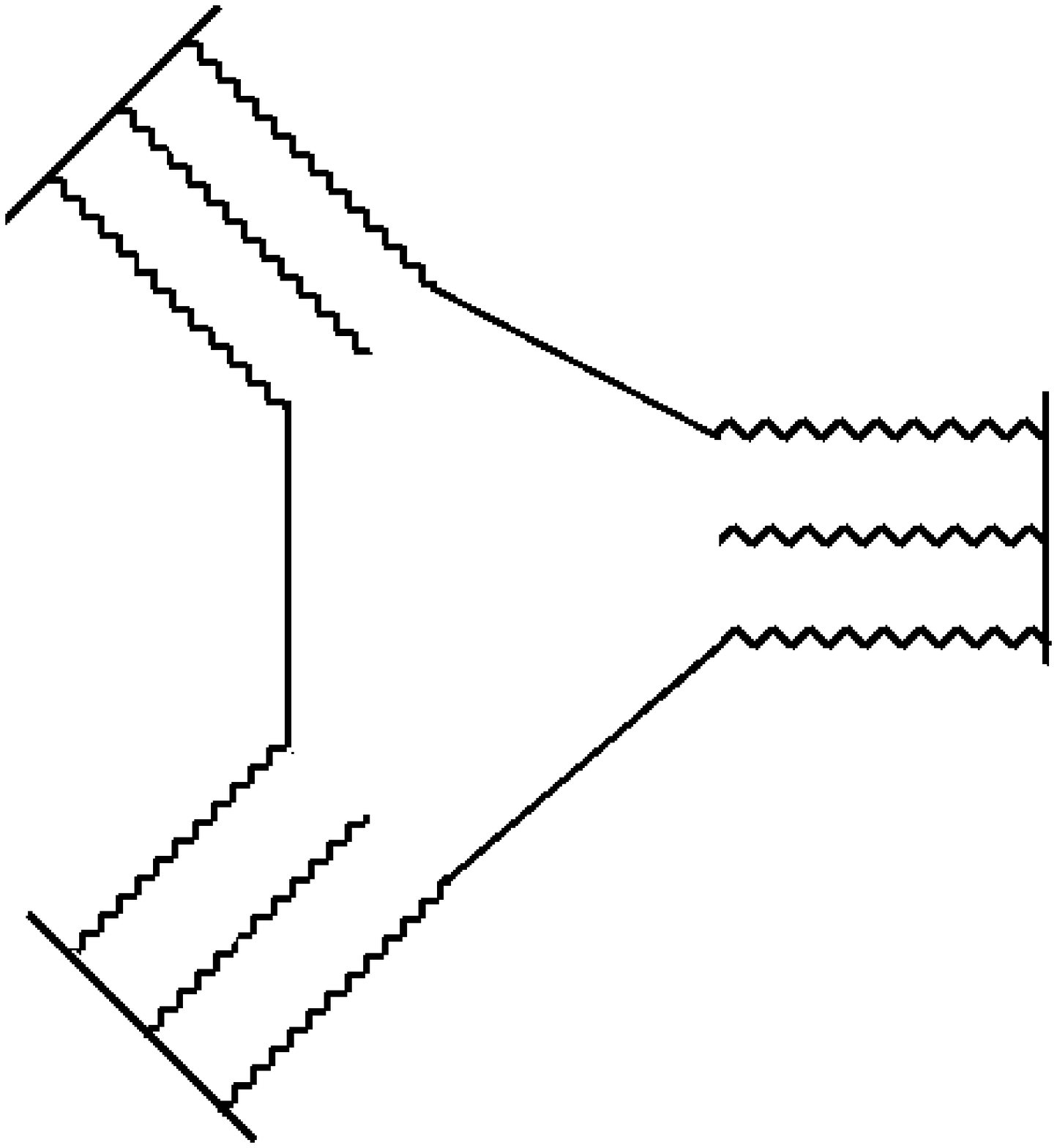}
\newline (a)
\end{center}}
\parbox{1.45in}{
\begin{center}
\epsfxsize=1.1in \epsffile{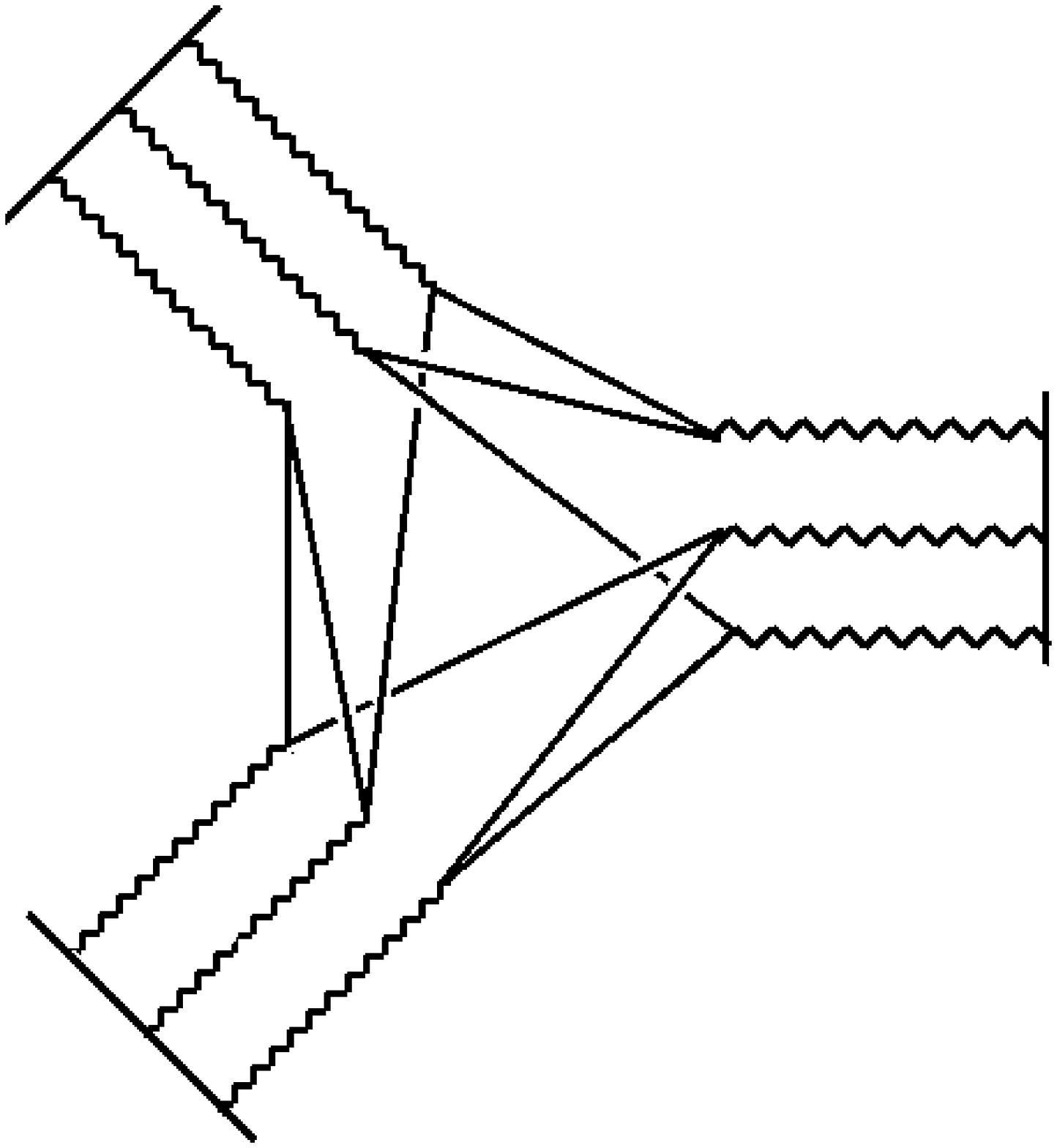}
\newline (b)
\end{center}}
\parbox{1.45in}{
\begin{center}
\epsfxsize=1.1in \epsffile{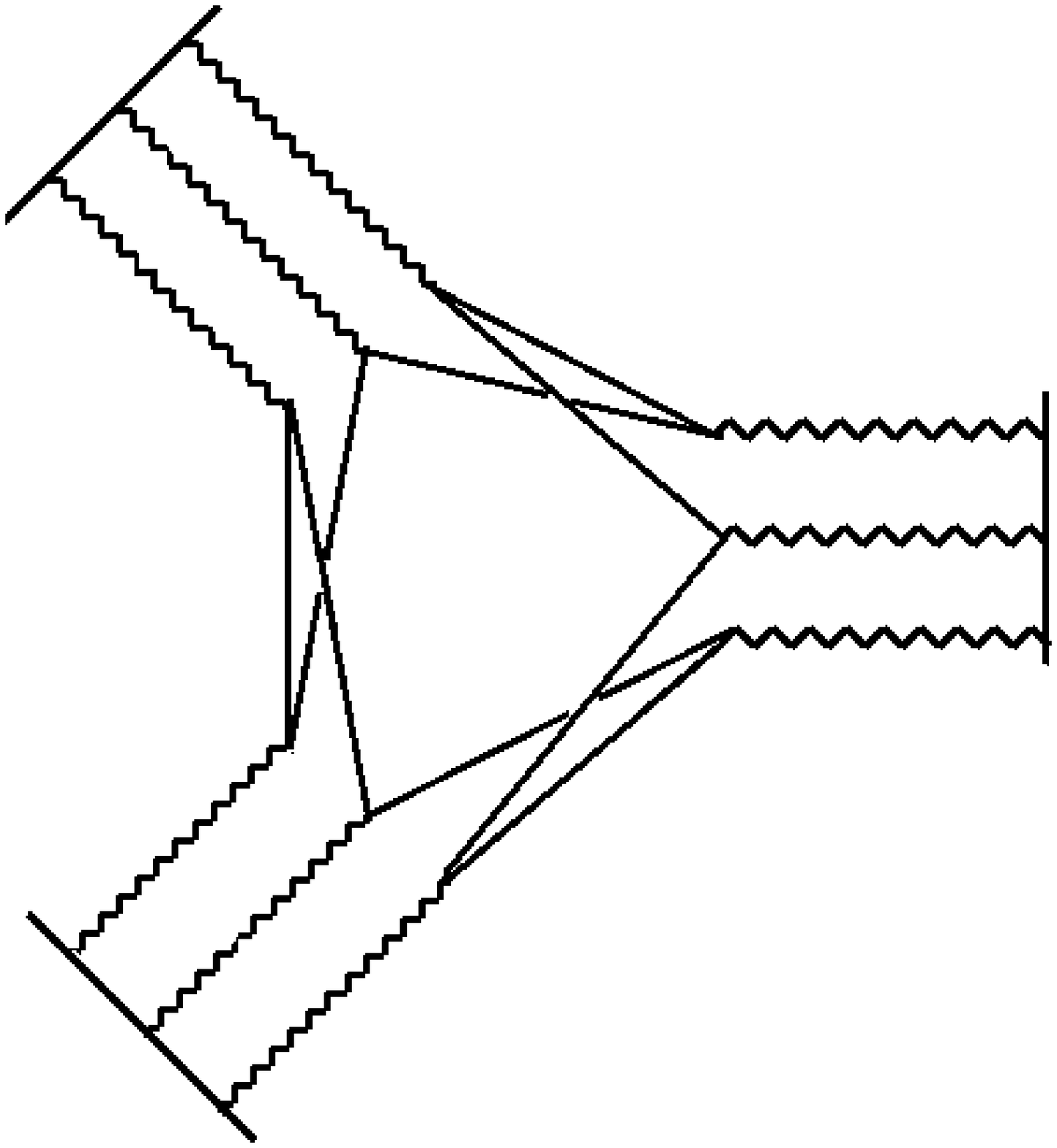}
\newline (c)
\end{center}}
\parbox{1.45in}{
\begin{center}
\epsfxsize=1.1in \epsffile{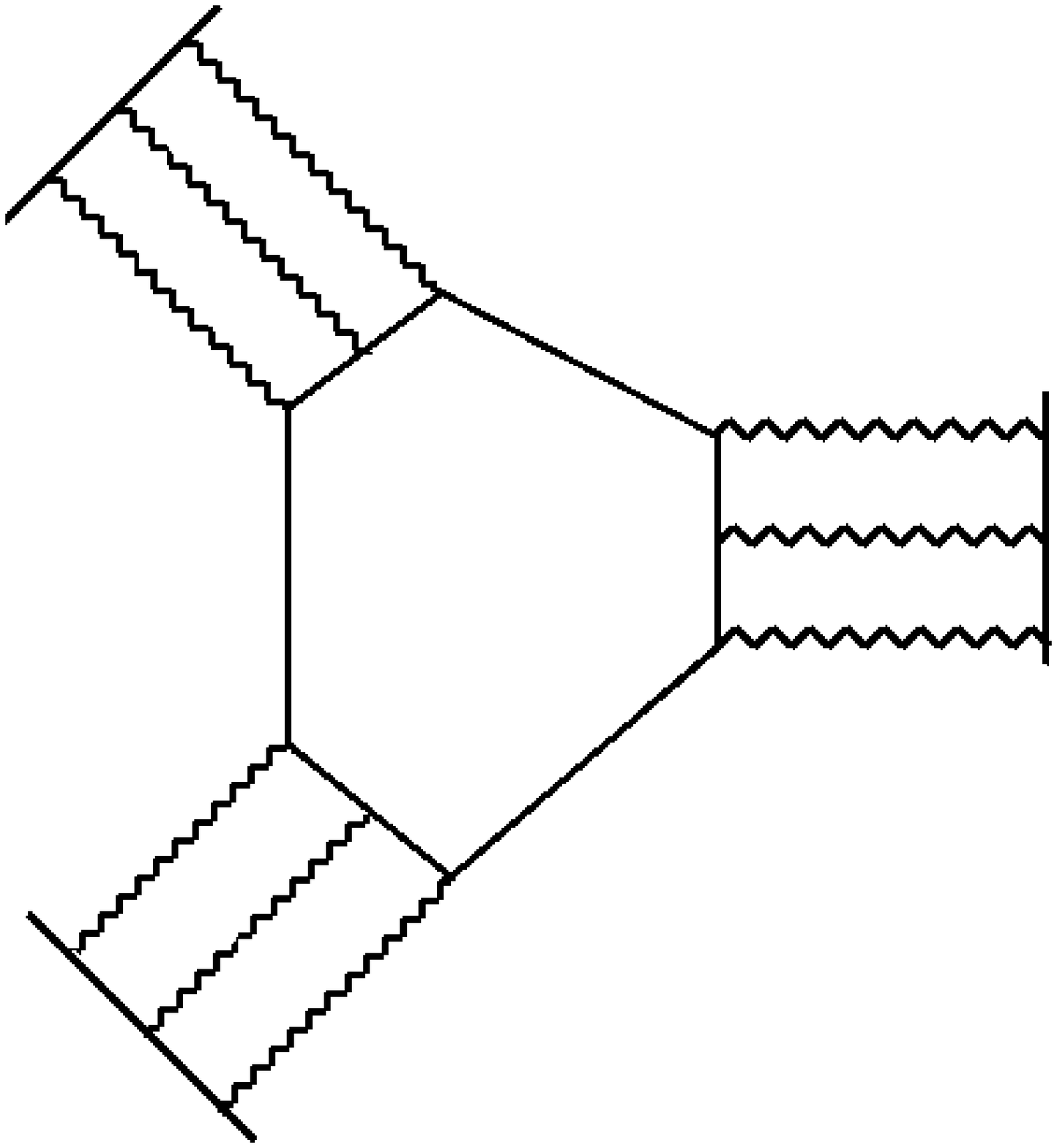}
\newline (d)
\end{center}}
\newline\centerline{
Fig.~5.4 (a) The ``bare'' diagram (b) - 
(d) Full diagrams.}

The diagram of Fig.~5.4(b) can be analysed very similarly
to our analysis of Fig.~3.7(a). 
As we described at the end of the last Section,
a pseudothreshold triple discontinuity 
will be present if the six non-exterior loop lines can be grouped
into three pairs, each associated with a particular discontinuity, 
such that the loop momentum flows across the discontinuity
line in the same direction for each pair.
In Fig.~5.5(a) we have drawn the appropriate
cuts of Fig.~5.4(b) and in Fig.~5.5(b) we have isolated the cut lines that
contribute to one discontinuity.
\newline \parbox{2.4in}{
\begin{center}
\epsfxsize=1.9in
\epsffile{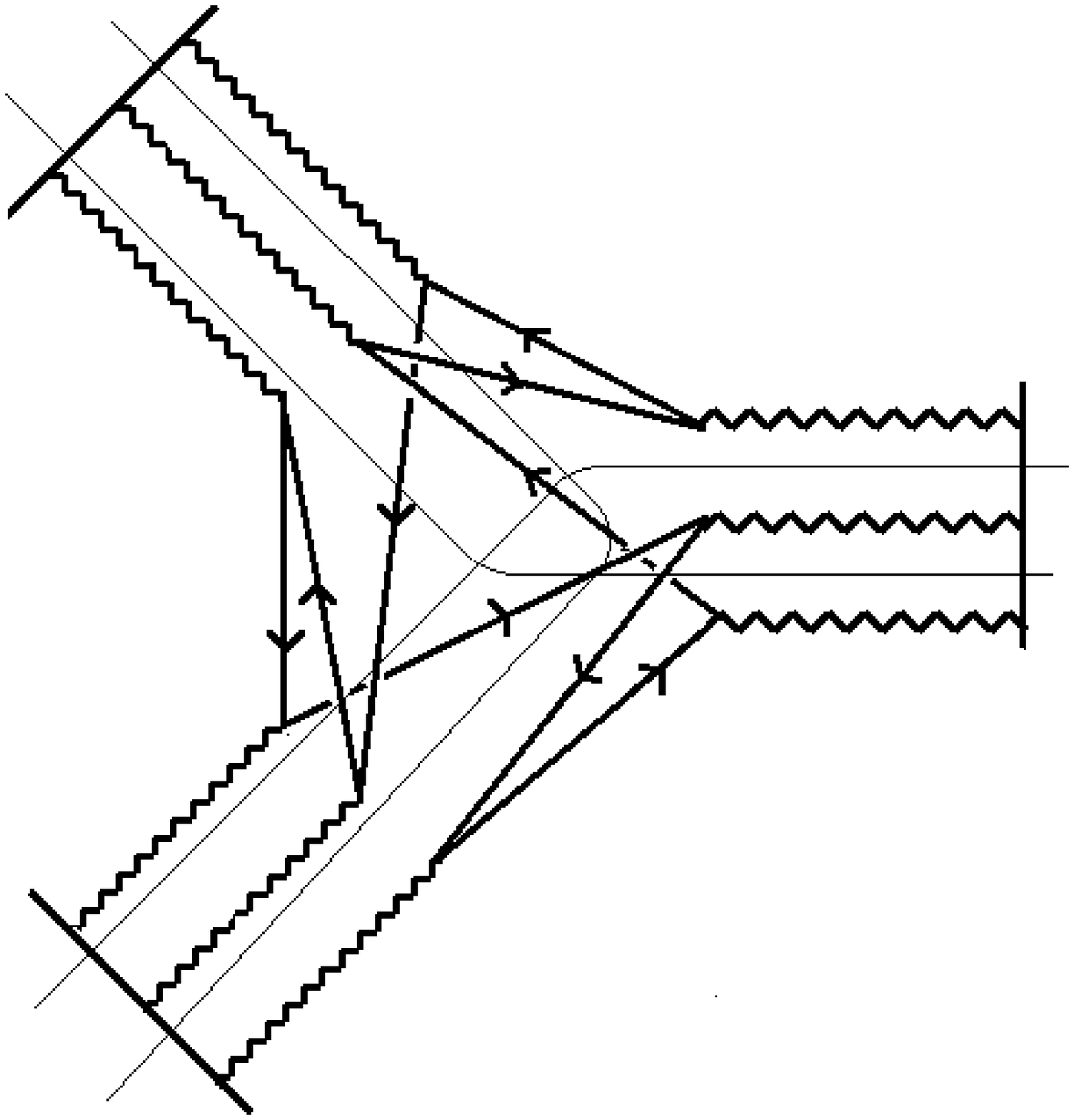}
\newline (a)
\end{center}}
\parbox{2.4in}{
\begin{center}
$~$
\newline $~$
\newline $~$
\epsfxsize=2in
\epsffile{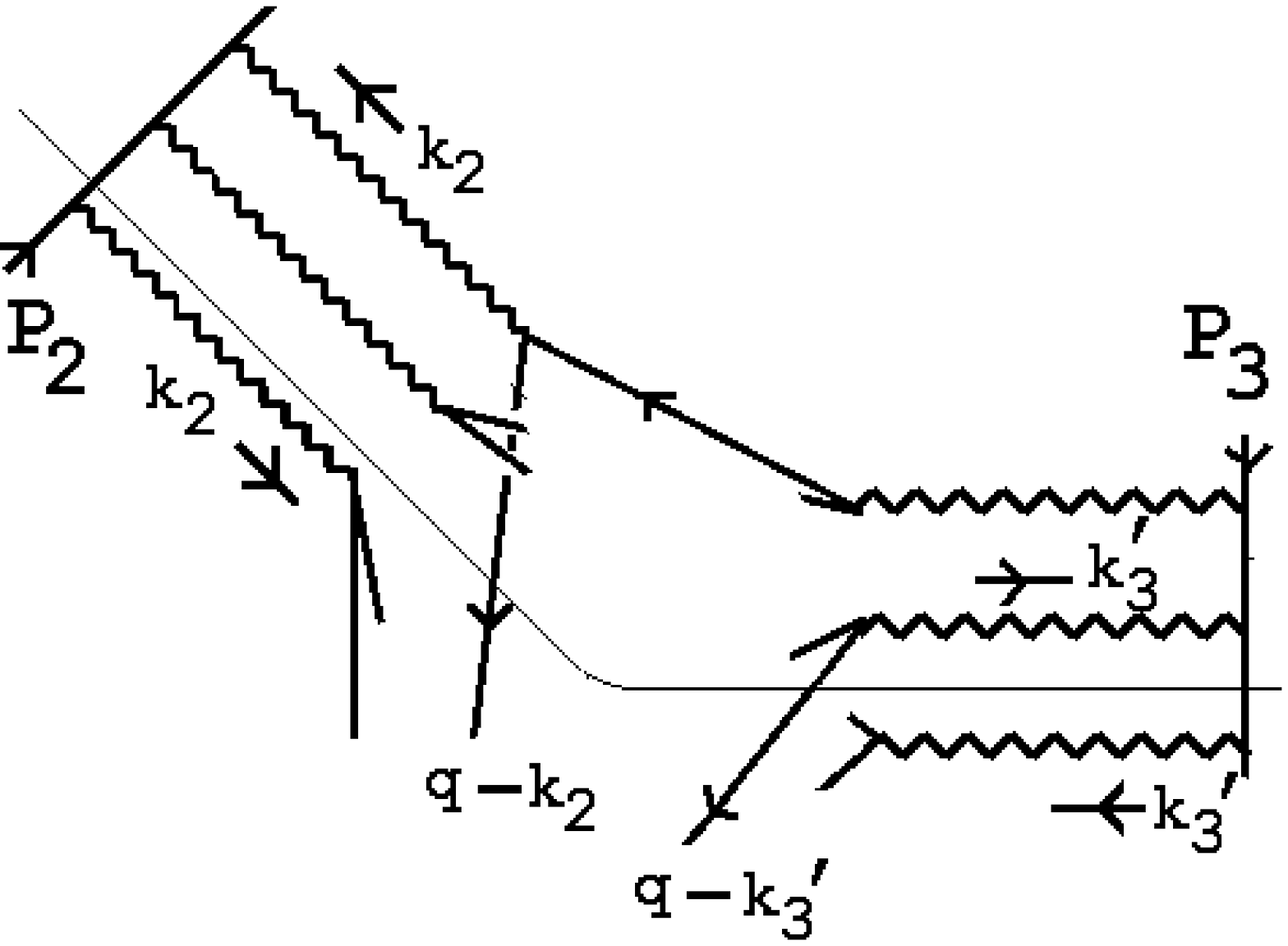}
\newline $~$
\newline (b)
\end{center}
}
\parbox{1.1in}{
\begin{center}
$~$
\newline $~$
\newline $~$
\newline $~$
\epsfxsize=0.75in
\epsffile{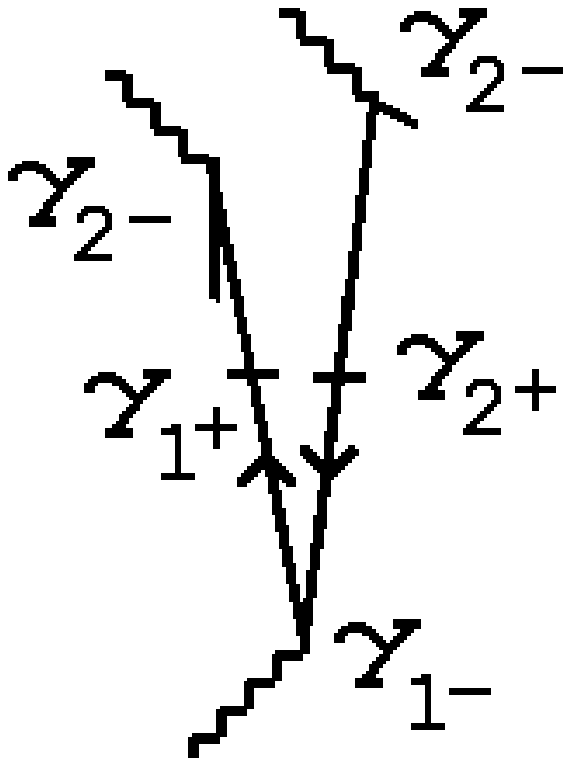}
\newline $~$
\newline $~$
\newline (c)
\end{center}
}
\begin{center}
Fig.~5.5 (a) Cuts (b) one discontinuity (c) a
$\gamma$-matrix vertex, of Fig.~5.3(b).  
\end{center}
Both criteria for a symmetric triple discontinuity are satisfied.
However, we must also consider the $\gamma$-matrix structure of the vertices
that appear in the triangle diagram that is obtained. In fact, we find 
products of $\gamma$-matrices of the form shown in Fig.~5.5(c), which
do not produce the $\gamma_5$ coupling needed for the anomaly. The diagrams 
of Fig.~5.4(c) and (d) clearly do not have sufficient non-planar
structure to give a pseudothreshold triple discontinuity. We conclude, 
therefore, that none of the additional diagrams of Fig.~5.4 can produce
an anomaly contribution to a reggeon vertex. 

\subhead{5.4 Color Factors and Signature}

Similarly to our discussion of the triple discontinuity of Fig.3.7(b),
a priori $R^9$ can contribute to 
vertices for fewer than nine reggeons. However, in \cite{arw01} we 
argued that the anomaly would cancel, after all integrations
over transverse momenta, unless each reggeon state has
anomalous color parity (not equal to the signature). 
When SU(3) color amplitudes are obtained by
first constructing the color superconducting theory with SU(2) color,
as in \cite{arw02},
the relevant reggeon anomaly interactions are those for SU(2) reggeon
states. In this case the simplest reggeon state with anomalous color
parity is the color zero, odd signature, three reggeon state. 
A reggeon state that is 
``vector-like'' in that it has (close to) unit angular momentum 
and appears in odd-signature amplitudes,
is composed of (at least) three gluons, and has abnormal color parity,
has all the quantum numbers of the anomaly current.
As a result, the ultraviolet anomaly discussed above will directly involve
interactions of the anomaly current.
It is somewhat remarkable that we are led directly to 
the anomaly current by looking for the infra-red anomaly within   
reggeon interactions.
For SU(3) color, a two reggeon even signature state with octet color and 
odd color parity would also be possible. For color zero, however, 
the three reggeon state is again the simplest possible. 

If each reggeon state must contain at least three reggeons, 
the lowest-order reggeon vertex that can contain the anomaly 
is the nine reggeon vertex. 
In fact, we showed in \cite{arw01} that 
the analyticity properties of amplitudes imply 
that the anomaly can only appear when 
signature conservation is also satisfied, which it is not
if all three reggeon states carry odd signature. However, this 
conservation rule should be satisfied only after 
all relevant diagrams have been added. This would
include the addition of all diagrams having the structure 
of Fig.~3.7(a) but with (one or two)
incoming and outgoing lines interchanged. 
To avoid this cancelation additional reggeons (reggeized gluons or quarks)
must be present. As we discuss in the next Section,
additional reggeons are also required for the 
infra-red anomaly to play the 
dynamical role we anticipate. 

Amplitudes giving vertices with four reggeons in each reggeon state 
no longer need to be completely symmetric, provided they collapse to give 
a symmetric triangle diagram. In fact, when four reggeons (or more)
are present in each
state a new subtlety arises in the process of taking a  
triple discontinuity. Consider the diagram 
shown in Fig.~5.6, which is a simple generalization of the
diagram of Fig.~3.7(a) that we have discussed so much. Two of the single 
reggeon lines in Fig.~3.7(a) are replaced by two reggeons,  with no
additional non-planarity. In Fig.~5.6 we have also drawn triplets of cuts 
through the diagram in four distinct ways. 
\newline \parbox{1.45in}{
\begin{center}
\epsfxsize=1.2in
\epsffile{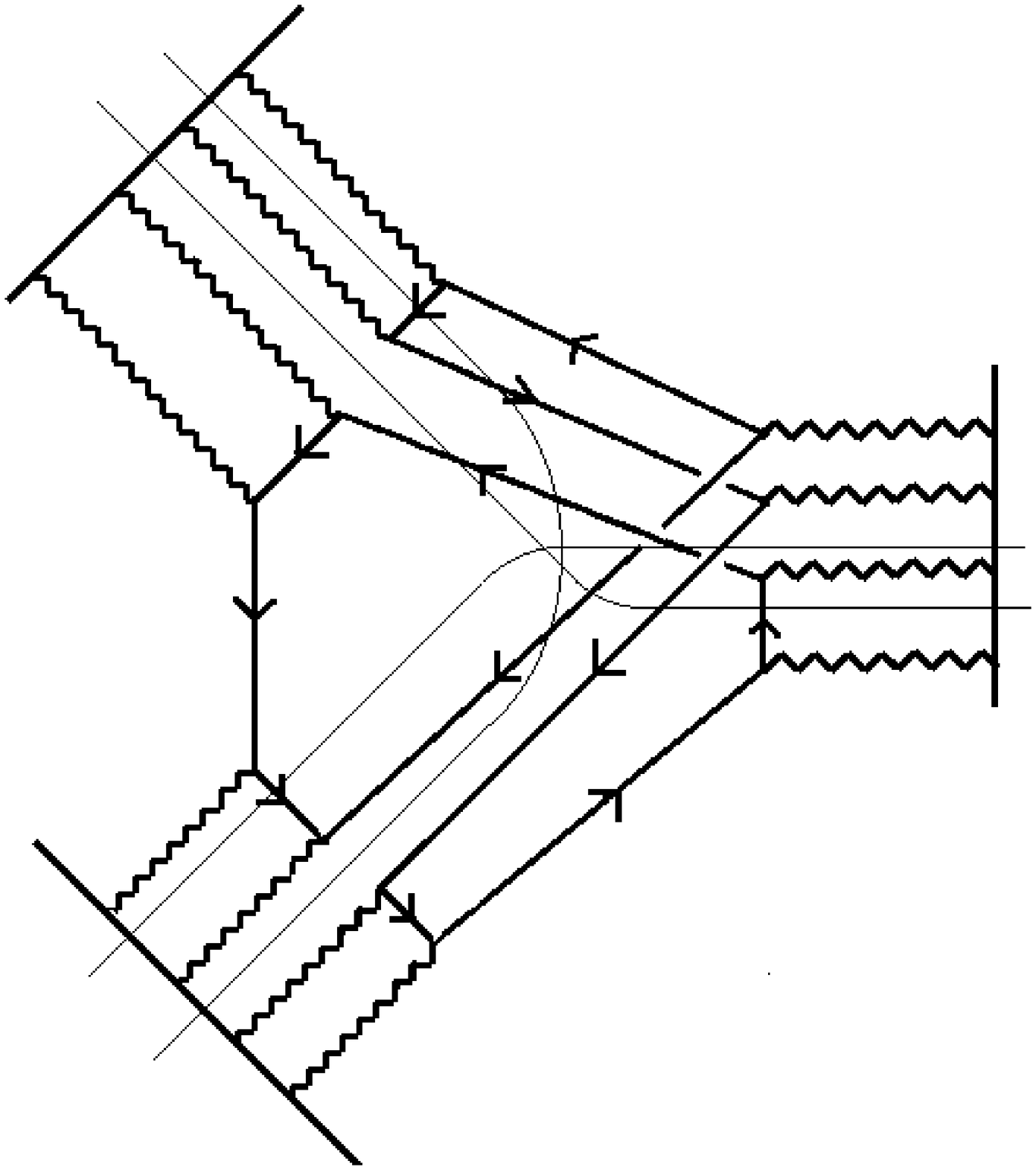}
\end{center}
}
\parbox{1.45in}{
\begin{center}
\epsfxsize=1.2in
\epsffile{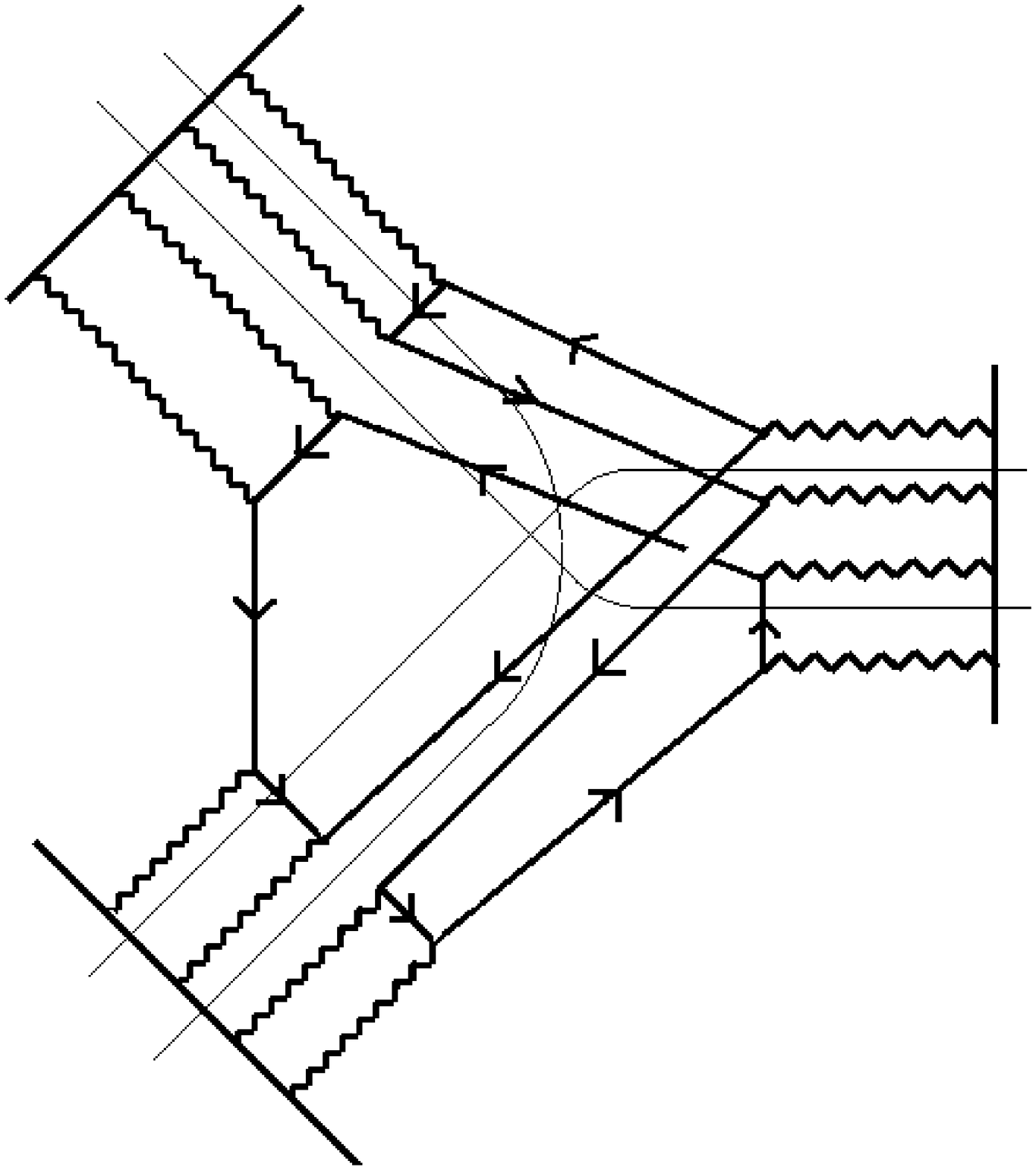}
\end{center}
}\parbox{1.45in}{
\begin{center}
\epsfxsize=1.2in
\epsffile{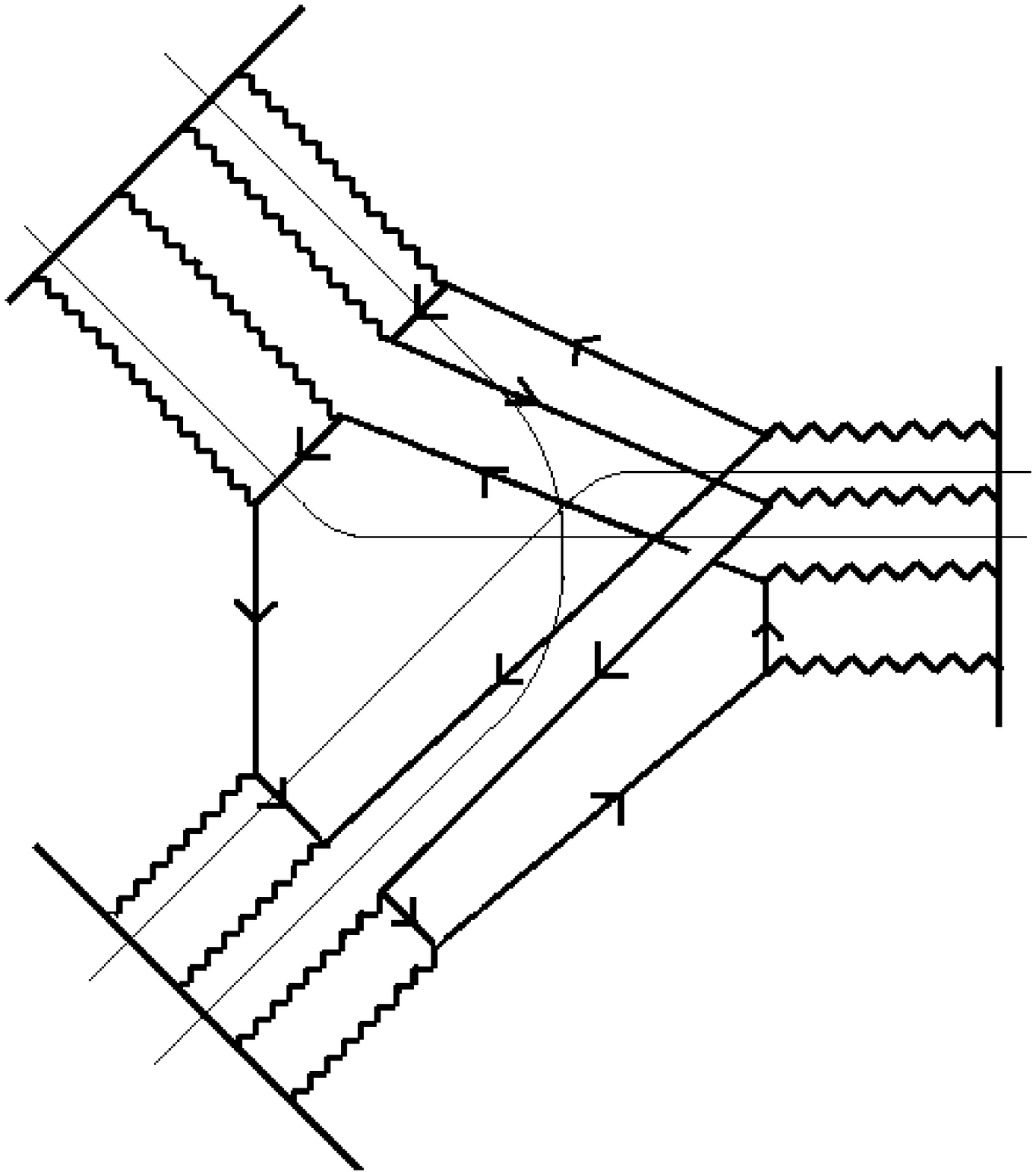}
\end{center}
}\parbox{1.45in}{
\begin{center}
\epsfxsize=1.2in
\epsffile{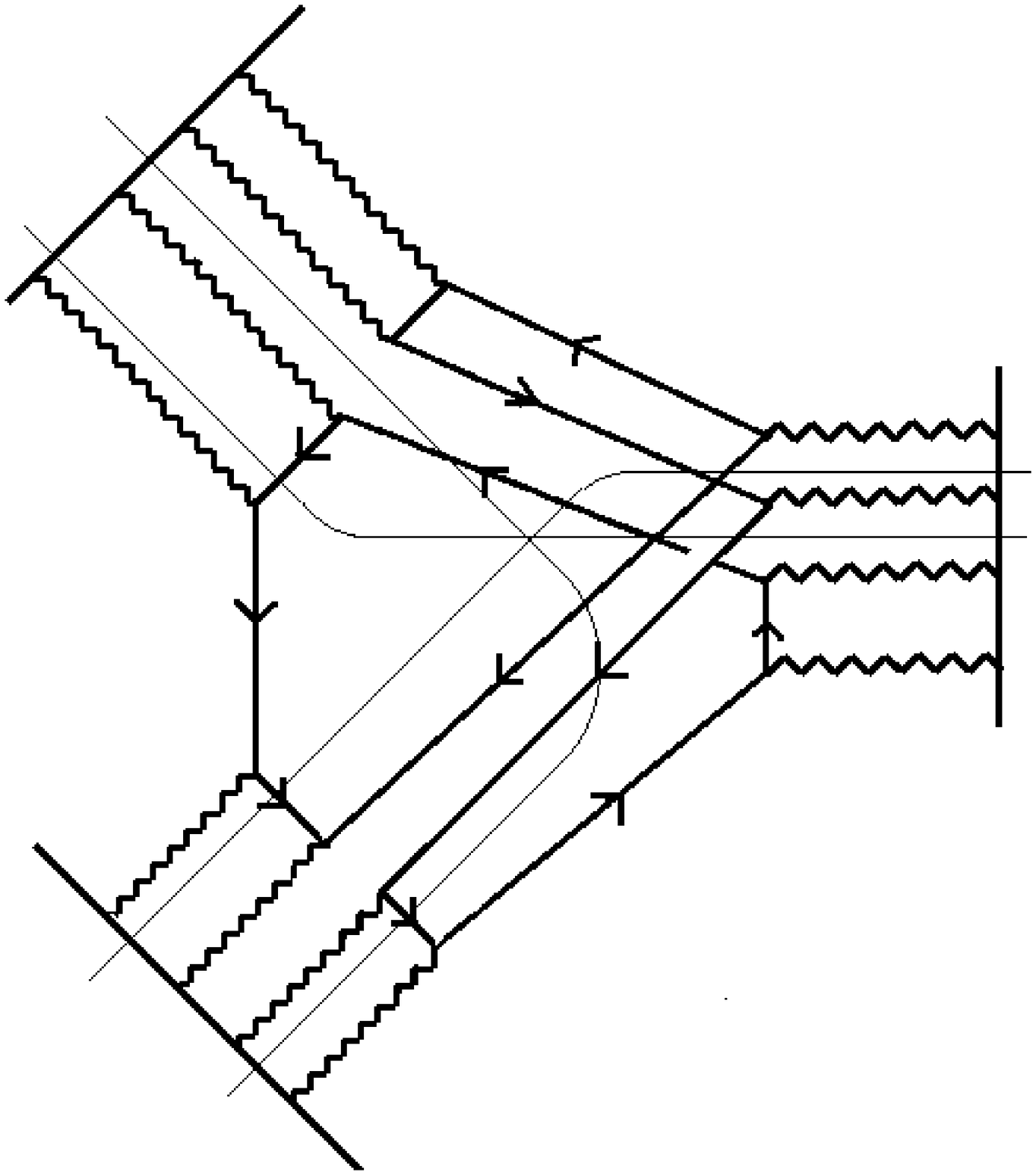}
\end{center}
}
\begin{center}
Fig.~5.6 Four triple cuts of a diagram for four reggeon states
\end{center}
These are the only possible triplets
if we require pairs of non-exterior loop lines to be
associated with each cut, 
such that the loop momentum flows across the cut 
line in the same direction for each pair.
However, if we consider just one triplet and 
take asymptotic discontinuities for each cut (as above)
by considering pairs of external logarithms, we do not obtain a 
complete triple discontinuity of the diagram. There are always 
three internal 
lines that are not put on shell. As a result, one or more, of the pinchings
does not give a complete, invariant, cut of the diagram.
To obtain a genuine  
triple discontinuity we have to combine all the pinchings of logaritms 
involved in
the four sets of cuts shown in Fig.~5.6. All internal lines are then on 
shell and a complete triple discontinuity is obtained. The vertices for the 
corresponding triangle are the rotationally
symmetric products of $\gamma$-matrices shown in 
Fig.~5.7 and so the extracted twelve reggeon vertex will contain the anomaly.
\begin{center}
\epsfxsize=1.5in
\epsffile{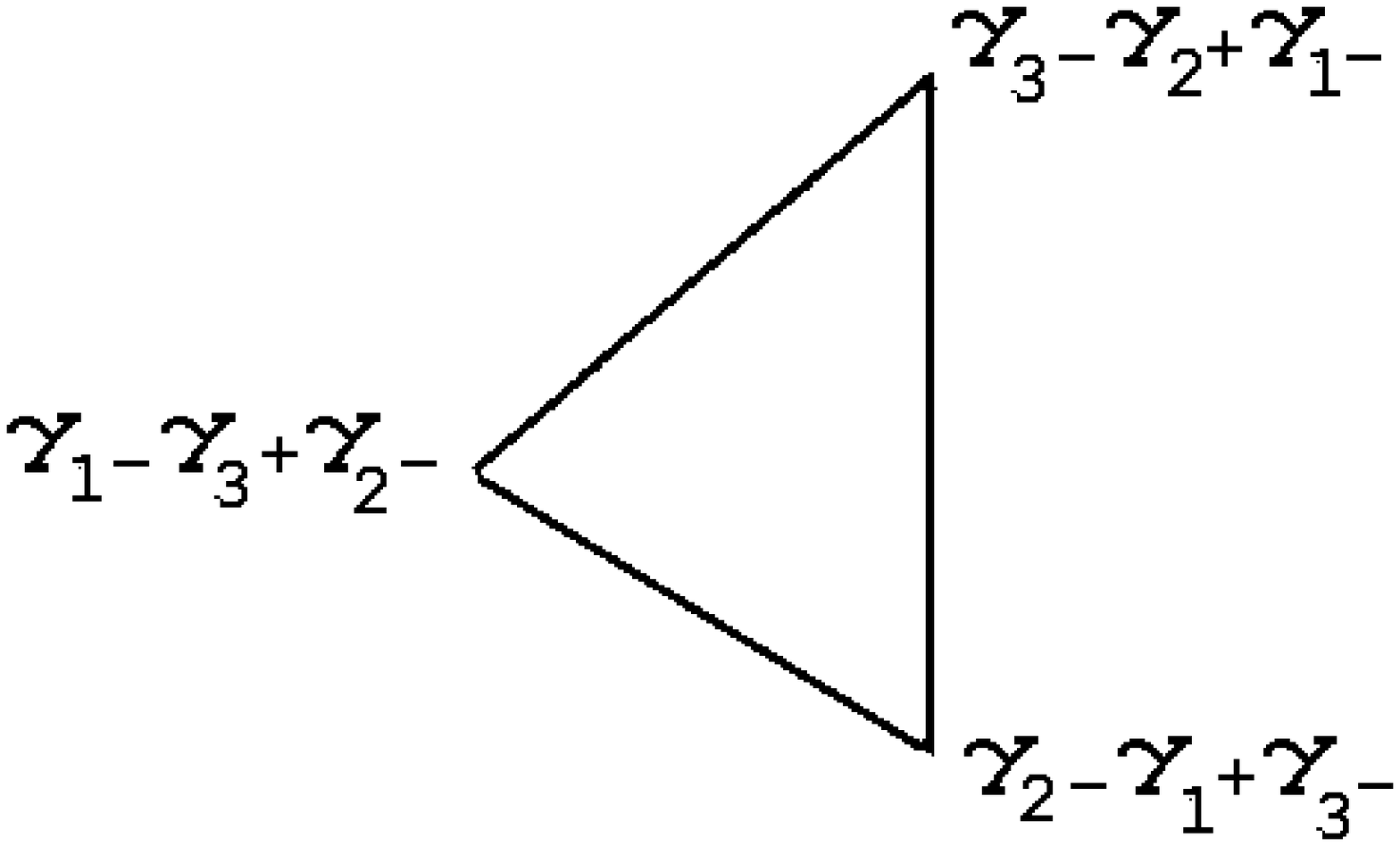}

Fig.~5.7 The $\gamma$-matrix vertices obtained from Fig.~5.6
\end{center}
 
We will postpone a systematic discussion of cancelations,
how and when the anomaly survives after all diagrams 
are summed etc.,  until following papers.
Our priority in this paper has been simply
to find diagrams in which an asymptotic 
discontinuity analysis determines 
that the anomaly is definitively present
in the extracted reggeon interaction. 

\newpage 

\mainhead{6. PION AND POMERON VERTICES IN COLOR SUPERCONDUCTING QCD}

For completeness, 
we briefly describe the physical pomeron and pion interactions
that appear in color superconducting QCD.   
Pion scattering is described in \cite{arw02} and we anticipate that
the corresponding multi-regge amplitudes are 
given by modifying the procedure described in \cite{arw98}
to incorporate the explicit structure of anomaly vertices that we have
since discovered. 
Here we give only enough details to show that 
a staightforward extension of the above 
analysis will demonstrate that such interactions contain the
anomaly. 
 
When the SU(3) gauge symmetry of QCD is broken to SU(2) 
the infra-red divergence\cite{arw02,arw98} that involves the
anomaly and that actually dominates bound-state interactions
occurs in diagrams that are very similar to the ones we have discussed.
The divergence is 
factorized off to give a wee-gluon condensate within both pion 
(i.e. Goldstone boson) bound states and the pomeron.
The pomeron is a single reggeon (i.e. a massive, SU(2) singlet, reggeized 
gluon) within the wee-gluon condensate and 
the pion is a quark/antiquark pair in the same condensate.
A diagram contributing to the 
triple-pomeron interaction is shown in Fig.~6.1(a)
and a class of diagrams contributing to multipomeron interactions.
is shown in Fig.~6.1(b).
\newline \parbox{2.9in}{\begin{center}
\epsfxsize=2.1in
\epsffile{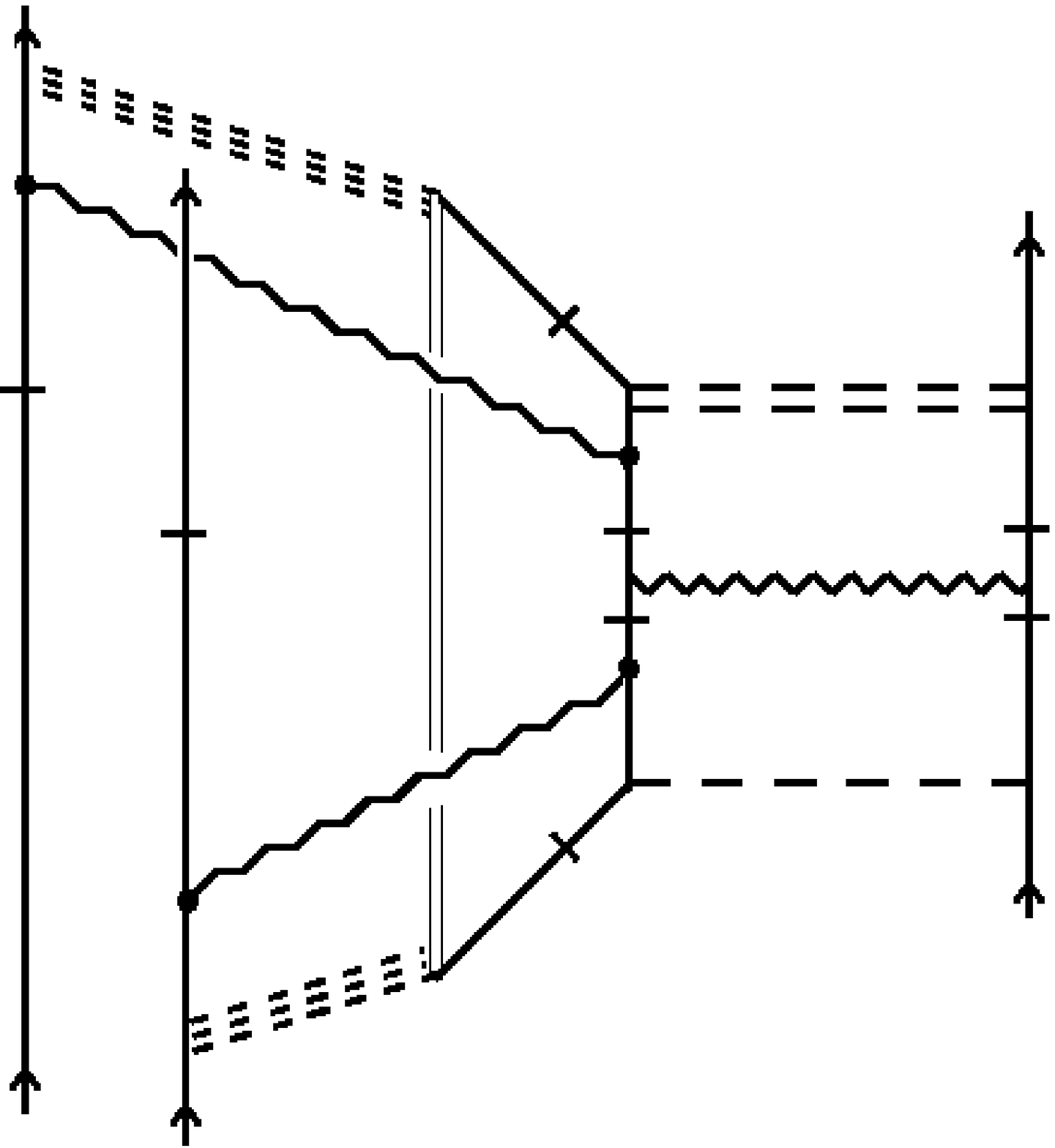}
\centerline{(a)}
\end{center}
}
\parbox{2.9in}{\begin{center}
\epsfxsize=2.1in
\epsffile{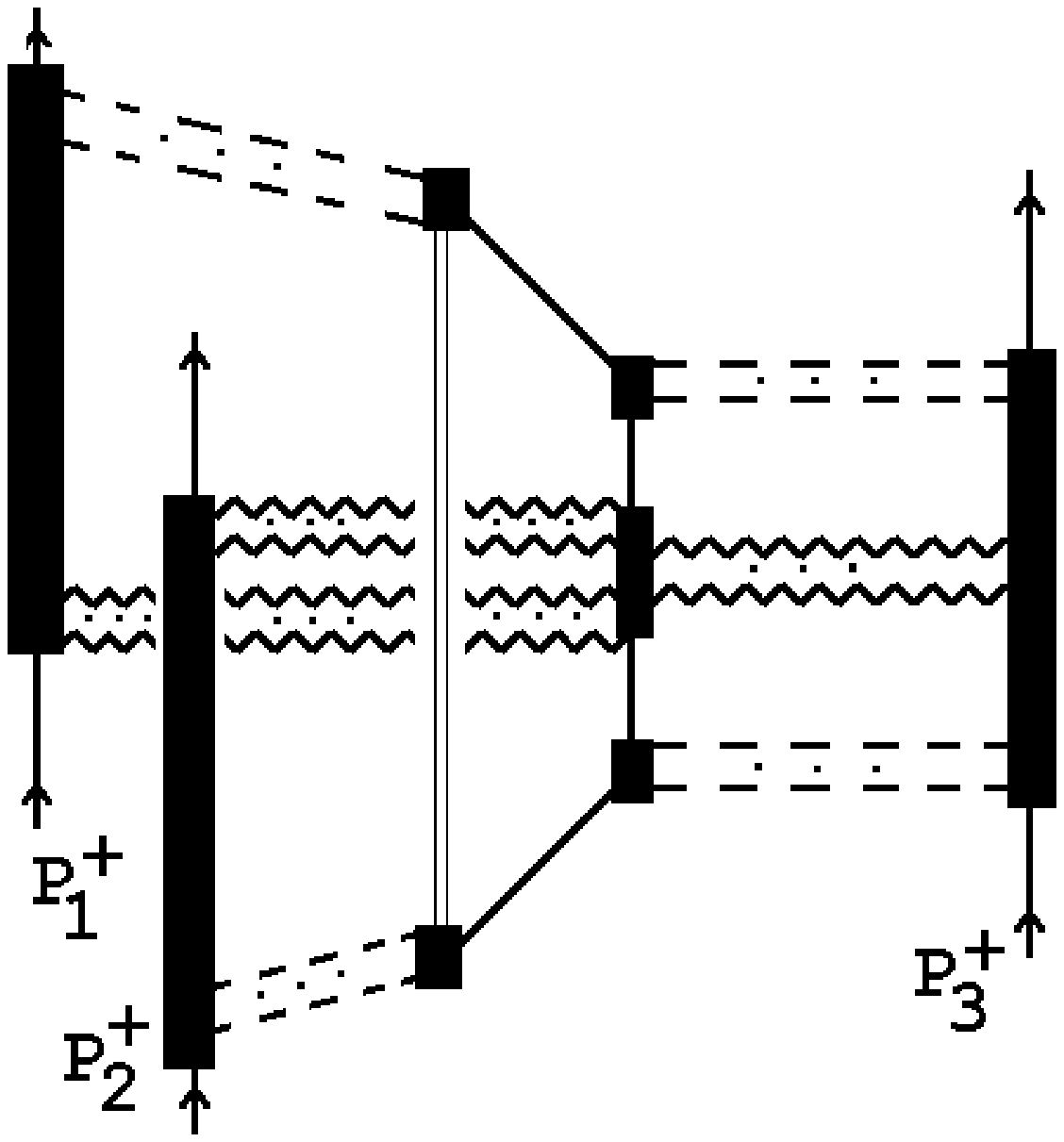}
\centerline{(b)}
\end{center}}
\begin{center}
Fig.~6.1 (a) The triple pomeron interaction (b) A multipomeron interaction. 
\end{center}
The scattering states are now pions and the solid, wavy, lines
are reggeons. The dashed lines represent massless 
gluons that carry zero transverse momentum and, in collaboration with
the anomaly, produce the divergence. The $\delta$-function due to
the anomaly produces transverse momentum conservation
at the vertex where the reggeons interact. 
 
We have drawn the diagrams as basic anomaly processes in Fig.~6.1, rather 
than in a form that exhibits their unphysical discontinuity
properties.
The triple pomeron process in Fig.~6.1(a) corresponds to a
diagram that is just a little more 
complicated than the diagram of Fig.~5.6. There is an additional reggeon
in each of the initial and final wee gluon configurations. The accompanying
reggeon state contains two gluons - which can give the imaginary part
of the single reggeon state that is anticipated to survive
in the pomeron\cite{arw02}.
In both Fig.~6.1(a) and (b) the three
multi-reggeon (pomeron) states that are interacting through the anomaly 
all have a wee-gluon component that participates in the divergence.
In the notation of (\ref{chm1})-(\ref{chm30}) the corresponding
basic anomaly process involves (as already discussed in Section 3) 
taking the limit $l \to 0$ 
while simultaneously making a boost $a_z(\zeta)$
such that $ l \cosh\zeta =n$ is kept finite.

The diagram that gives the
pion/pomeron coupling utilized in \cite{arw02} is 
shown in Fig.~6.2(a). The corresponding
basic anomaly process is shown in Fig.~6.2(b)
\newline \noindent \parbox{2.9in}{
\begin{center}
\epsfxsize=1.8in
\epsffile{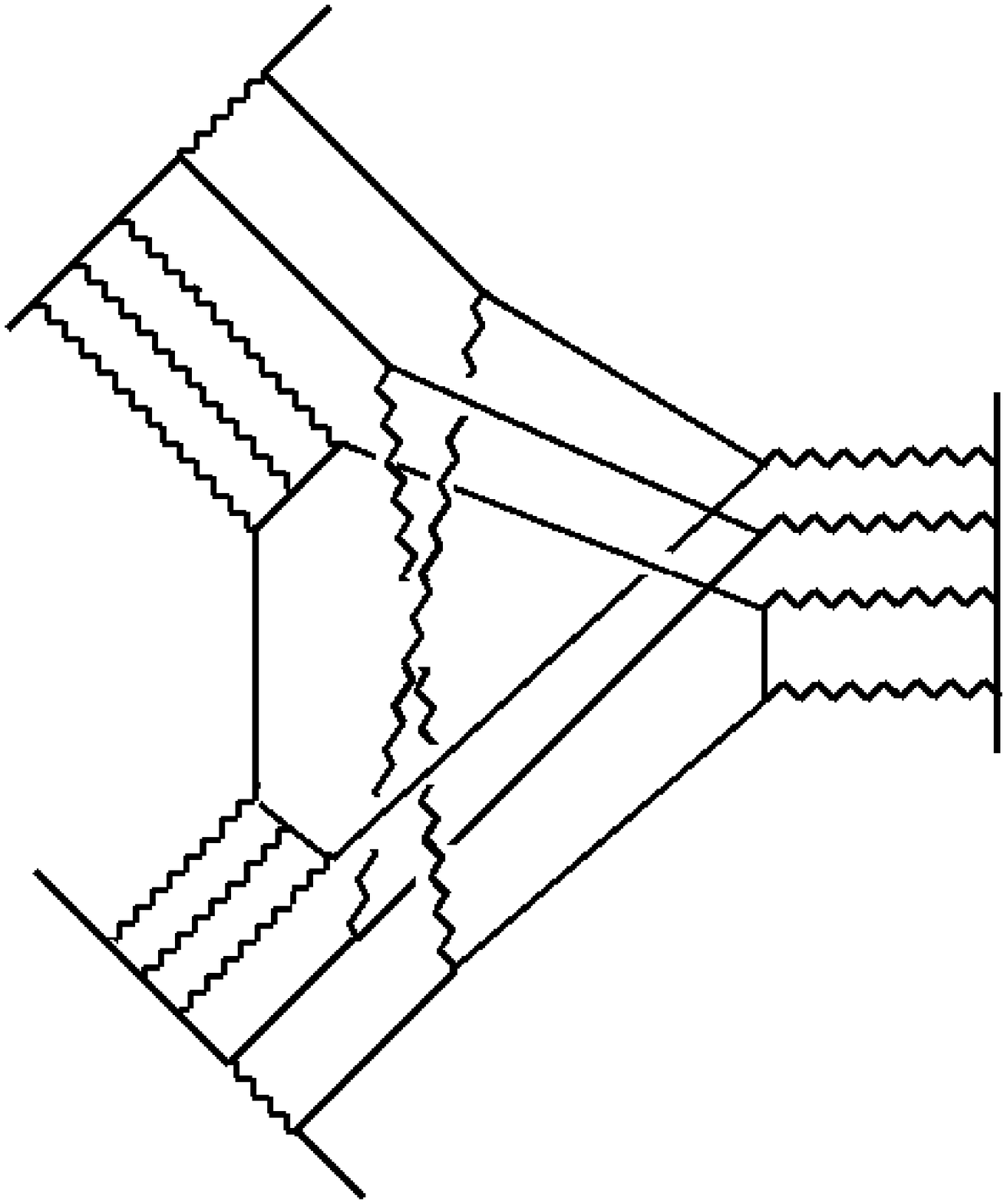}
\newline (a)
\end{center}}
\parbox{2.9in}{
\begin{center}
\epsfxsize=1.9in
\epsffile{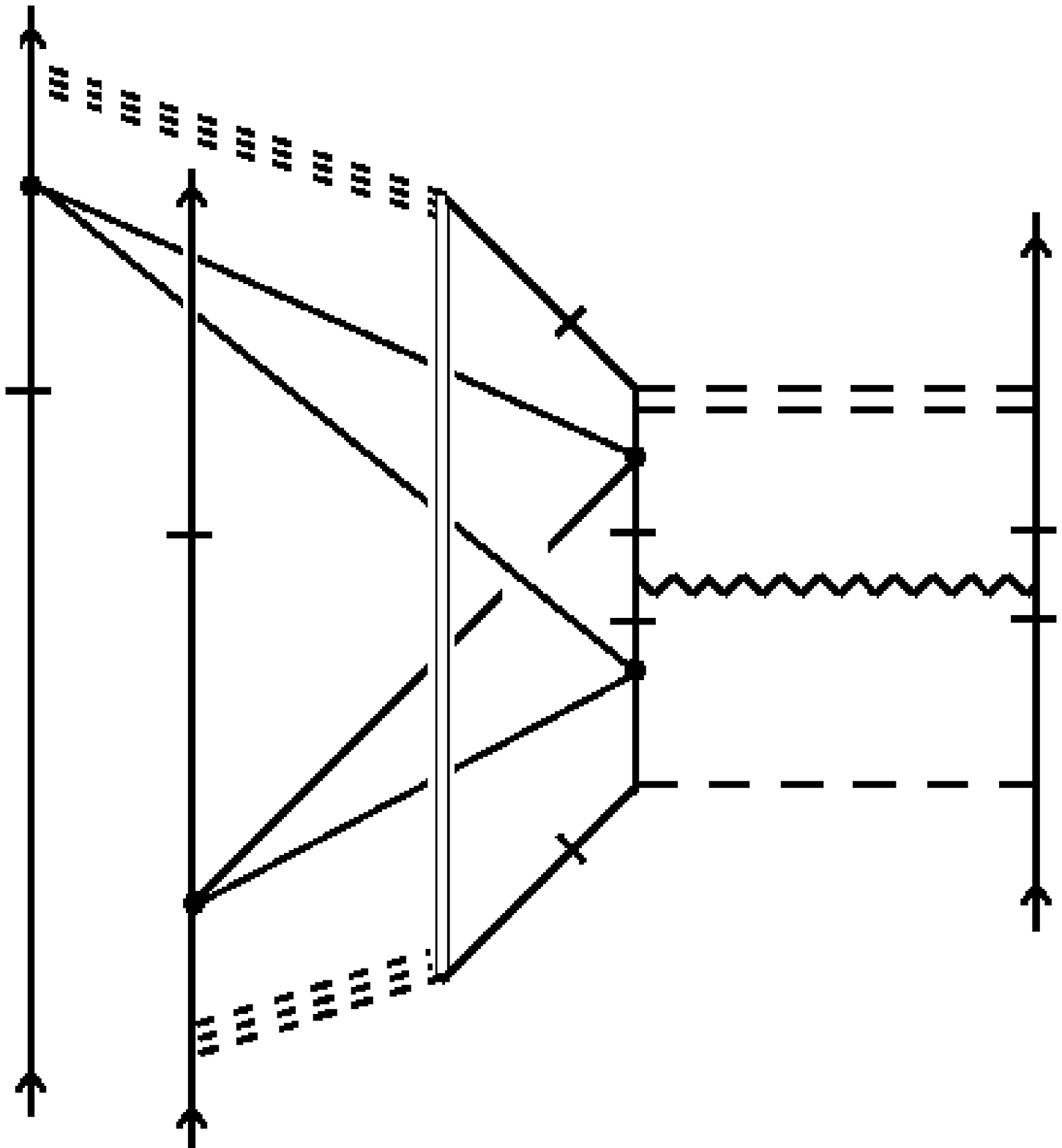}
\newline (b)
\end{center}}
\begin{center}
Fig.~6.2 The Pion/Pomeron Coupling (a) the feynman diagram (b) the
basic anomaly process
\end{center}
The diagram of Fig.~6.2(a) has a triple discontinuity
structure very similar to that of Fig.~5.6.

\newpage

\renewcommand{\theequation}{A.\arabic{equation}}
\setcounter{equation}{0}
\vskip 1cm \noindent
\noindent {\bf APPENDIX:~ ASYMPTOTIC DISCONTINUITY ANALYSIS }
\vskip 3mm \noindent

In Section 4 we analyse triple-regge asymptotic discontinuities
using a generalization of the simple light-cone analysis
that we develop in the following.

Consider the box-diagram illustrated in Fig.~A1. 
\begin{center} 
\leavevmode
\epsfxsize=2.5in
\epsffile{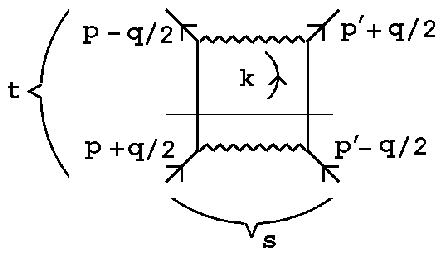}

Fig.~A1 The box diagram.
\end{center}
Initially we ignore the role played by 
numerators and so we consider, in the notation shown,
$$
\eqalign{             
I(s,t,m^2) = &\int d^4k \left[k^2-m^2+i\epsilon\right]^{-1}
\left[\left(p-{q\over 2}+k\right)^2-m^2+i\epsilon\right]^{-1}\cr
&\times \left[(q-k)^2-m^2+i\epsilon\right]^{-1}\left[\left(
p' +{q\over 2}-k\right)^2-m^2+i\epsilon\right]^{-1}.}
\auto\label{lcan1}
$$
This integral is, of course, a function of invariants only even though it is
specified using four momenta. Indeed, we can evaluate the integral using
complex, unphysical, momenta that give physical values of the invariants,
provided we are careful to define the integral via analytic continuation from
the appropriate physical momentum region. Our purpose in this Section is to
discuss momentum dependence of this kind for the simplifying
case of the leading asymptotic behavior, in a manner that we apply 
to much more complicated diagrams in Section 4.

For illustrative purposes we set both $q=0$ and $m=0$ in (\ref{lcan1}) 
and ignore infra-red divergences. We can then write
$$
I(s) ~= ~\int d^4k \left[k^2+i\epsilon\right]^{-2}
\left[\left(p +k\right)^2 +i\epsilon\right]^{-1}
\left[\left(p'  -k\right)^2+i\epsilon\right]^{-1}
\auto\label{lcan2}
$$
We choose a particular Lorentz frame and introduce light-cone co-ordinates 
such that 
$$
\eqalign{
p&~= ~\left({P_+ \over 2}~,{P_+ \over 2}~,~\til{0}\right) 
~ +O\left({1\over  s}\right), ~~~~~P_+ ~\sim ~ s ~\to ~ \infty \cr
p'&~=~  \left({P_+'+ P_-' \over 2}~, {P_+'- P_-' \over 2}~,~
\underline{p}_{\perp}'~\right)   }
\auto\label{lcan3}
$$
so that $s~ = P_+P_-'~[1 + O(1/ s)]~$. We can then write
$$
\eqalign{I(s)~\centerunder{$\large\sim$} {\raisebox{-3mm} 
{$\scriptstyle s\to \infty$}}~~
{1\over 2}\int &                                        
d^2\underline{k}_{\,\perp} dk_+dk_- \left[ k_+k_- -k^2_{\,\perp} + 
i\epsilon\right]^{-2}~
\bigg[ \left( k_+ + P_+ \,\right) k_-
- \underline{k}^2_{\,\perp}
+i\epsilon\bigg]^{-1}  \cr
& \times~
\left[\left(k_+ -P_+' \right)\left(k_- -P_-' \right)- 
(\underline{k}_{\,\perp} - \underline{p}_{\,\perp}')^2
+i\epsilon\right]^{-1} }
\auto\label{lcan4}
$$

To obtain a non-zero answer by closing the $k_+$ contour, with $k_-$ and 
$k_{\perp}$ fixed, the three poles given by the three square brackets of 
(\ref{lcan4}) must
not be on the same side of the contour. This requires $ 0 <k_-< P_-'$ and,
in this case, the $k_+$ contour can be closed to pick up only
the pole in the last bracket. This gives 
$$
k_+ ~=~ P_+' +
{(\underline{k}_{\,\perp} - \underline{p}_{\,\perp}')^2
- i\epsilon \over \left(k_- -P_-' \right)}
\auto\label{lcan40}
$$
which is finite and so can be neglected compared to $P_+$. Note also that 
$$
k_- \sim 0,~~ {k_{\perp}}^2 \sim 0 ~~=> ~~k_+~\sim ~2k_0~ \sim 
~{{p'}^2 \over P_-'}
\auto\label{5an}
$$
(we use this approximation in the analysis of 
Section 4). We thus obtain, 
$$
I(s)~\centerunder{$\large\sim$} {\raisebox{-3mm} 
{$\scriptstyle s\to \infty$}}~~\pi i\int 
d^2\underline{k}_{\perp} \left[ -k^2_{\perp} + 
i\epsilon \right]^{-2} ~\int_0^{P_-'} dk_-  
\left[k_- -P_-' \right]^{-1}\left[ P_+ k_-
- \underline{k}^2_{\perp}
+i\epsilon \right]^{-1} 
\auto\label{lcan5}
$$

We are specifically interested in the leading real and imaginary parts of 
(\ref{lcan5}). They are given by the logarithm generated by 
the pole factor containing $P_+$ as it approaches the $k_- = 0$ end-point of 
the integration. If we keep only the integration over $0 < k_- < \lambda P_-'$ 
and take $\lambda << 1$ so that we can 
make the approximation $k_- / P_-'~ \sim 0$ we obtain
$$
\eqalign{ I(s)~& \centerunder{$\large\sim$} {\raisebox{-3mm} 
{$\scriptstyle s\to \infty$}}~~
\pi i\int 
d^2\underline{k}_{\perp} \left[ -\underline{k}^2_{\perp} + 
i\epsilon \right]^{-2} ~
{1 \over P_-'} ~\int_0^{\lambda P_-'} dk_-  
\left( P_+ k_-
- \underline{k}^2_{\perp}
+i\epsilon \right)^{-1} \cr 
&~\sim ~{1 \over P_+ P_-'} ~[\log{(P_+P_-'\lambda  
 -\underline{k}^2_{\perp} + i\epsilon]~J_1(0)} \cr 
&~\sim ~{1 \over s} ~[\log{(s\lambda + i\epsilon]~J_1(0)}
~ \sim ~{1 \over s} 
~[\log{s} + i\pi]~J_1(0) }
\auto\label{lcan60}
$$
where $J_1(0) ~\sim ~ \int 
d^2\underline{k}_{\perp} \left[ -\underline{k}^2_{\perp} + 
i\epsilon \right]^{-2}$ is infinite, but would be finite if we added a 
mass to the particle propagators. 

As we have indicated, the sign of the imaginary part in (\ref{lcan60}) arises 
directly from the $i\epsilon$ prescription. To obtain the leading imaginary 
part or, equivalently, the leading behavior of the discontinuity in $s$,
it suffices to keep the $i\epsilon$ dependence while dropping the 
$ -\underline{k}^2_{\perp}$ dependence in the $k_-$ integral. 
(\ref{lcan60}) is, of course, independent of $\lambda$. It will, however, be 
useful to note the role of $\lambda$ with respect to the 
analytic structure of $I(s)$ in the $s$-plane. As illustrated in Fig.~A2,
the finite end of the branch-cut asociated with the logarithm in 
(\ref{lcan60}) moves out as $\lambda \to 0$.
\begin{center}
\leavevmode
\epsfxsize=1.6in
\epsffile{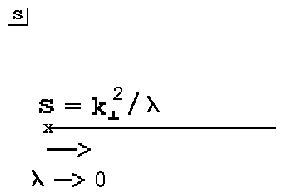}

Fig.~A2 $\lambda$-dependence of the branch cut.

\end{center}
This is irrelevant to the asymptotic behavior and the 
``asymptotic discontinuity'' clearly remains unchanged. We, nevertheless, 
exploit this 
simple feature in evaluating multiple discontinuities in Section 4.
Also, although (\ref{lcan60}) is an 
invariant result, for
our purposes it will be useful to keep the dependence on both $P_+$ and $P_-'$
and discuss the dependence of the phase on $P_+$.

The initial $k_-$ integration contour for (\ref{lcan60})
is as shown in Fig.~A3(a) with the pole at 
$ k_- = \underline{k}_{\perp}^2/ P_+$ indicated by a dot.
\begin{center}
\leavevmode
\epsfxsize=4.5in
\epsffile{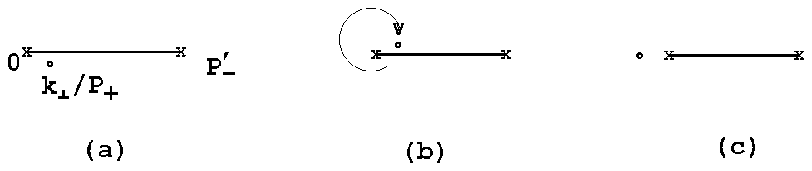}

Fig.~A3 Integration Contours for (a) (\ref{lcan5})~ (b) 
~$P_+ \to~ e^{2\pi i}P_+$ ~(c) Fig.~4.4.
\end{center}
As $P_+$ (and therefore $s$)
completes a circle in the complex plane the pole 
moves around the end-point as illustrated in Fig.~A3(b). The result is 
that the phase of the logarithm in (\ref{lcan60}) changes from $\pi$ 
to $-\pi$ and there is a net discontinuity of $2\pi i / s$, as is given 
directly by (\ref{lcan60}).
This is also the result that would be 
obtained by applying directly the standard
cutting rules to Fig.~A1, cut by the thin line, if 
the $k_+$ and $k_-$ integrations are used to put the vertical lines 
on shell. The above discussion is simply an asymptotic analysis of 
how the two cut propagators pinch the integration region to generate 
a branch-point in $s$. Introducing $\lambda$ limits the integration 
region for the original integral such that the pinching only takes place for
$s \sim P_+ ~> 1 /\lambda$. Note also that the residue function $J_1(0)$,
multipying the logarithm in (\ref{lcan60}),
is directly obtained from the original box
diagram by putting the cut lines giving the discontinuity on-shell using 
the longitudinal momentum integrations. This is a very simple example 
(the simplest) of the relationship between a discontinuity and asymptotic
behavior.  

In evaluating unphysical (multiple) discontinuities in 
Section 4 we do not assume 
that the standard cutting rules apply. Instead we directly analyse 
the discontinuities produced by logarithms. To understand how a discontinuity
generated by a logarithm can provide leading asymptotic behavior 
we note that the twisted diagram of Fig.~A4, for
$q=0$, differs from that of Fig.~A1 only by $P_+ \to -P_+$.
\begin{center}
\leavevmode
\epsfxsize=2in
\epsffile{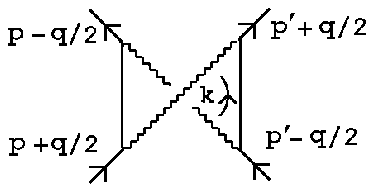}

Fig.~A4 The twisted box diagram.
\end{center}
As a result,
the integration contour and pole position of Fig.~A3(a) is replaced by 
that of Fig.~A3(c). In this case a discontinuity is generated 
for $s< 0$. For
$s>0$ there is no phase generated by Fig.A4 and only the real
logarithms cancel when this diagram is added to that of Fig.~A1. 
The leading behavior 
of the discontinuity in $s$, i.e the imaginary part, produced by the
diagram of Fig.~A1 remains. 
This cancelation of the logarithms is very well-known, of course. 
It is also well known
that the cancelation fails when a non-abelian symmetry group is present
and that a consequence is the reggeization of the gluon.

We can briefly summarize the effect of adding numerators to (\ref{lcan1})
as follows.
First we note that the numerator of the internal fermion propagator 
carrying $P_+$ gives an additional $P_+$ factor of the form $\gamma_- P_+$. 
As a consequence, in (\ref{lcan60}), there is the replacement 
$$
~\int_0 dk_- ~
\left( P_+ k_- + \cdots \right)^{-1} ~~ \to ~\gamma_- P_+ ~ 
\int_0 dk_-  \left( P_+ k_- + \cdots \right)^{-1} ~\sim~ \log{P_+}
\auto\label{lcan61}
$$
and there is no inverse power of $P_+$. Also, each
coupling to a gluon gives a $\gamma$ matrix factor and since the external 
fermion lines are on-shell we can use the asymptotic form of 
the Dirac equation (i.e. $ \gamma_- P_+ \psi ~\sim m ~\psi $) to write
$$
\eqalign{ <P_+|\gamma_{\mu}\gamma_- \gamma_{\nu}|P_+>~& \sim ~
<P_+|{\gamma_- P_+ \over m} ~\gamma_{\mu}\gamma_- \gamma_{\nu}~
{\gamma_- P_+ \over m}|P_+> \cr
&  =~ <P_+|P_+ \gamma_- P_+ |P_+> / m^2~~
\sim P_+ ~/m }
\auto\label{coup}
$$
This gives another power of $P_+$ ($\sim s$) provided that the corresponding
factor of $P_-'$ is present in the finite momentum part of the scattering 
process. Not surprisingly this factor emerges from that part 
which would dominate if $P_-'$ were large. However, we want to 
emphasize that this selection is made only by the need to form a Lorentz 
invariant amplitude from the non-invariant large momentum process.

Finally we note that the above analysis goes through with very little
modification if we take both $m^2$ and $q$ to
be non-zero so that (\ref{lcan2}) will not be infra-red divergent.

\end{document}